\documentclass[twocolumn,a4,showpacs,preprintnumbers,amsmath,amssymb]{revtex4}
\usepackage{hyperref}
\usepackage{amssymb}
\usepackage{amsmath}
\usepackage{multirow}
\usepackage{color}
\usepackage{tipa}
\usepackage{mathrsfs}
\usepackage{longtable}

\usepackage{graphicx}
\usepackage{dcolumn}
\usepackage{bm}
\usepackage{mathptmx, courier, pifont}
\usepackage[scaled=0.92]{helvet}
\usepackage[T1]{fontenc}
\usepackage{textcomp}

\begin{document}


\title{Systematical shell-model calculation in the pairing-plus-multipole
Hamiltonian with a monopole interaction for the $pf_{5/2}g_{9/2}$
shell}

\author{K.~Kaneko}
\email{kaneko@ip.kyusan-u.ac.jp} \affiliation{Department of Physics,
Kyushu Sangyo University, Fukuoka 813-8503, Japan}
\author{T.~Mizusaki}
\affiliation{Institute of Natural Sciences, Senshu University, Tokyo
101-8425, Japan}
\author{Y.~Sun}
\email{sunyang@sjtu.edu.cn} \affiliation{Department of Physics and
Astronomy, Shanghai Jiao Tong University, Shanghai 200240, China}
\affiliation{Collaborative Innovation Center of IFSA (CICIFSA),
Shanghai Jiao Tong University, Shanghai 200240, China}
\affiliation{State Key Laboratory of Theoretical Physics, Institute
of Theoretical Physics, Chinese Academy of Sciences, Beijing 100190,
China }
\author{S.~Tazaki}
\affiliation{Department of Applied Physics, Fukuoka University, Fukuoka
814-0180, Japan}

\begin{abstract}

The recently-proposed effective shell-model interaction, the
pairing-plus-multipole Hamiltonian with the monopole interaction
obtained by empirical fits starting from the monopole-based
universal force (PMMU), is systematically applied to nuclei of the
$pf_{5/2}g_{9/2}$ shell region. It is demonstrated that the
calculation describes reasonably well a wide range of experimental
data, including not only the low-lying and high-excitation spectra,
$E2$ transitions, quadrupole moments, and magnetic moments, but also
the binding energies, for Ni, Cu, Zn, Ga, Ge, As, and Se isotopes
with $A=64-80$. In particular, a structure of the neutron-rich Ge
and Se isotopes is discussed in detail.

\end{abstract}

\pacs{21.10.Dr, 21.60.Cs, 21.60.Jz, 21.10.Re}

\maketitle

\section{Introduction}\label{sec1}

Nuclei in the $pf_{5/2}g_{9/2}$ shell, i.e. those with nucleon
numbers between the magic numbers 28 and 50, show at least two
attractive properties. The first one is a structure variation
around the semi-magic shell-closure at $N=40$. It has been known
that the $Z=28$ isotone $^{68}$Ni has one of the main characters
of a double closed-shell nucleus \cite{Bernas82,Broda95}, and isomers
have been found in the neighborhood of this nucleus. The lowered
first excited $0_{2}^{+}$ state in $^{68}$Ni is interpreted so
as to support the magicity. However, $^{68}$Ni does not exhibit
an irregularity in two-neutron separation energy as expected for
magic nuclei \cite{Guenaut07}. It has been shown that this
contradiction can be understood by the parity change across the
$N=40$ shell gap \cite{Guenaut07,Mueller99, Sorlin02,Langanke03}.
Moving to heavier isotones of $N=40$ with increasing proton number,
there have been many experimental evidences suggesting an increasing
collectivity, for example those in the Ge \cite{Luttke12,Gurdal13}
and Se isotopes \cite{Hurst07,Ljungvall08,McCutchan11,McCutchan13}.
With $Z$ approaching 40, the $N=Z$ isotone $^{80}$Zr is known to be
strongly deformed \cite{Lister87}, and signs of a shell closure
disappear completely and the nucleus can be regarded as a good rotor.

Another interesting property to be mentioned for the
$pf_{5/2}g_{9/2}$ shell is the shape evolution in the mass-70 region
\cite{Ljungvall08,McCutchan11,McCutchan13,Doring98,Ennis91,Fischer00}.
In particular, the nuclei around the $N=Z$ line exhibit a variety of
nuclear shapes \cite{Kaneko04,Obertelli09}. Shape changes are
sensitive probes of the structure, and the corresponding nuclei can
serve as good testing grounds for shell-model interactions. For
example, $^{64,66,68}$Ge were studied using large-scale shell model
calculations \cite{Kaneko02,Hasegawa07,Honma09}. For
$^{70,72,74,76}$Ge, shape changes were discussed with the observed
spectroscopic quadrupole moment \cite{Robinson11}, which were
attributed to the large shell gaps at the prolate and oblate
deformation for proton and neutron numbers 34 and 36 as seen in the
Nilsson diagram. For $^{68}$Se, it was found that the oblate
ground-state band coexists with a prolate excited band, forming a
shape coexistence phenomenon \cite{Fischer00,Obertelli09}. Shape
coexistence was also investigated for $^{70,72,74}$Se
\cite{Ljungvall08,McCutchan11,McCutchan13}. It was reported that in
$^{70}$Se, the oblate shape near the ground state evolves quickly to
a prolate shape at higher spins \cite{Ljungvall08}. Further
experiments suggested that in $^{72}$Se, there is no well-defined
shape for the lowest levels due to shape mixings \cite{McCutchan11},
and the low-lying states in $^{74}$Se show coexistence of spherical
and prolate shapes \cite{McCutchan13}.

Various theoretical approaches based on deformed mean-fields have
been applied to this mass region
\cite{Petrovici02,Sun04,Sun05,Hinohara09,Yang10}. On the other hand,
shell-model calculations using different effective interactions,
such as JUN45 \cite{Honma09} and JJ4B
\cite{Padilla05,Lisetskiy04,jj4b} for the $pf_{5/2}g_{9/2}$ model
space, have also been reported. In principle, interactions for
nuclear shell-model calculations should be derived microscopically
from the free nucleon-nucleon forces \cite{Kuo68,Jensen95} and also
the three-body nucleon forces \cite{Epelbaum09}. However, it has
been shown that such effective interactions fail to describe nuclear
properties such as binding energies, excitation spectra, and $E2$
transitions once the number of valence nucleons becomes large.
Therefore, considerable effort has been put forward to construct
effective interactions for different shell-model spaces
\cite{Brown88,Brown06,Poves01,Honma04,Padilla05,
Lisetskiy04,jj4b,Sorlin02,LNPS}. For the JUN45 and JJ4B effective
interactions, the starting point for the fitting procedure is a
realistic G-matrix interaction based on the Bonn-C $NN$ potential
and core-polarization corrections with a $^{56}$Ni core. With the
inclusion of the proton $f_{7/2}$ and neutron $d_{5/2}$ orbitals, in
addition, the LNPS \cite{Sorlin02,LNPS} interaction based on the
Kuo-Brown interaction including the Kahana-Lee-Scott potential
\cite{Kahana69} has also been proposed.

Among the effective interactions, the pairing plus quadrupole-quadrupole
($P+QQ$) interaction \cite{Kisslinger63,Bes69} has been widely
applied to describe various nuclear properties, such as excitation
energies, moments, transitions, and reaction rates, for a wide range
of nuclei in the medium to heavy mass regions. This interaction is
represented by two basic components, the pairing and quadrupole
forces as the short and long range parts of the interaction,
respectively. Dufour and Zuker have shown that any
realistic effective interaction is dominated by the $P+QQ$
interaction with the monopole terms \cite{Dufour96}. It has been
understood that, while the pairing and quadrupole terms take care of
the main and smooth part of the structure properties, the monopole
terms play important roles for the shell evolution and are often
responsible for explaining anomalous behaviors in spectra and
transitions. It has been shown that the extended $P+QQ$ model with
the monopole interaction ($EPQQM$) works surprisingly well for the
$N\approx Z$ nuclei \cite{Hasegawa01,Kaneko02}, where the monopole
terms are treated as corrections. The strength parameters in $EPQQM$
have been chosen so as to fit the known data. In a series of
publications, we have found that the monopole terms are important to
account for the unusual shell evolution in the neutron-rich region.
The model has also demonstrated its capability of describing the
microscopic structure in different $N\approx Z$ nuclei, as for
instance, in the $pf$-shell \cite{Hasegawa01} and the
$pf_{5/2}g_{9/2}$-shell regions \cite{Kaneko02}. It has been shown
that this model is also applicable for the neutron-rich nuclei in
the $fpg$-shell region \cite{Kaneko08} and the $sd$-$pf$ shell
region \cite{Kaneko11}. Quite recently, the $EPQQM$ model has also
been successfully applied to the neutron-rich nuclei around
$^{132}$Sn \cite{Jin11,Wang13,Wang14,Wang15}.

One of the important issues in nuclear structure is persistence or
disappearance of the traditional magic numbers when moving away from
the $\beta$-stability line. It has been known that the conventional
magic numbers disappear in some cases of the neutron-rich region, but
new magic numbers may emerge. For example, the neutron-rich nuclei
$^{12}$Be, $^{32}$Mg, and $^{42}$Si were found to exhibit large
collectivity in spite of the corresponding neutron magic numbers
$N=$ 8, 20, and 28. The monopole interaction is the key ingredient
for explaining the binding energies, the emergent magic numbers,
and the shell evolution in the neutron-rich region. The connection
between the monopole interaction and the tensor force \cite{Otsuka01}
was confirmed within the self-consistent mean-field model using the
Gogny force \cite{Otsuka05}. Thus one of the physical origins of the
monopole interaction was attributed to the tensor force \cite{Otsuka01}
which explains the shell evolution. This explanation is an
important development in understanding the structure of unknown mass
regions. Recently, general properties of the monopole components of
the effective interactions have been presented through introducing
the monopole-based universal interaction \cite{Otsuka10b}, which
consists of simple central and tensor forces and can produce a
variety of shell evolution across different mass regions. It has
been shown that the monopole matrix elements obtained from this
interaction are in good agreement with those of the SDFP-M in the
$sd$ shell and of the GXPF1A in the $pf$ shell \cite{Honma04}. The
monopole-based universal interaction seems to be really a universal
one, applicable for different nuclear shell regions. Thus it is of a
great interest to perform shell-model calculations with the pairing
and multipole interactions, starting from the monopole part constructed
from the monopole-based universal interaction.

The present authors have recently proposed a unified realistic
shell-model Hamiltonian called PMMU \cite{Kaneko14}, employing the
$P+QQ$ Hamiltonian with the monopole interaction
$V_{m}^{MU}$ constructed from the monopole-based universal force. It
was demonstrated \cite{Kaneko14} that PMMU describes well nuclear
properties of the $pf$ and $pf_{5/2}g_{9/2}$ shell nuclei, such as
systematics of the first excited $2^{+}$ states and $B(E2)$ values,
and detailed energy spectrum for $^{56}$Ni, $^{72}$Ge, $^{55}$Co,
and $^{69}$Ge. It is now important to investigate general nuclear
properties for a considerable amount of nuclei including binding
energies, detailed energy levels, and $E2$ transitions,
to confirm further the reliability of the PMMU model. The present
article is a comprehensive work following Ref. \cite{Kaneko14}, and
we perform large-scale shell-model calculations systematically for
nuclei in the $pf_{5/2}g_{9/2}$ model space. We shall show that
starting from the monopole interaction $V_{m}^{MU}$, a set of 28
parameters including four $P+QQ$ force strengths, 14 monopole matrix
elements, and four single-particle energies are determined by
refitting to the experimental binding energies of 91 nuclei with the
mass $A=64-80$. Data for excited states are also used to fit these
parameters. We will discuss the shell evolution, binding energies,
excited energy spectrum and $E2$ transitions. Thus, the
main purpose of the present work is to test the PMMU model for a
wide range of nuclei in the $pf_{5/2}g_{9/2}$ shell region, with a
particular attention paid to the structure of Ge and Se isotopes.

The paper is arranged as follows. In Sec.~\ref{sec2}, we outline our
model. In Section~\ref{sec3}, we present the model parameters with
the detailed fitting procedure and description of binding energies.
The rest sections are devoted to discussion of the results: the
structure of low-lying states are discussed in Section~\ref{sec4}
and structure of highly excited states in Ge and Se isotopes are
presented in Section~\ref{sec5}. Finally, summary and conclusions
are given in Section~\ref{sec6}.

\section{The PMMU model}\label{sec2}

The Hamiltonian of the PMMU model, proposed by the present authors
in Ref. \cite{Kaneko14}, takes the following form
\begin{eqnarray}
 H & = & H_{0} + H_{PM}  + H_{m},  \\ \nonumber \\
          \label{eq:1}
 H_{0} & = & \sum_{\alpha} \varepsilon_a c_\alpha^\dag c_\alpha, \nonumber \\
 H_{PM}  & = &
    -  \sum_{J=0,2} \frac{1}{2} g_J \sum_{M\kappa} P^\dag_{JM1\kappa} P_{JM1\kappa}  \nonumber \\
 && - \frac{1}{2} \chi_2 \sum_M :Q^\dag_{2M} Q_{2M}:
         - \frac{1}{2} \chi_3 \sum_M :O^\dag_{3M} O_{3M}:  \nonumber \\
 H_{m}  & = &  \sum_{a \leq b, T} V_{m}(ab,T) \sum_{JMK}A^\dagger_{JMTK}(ab) A_{JMTK}(ab), \nonumber
\end{eqnarray}
where $H_{0}$ is the single-particle Hamiltonian, $H_{PM}$ the
pairing plus multipole term, and $H_{m}$ the monopole term. For
$H_{PM}$, we take the $J=0$ and $J=2$ forces in the pairing channel,
and the quadrupole-quadrupole ($QQ$) and octupole-octupole ($OO$)
forces in the particle-hole channel \cite{Hasegawa01,Kaneko02}.
Higher order pairing and multipole terms can be added if necessary.
The monopole matrix elements $V_{m}(ab,T)$ in $H_{m}$ are defined as
\cite{Poves81}
\begin{eqnarray}
V_{m}(ab,T) & = & \frac{\sum_{J}(2J+1)V_{ab,ab}^{JT}}{\sum_{J}(2J+1)},
          \label{eq:2}
\end{eqnarray}
where $V_{ab,ab}^{JT}$ are the diagonal matrix elements between the
two-nucleon states coupled to angular momentum $J$ and isospin $T$.
The monopole interaction has connections with the tensor force,
which generally explains the shell evolution \cite{Otsuka05}. It was
pointed out that the three-nucleon forces in the monopole
interaction are also important for neutron-rich nuclei
\cite{Zuker03,Otsuka10b,Sieja12}. Recently, Otsuka {\it et al.} have
discussed universal properties of the monopole interaction in the
effective interaction \cite{Otsuka10a} and suggested the so-called
monopole-based universal force \cite{Otsuka10b} as an $NN$
potential, $V_{MU}$, which consists of the Gaussian central force
and the tensor force. The $V_{MU}$ force has been successfully
applied to light nuclei \cite{Yuan12,Utsuno12} and somewhat heavier
ones \cite{Togashi15}.

\begin{figure}[b]
\includegraphics[totalheight=6.0cm]{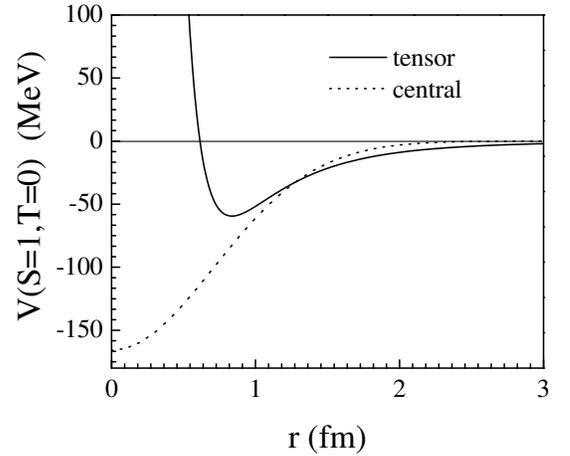}
  \caption{Triplet-even potential due to the central force and the tensor force.}
  \label{fig1}
\end{figure}

In a recent publication \cite{Kaneko14}, we have proposed a unified
realistic shell-model Hamiltonian employing the pairing plus
multipole Hamiltonian with the monopole interaction
constructed from $V_{MU}$, subject to modifications through fitting
the known experimental data. The monopole-based universal force
consists of two terms, the Gaussian central force and the tensor
force
\begin{equation}
V_{MU} = V_{\rm central} + V_{\rm tensor},
 \label{eq:3}
\end{equation}
with
\begin{eqnarray}
V_{\rm central} & = & \sum_{S,T}f_{S,T}P_{S,T}{\rm exp}(-(r/\mu)^{2}), \nonumber \\
V_{\rm tensor} & = & (\vec{\tau}_{1}\cdot \vec{\tau}_{2})
([\vec{s}_{1} \vec{s}_{2}]^{(2)}\cdot Y^{(2)})g(r), \nonumber
\end{eqnarray}
where $S(T)$ means spin (isospin), $P$ denotes the projection
operator onto the channels $(S, T)$ with strength $f$, and $r$ and
$\mu$ are the internucleon distance and Gaussian parameter,
respectively. Here $\vec{\tau}_{1,2}$ $(\vec{s}_{1,2})$ denotes the
isospin (spin) of nucleons. In the central force, the Gaussian
parameter is fixed to be $\mu =1.0$ fm, and the strength parameters
are $f_{0,0}=f_{1,0}=-166$, $f_{0,1}=-99.6$, and $f_{1,1}=132.8$
(all in MeV). The $\pi + \rho$ meson exchange force is used for the
strength $g(r)$ \cite{Otsuka05}. Figure \ref{fig1} illustrates the
triplet-even potential due to the Gaussian central force $V_{\rm
central}$ and the tensor force $V_{\rm tensor}$ with the above
parameters. The monopole Hamiltonian $H_{m}$ in Eq. (1) can
then be rewritten in the known form \cite{Bansal64,Poves81}
\begin{eqnarray}
H_{m}= \sum_{a \leq b} \left[  r_{ab}\hat{n}_{a}(\hat{n}_{b}-\delta_{ab})
 + s_{ab}(\hat{T}_{a}\cdot \hat{T}_{b}-\frac{3}{4}\hat{n}_{a}\delta_{ab}) \right],
          \label{eq:4}
\end{eqnarray}
with
\begin{eqnarray}
r_{ab} & = & \frac{3V_{m}(ab,T=1)+V_{m}(ab,T=0)}{4(1+\delta_{ab})}, \nonumber \\
s_{ab} & = & \frac{V_{m}(ab,T=1)-V_{m}(ab,T=0)}{1+\delta_{ab}}. \nonumber
          \label{eq:5}
\end{eqnarray}

\section{The PMMU interaction and the ground state description}
\label{sec3}

In this section and the two subsequent sections, we present the
results obtained from systematical shell-model calculations with the
PMMU model, and compare the results with available experimental
data. The present section discusses how the PMMU interaction parameters
are determined. The two subsequent sections are devoted to discussions
of systematical calculations for low-lying states and for states of
high excitations, respectively. Most of the shell-model calculations
shown in the present paper are carried out by using the shell-model
code MSHELL64 \cite{Mizusaki}, which enables calculations with a
M-scheme dimension up to $\sim 3\times 10^{9}$.

\subsection{Determination of the PMMU interaction}

In Ref. \cite{Kaneko14}, we have shown that the PMMU model is
successful in describing energy levels and $E2$
transitions for a large number of nuclei in the $pf$- and
$pf_{5/2}g_{9/2}$-shell regions. In this paper, we consider
$pf_{5/2}g_{9/2}$ as the model space. Starting from the monopole
interaction $V_{m}^{MU}$, a set of 28 parameters, which include the
four strengths for the pairing-plus-multipole forces, twenty for the
monopole matrix elements, and four single-particle energies, are
determined by fitting to the known experimental data of 91 nuclei.
These nuclei are taken from the mass region with $A=64-80$,
including $^{64-76}$Ni, $^{64-78}$Cu, $^{65-80}$Zn, $^{66-80}$Ga,
$^{69-80}$Ge, $^{67-78}$As, and $^{73-80}$Se. For the Ni, Cu, Zn,
Ga, and Ge isotopes, calculations are performed without any
truncation, while for As and Se isotopes, some truncations have to
be introduced.

\begin{figure}[t]
\includegraphics[totalheight=7.0cm]{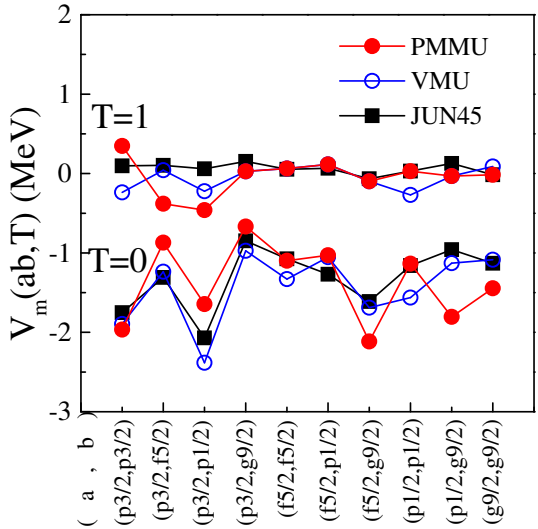}
  \caption{(Color online) Comparison of the monopole matrix elements
$V_{m}(ab,T)$ among the effective interactions JUN45, VMU, and PMMU.
}
  \label{fig2}
\end{figure}

\begin{figure}[t]
\includegraphics[totalheight=9.0cm]{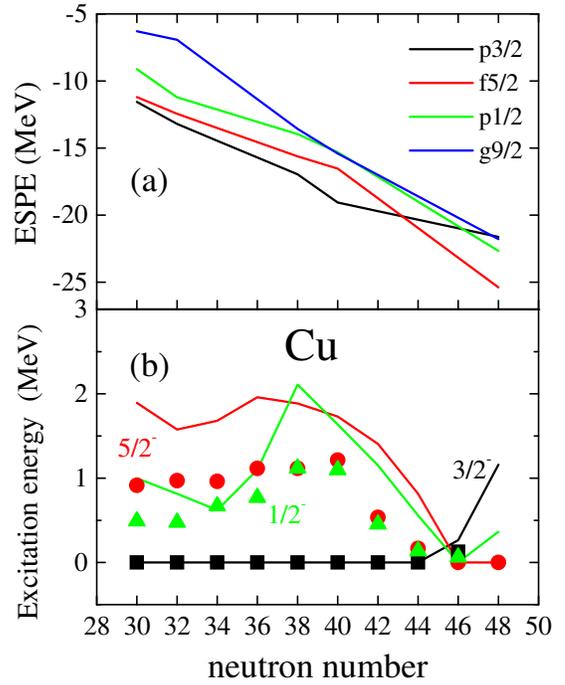}
  \caption{(Color online) (a) Effective single-particle energies of proton orbits
for Cu isotopes. (b) Comparison of the energy levels of the lowest
$3/2^{-}$, $5/2^{-}$, and $1/2^{-}$ states for odd-mass Cu isotopes.
The filled symbols and the lines are the experimental data and the
shell-model results, respectively. Experimental data are taken from
Ref. \cite{ENDSF}.}
  \label{fig3}
\end{figure}

The detailed fitting procedure carried out in Ref. \cite{Kaneko14}
is as follows. We take the pairing plus multipole Hamiltonian, and
for the monopole part, we start from $V^{MU}_{m}(ab,T)$ constructed
by the monopole-based universal force of Otsuka {\it et al.}
\cite{Otsuka10b} and modify the monopole matrix elements so as to
fit the experimental data. The experimental yrast states of the
even-even Zn, Ge and Se isotopes and the low-lying $1/2^{-}$,
$3/2^{-}$, $5/2^{-}$, and $9/2^{+}$ states of the odd-mass Ni, Zn,
Ge, Se are considered for the fitting. In total, 91 binding energy
data and 186 experimental energies of excited states from these 91
nuclei are taken in the fitting procedure. For example, the four
pairing-plus-multipole force strengths, the four single-particle
energies, and the $T=1$ matrix elements
$V_{m}^{MU}(p_{3/2},f_{5/2},T=1)$,
$V_{m}^{MU}(p_{3/2},p_{3/2},T=1)$, and
$V_{m}^{MU}(p_{3/2},p_{1/2},T=1)$ are modified to fit mainly the
yrast energy levels for $^{60-64}$Zn and $^{64-68}$Ge. The $T=0$
matrix elements are adjusted to fit the binding energies of the
nuclei with $28 \le Z \le 34$ and $30 \le N \le 50$. As a result,
the rms deviations for binding and excitation energies (see
discussions below) are 691 keV and 256 keV, respectively.

All together, a total of 14 monopole terms are modified from the
original $V^{MU}_{m}$, and all the monopole matrix elements are
scaled by a factor $(58/A)^{0.3}$ for the calculation with the
present model space. The modified monopole matrix elements, denoted
as $V^{\rm PMMU}_{m}$, are shown in Fig. \ref{fig2}. The original
monopole matrix elements, $V^{MU}_{m}$, are also displayed in the
same figure, which are found for most cases to be closer to those of
the JUN45 effective force \cite{Honma09}. Thus from Fig. \ref{fig2},
it can be seen that our fitted $V^{\rm PMMU}_{m}$ is clearly
different from the original one in the following matrix elements:
$V_{m}^{MU}(p_{3/2},p_{1/2},T=0)$,
$V_{m}^{MU}(f_{5/2},g_{9/2},T=0)$,
$V_{m}^{MU}(p_{3/2},f_{5/2},T=0)$,
$V_{m}^{MU}(g_{9/2},g_{9/2},T=0)$, $V_{m}^{MU}(p_{1/2},p_{1/2},T=0)$
and $V_{m}^{MU}(p_{1/2},g_{9/2},T=0)$ for $T=0$, and
$V_{m}^{MU}(p_{3/2},f_{5/2},T=1)$, $V_{m}^{MU}(p_{3/2},p_{3/2},T=1)$
and $V_{m}^{MU}(p_{3/2},p_{1/2},T=1)$ for $T=1$.

The modifications of the $T=0$ monopole matrix elements between the $fp$
shell and the $g_{9/2}$ orbit are particularly important for the shell
evolutions seen in the shell-structure changes due to the filling of
specific single-particle orbits. The single-particle energies and
interaction strengths in Eq. (1) are taken from Ref. \cite{Kaneko14}
as follows (all in MeV)
\begin{eqnarray}
 & {} &  \varepsilon_{p3/2} =-9.40, \varepsilon_{f5/2} =-8.29,\varepsilon_{p1/2} =-7.49, \nonumber \\
 & {} &  \varepsilon_{g9/2} =-5.70,       \nonumber \\  \nonumber \\
 & {} &  g_0 = 18.0/A, \quad g_2 = 0.0,            \nonumber \\
 & {} &  \chi_2 = 334.0/A^{5/3},  \quad \chi_3 = 259.2/A^{2}.
\label{eq:8}
\end{eqnarray}

The modifications for the $T=0$ monopole matrix elements
$V_{m}^{\rm PMMU}(f_{5/2},g_{9/2},T=0)$ and
$V_{m}^{\rm PMMU}(p_{1/2},g_{9/2},T=0)$ cause significant shell-variations
with increasing neutron number. The proton effective single-particle
energies (ESPE) as a function of neutron number are shown in Fig.
\ref{fig3} (a). Due to the large difference between the $T=0$ matrix
elements $V_{m}^{\rm PMMU}(p_{3/2},g_{9/2},T=0)$ and
$V_{m}^{\rm PMMU}(f_{5/2},g_{9/2},T=0)$, the ESPE of the proton $f_{5/2}$
orbit is pushed down drastically relative to the $p_{3/2}$ orbit as
the neutron $g_{9/2}$ orbit is occupied for $N > 40$, and becomes
lower than the other orbits for $N > 46$. This shell variation
reflects the behavior of the low-lying $5/2^{-}$ energy level of Cu
isotopes as shown in Fig. \ref{fig3} (b), where the calculation
reproduces the experimental data reasonably. As another example, the
experimentally observed feature of the near-degeneracy of the
$5/2^{-}$ and $1/2^{-}$ states for $N > 40$ is reproduced very well.
This is due to the large $T=0$ matrix element
$V_{m}^{\rm PMMU}(p_{1/2},g_{9/2},T=0)$, which is adjusted to have a
similar magnitude to the one of $V_{m}^{\rm PMMU}(f_{5/2},g_{9/2},T=0)$.

\subsection{Description of binding energies}

\begin{figure}[t]
\includegraphics[totalheight=6.0cm]{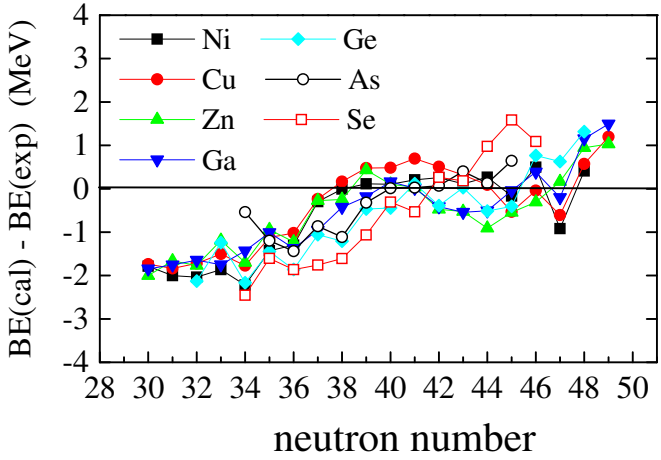}
  \caption{(Color online) Deviations of the calculated binding energies
  from the experimental data for $Z=28-34$ as functions of neutron number.
  Data are taken from Ref. \cite{Audi03}.}
  \label{fig4}
\end{figure}

The total binding energies are expressed as
\begin{eqnarray}
 & {} &  B(Z,N) = E_{SM}(Z,N) + E_{C}(Z,N) + BE(^{56}{\rm Ni}),  \nonumber \\
\label{eq:9}
\end{eqnarray}
where $E_{SM}(Z,N)$ is the shell-model ground-state energy
calculated by the present PMMU model, $E_{C}(Z,N)$ the Coulomb
energy, and $BE(^{56}{\rm Ni})$ the binding energy of the $^{56}{\rm
Ni}$ core \cite{Audi03}. The Coulomb energy is evaluated by using
the empirical formula
\begin{eqnarray}
 & {} &  E_C(\pi,\nu) = \varepsilon_{C}\pi + \frac{\pi(\pi - 1)}{2}V_{C} +
\left( \frac{1}{2}\pi \right)b_{C} + \Delta_{np}\pi\nu,  \nonumber \\
\label{eq:10}
\end{eqnarray}
where $\pi$ and $\nu$ denote the numbers of valence protons and
neutrons, respectively. We use the parameter set 2 in Table I of
Ref. \cite{Cole99}, where the values of $\varepsilon_{C}$, $V_{C}$,
$b_{C}$, and $\Delta_{np}$ in Eq. (\ref{eq:10}) are determined so as
to fit the observed Coulomb displacement energies for the mass
region of $20 < Z < 42$ and $32 < N < 50$.

In Fig. \ref{fig4}, deviations of the calculated binding energies
from the experimental values are shown for various isotopic chains
as functions of neutron number $N$. All the results are obtained by
the shell-model calculations without any truncation for Ni, Cu, Zn, Ga,
Ge, As, and Se isotopes. The binding energies of As and Se isotopes
are calculated for $N=$ 33, 34, 48, 49 and $N=$ 34, 35, 48, 49,
respectively. The overall agreement with data is quite good for the
entire mass range included in the fitting ($A=58-83$), except that
one finds relatively large deviations for nuclei below $N=36$ where
the calculations give an overbinding. One can expect that the
calculation may also be applicable to other nuclei that have not
been included in the fitting procedure.

\section{Structures of low-lying states}\label{sec4}

To test the validity of a new effective shell-model interaction, it
is important to examine it through systematical calculations for the
low-lying states of many nuclei. In this section, the first excited
$0_{2}^{+}$, $2_{1}^{+}$, $4_{1}^{+}$ states, and other low-lying
states are discussed.

\begin{figure}[b]
\includegraphics[totalheight=10.0cm]{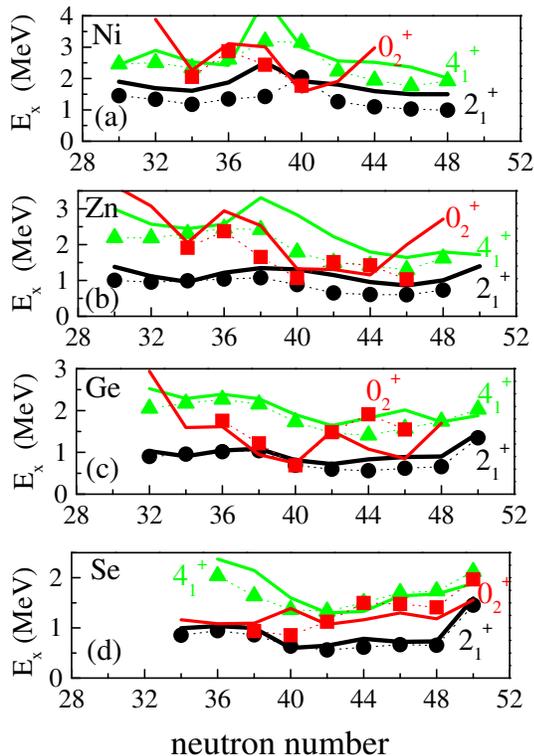}
  \caption{(Color online) Systematics in the excitation energies of the first
excited $0^{+}$, $2^{+}$, and $4^{+}$ states for Ni, Zn, Ge, and Se
isotopes. The $0^{+}$, $2^{+}$, and $4^{+}$ states are indicated by
the squares, circles, and triangles, respectively. The calculated
results are shown by the solid lines with the same colors, and compared
with the experimental data taken from Ref. \cite{ENDSF}.}
  \label{fig5}
\end{figure}

\begin{figure}[t]
\includegraphics[totalheight=10.0cm]{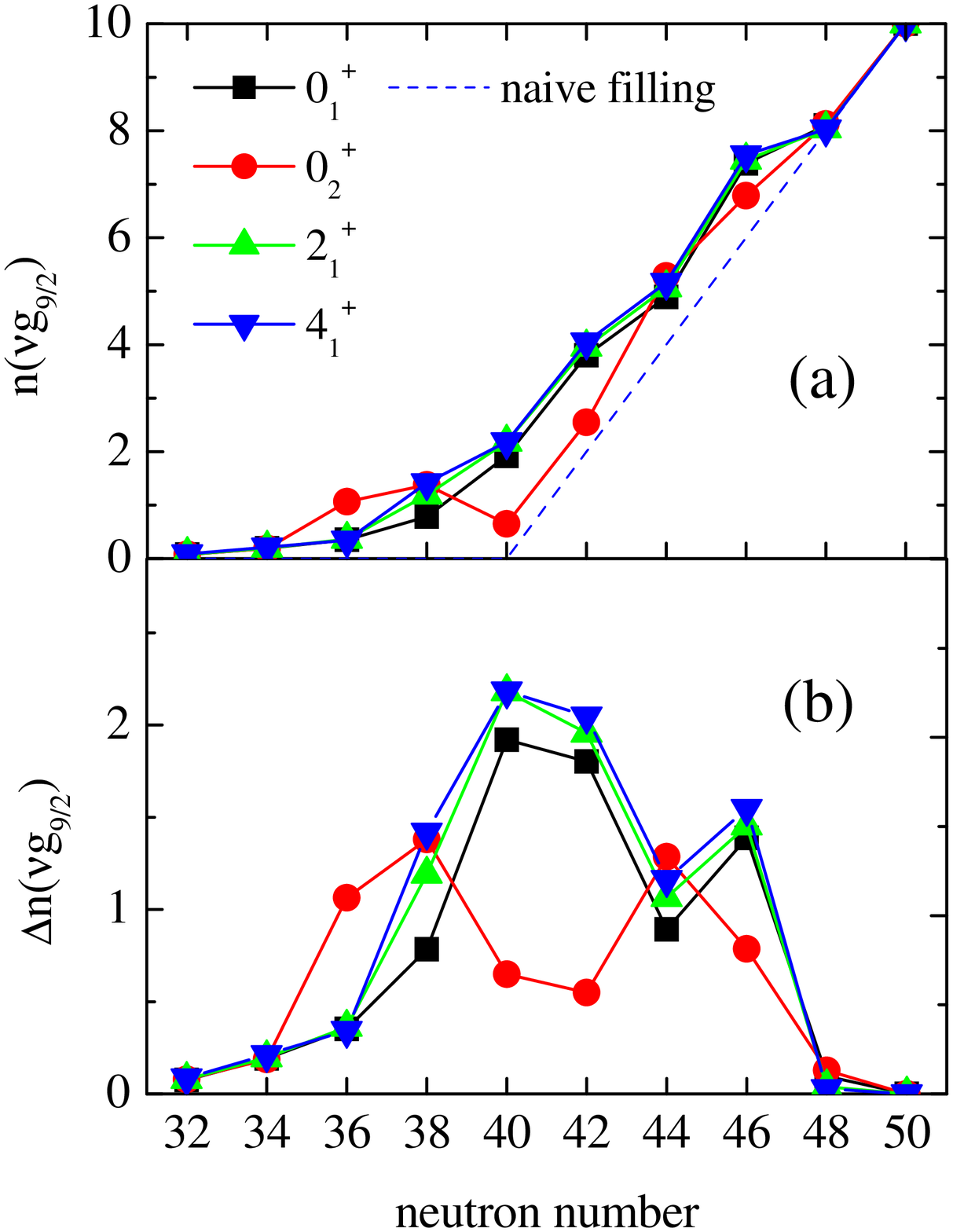}
  \caption{(Color online) Neutron $g_{9/2}$ occupancies for the low-lying states
in the Ge isotopes. (a) The occupation numbers $n(\nu g_{9/2})$ of
the low-lying states in the Ge isotopes. The broken line indicates
the naive filling configuration. (b) The differences between the
neutron occupation numbers and the naive filling configurations.}
  \label{fig6a}
\end{figure}

\begin{figure}[t]
\includegraphics[totalheight=10.0cm]{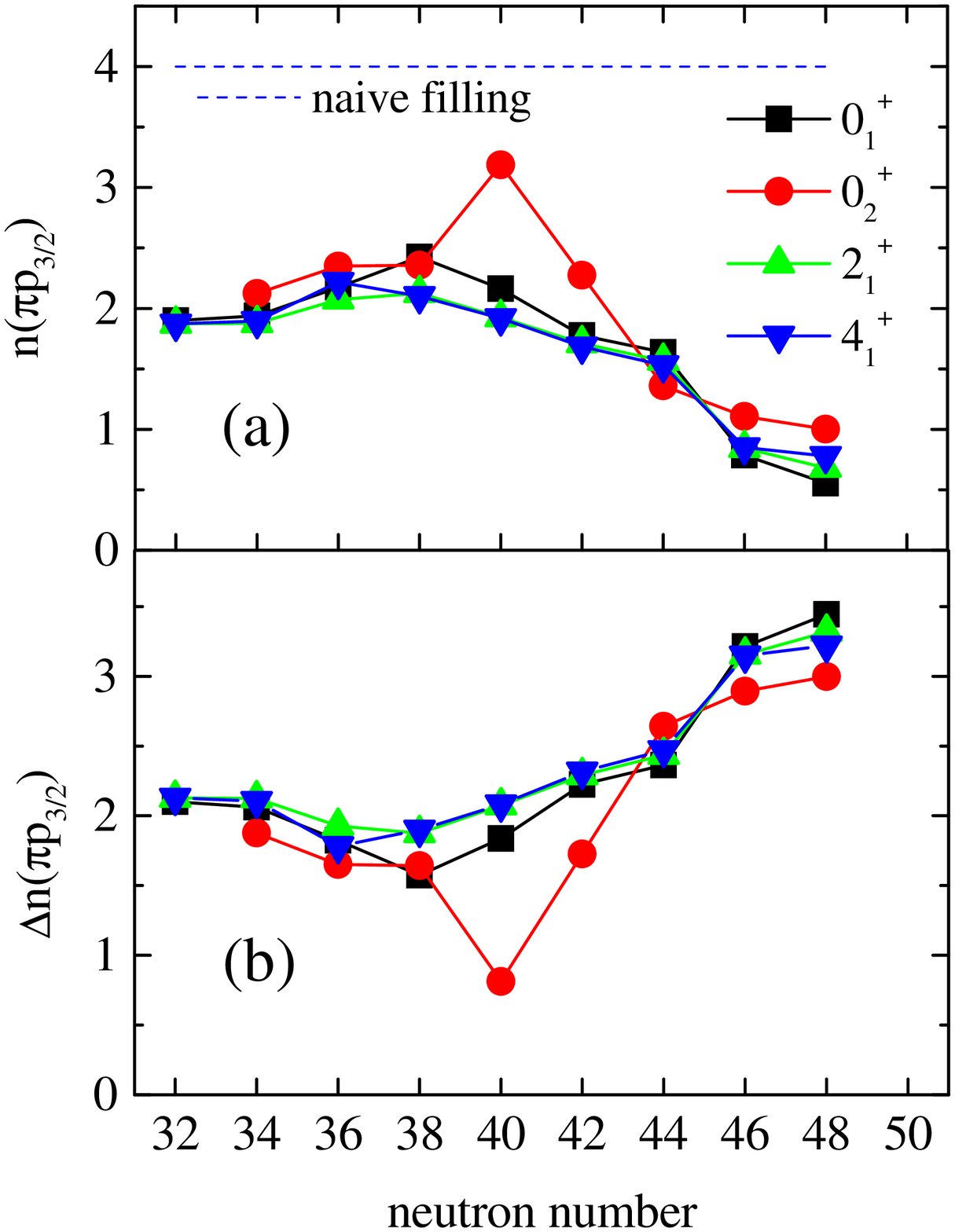}
  \caption{(Color online) Proton $p_{3/2}$ occupancies for the low-lying states
in the Ge isotopes. (a) The occupation numbers $n(\nu p_{3/2})$ of
the low-lying states in the Ge isotopes. The broken line indicates
the naive filling configuration. (b) The differences between the
neutron occupation numbers and the naive filling configurations.}
  \label{fig6b}
\end{figure}

\begin{figure}[b]
\includegraphics[totalheight=10.0cm]{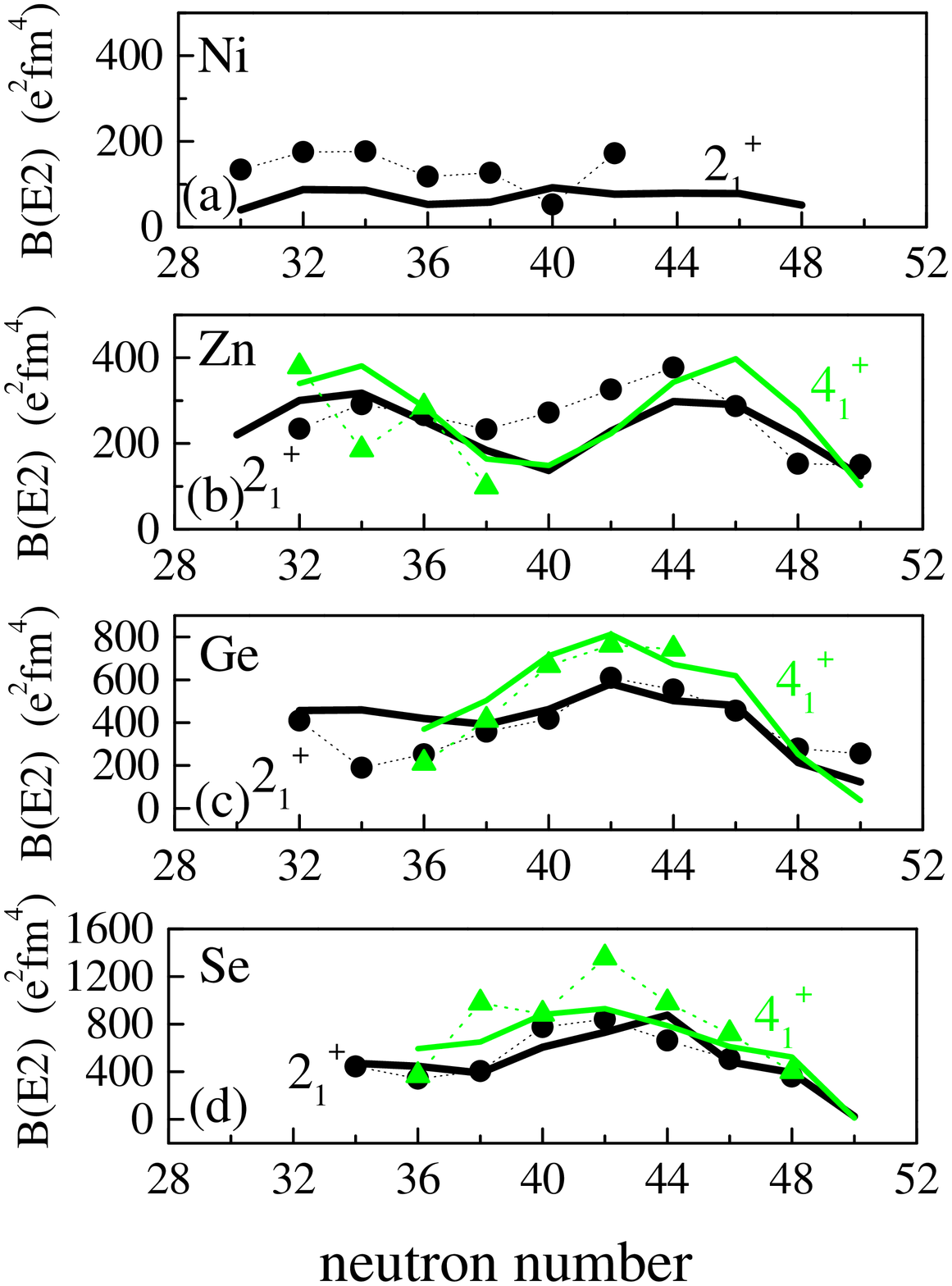}
  \caption{(Color online) Systematics in the $B(E2;2_{1}^{+}\rightarrow 0_{1}^{+})$ and
$B(E2;4_{1}^{+}\rightarrow 2_{1}^{+})$ values for Ni, Zn, Ge, and Se
isotopes. The calculated results are compared with the experimental
data taken from Ref. \cite{ENDSF}, where the $2_{1}^{+}$ and
$4_{1}^{+}$ data are indicated by circles and triangles,
respectively. The effective charges for proton and neutron are taken
as 1.5 and 1.1, respectively.}
  \label{fig6}
\end{figure}

\subsection{Systematics of excited
$2_{1}^{+}$, $4_{1}^{+}$, and $0_{2}^{+}$ states}

The systematics of the $2_{1}^{+}$, $4_{1}^{+}$ and $0_{2}^{+}$
states are studied for a wide range of even-even nuclei in this mass
region. Figure \ref{fig5} shows the excitation energies of these
states for Ni, Zn, Ge, and Se isotopes. The lightest isotope in each
isotopic chain is taken to be the one with $N=Z$. Overall, the
calculations reproduce the experimental data fairly well. The common
discrepancy of the calculation is that beyond $N=40$ in each
isotopic chain, the theoretical $2_{1}^{+}$ energies are higher than
the experimental ones. This would suggest that the present model
space is insufficient to describe the systematic behavior beyond
$N=40$, possibly due to the missing $d_{5/2}$ orbit. Other
discrepancies are seen in the description of the $2_{1}^{+}$ states
of the Ni isotopes and the $4_{1}^{+}$ states of the Zn isotopes. As
for the Ni isotopes, the observed excitation energy
$E_{x}(2_{1}^{+})$ takes the largest value at $N=40$ while the
calculation shows a peak at $N=38$. These could be attributed to the
missing $f_{7/2}$ orbit in the present model space. If the $f_{7/2}$
orbit had been included, the attractive $T=0$ monopole force between
the proton $f_{7/2}$ and neutron $p_{1/2}$ orbits would push down
the neutron $p_{1/2}$ orbit, causing an increase of the shell gap
between the neutron $g_{9/2}$ and $p_{1/2}$ orbits, which could
result in a large excitation energy $E_x(2_1^{+})$ at $N=40$ as seen
in the experiment.

We now focus the discussion on the interesting behavior of the first
excited $0_{2}^{+}$ state. As can be seen in Fig. \ref{fig5}, the
common feature is that the excitation energy of the $0_{2}^{+}$
states changes drastically along each isotopic chain and reaches the
minimum at $N=40$. In particular for $^{72}$Ge, the $0_{2}^{+}$
state lies below the $2_{1}^{+}$ state. Thus this is a mysterious
$0_{2}^{+}$ state for the Ge isotopes. There have been many
theoretical attempts to explain this behavior. However, most shell
model approaches did not succeed to answer this question. The
shell-model calculation using the JUN45 effective interaction
successfully reproduced the irregular behavior of the $0_{2}^{+}$
state \cite{Honma09}. Comparing the theoretical results and the
experimental data shown in Fig. \ref{fig5}, one can see that the
present calculation describes correctly the trend of systematics of
the $0_{2}^{+}$ states for all the Ni, Zn, Ge, and Se isotopes.

We can explain how this systematical trend is obtained in our
calculation. Figure \ref{fig6a} shows the occupation numbers, $n(\nu
g_{9/2})$, of the neutron $g_{9/2}$ orbit in the yrast $0_{1}^{+}$,
$2_{1}^{+}$, and $4_{1}^{+}$ states, and in the excited $0_{2}^{+}$
state. As one can see from Fig. \ref{fig6a} (a), $n(\nu g_{9/2})$ in
all the three yrast states increase monotonously with the neutron
number. On the other hand, $n(\nu g_{9/2})$ of the $0_{2}^{+}$ state
shows a different behavior: the increasing trend turns back at
$N=40$ where the minimum is seen. Moreover, the $n(\nu g_{9/2})$
curve of the $0_{2}^{+}$ state deviates clearly from the yrast ones
at $N=40$ and 42, but behaves as the broken line in Fig. \ref{fig6a}
(a) that represents the naive filling configuration. The additional
occupancy of the $g_{9/2}$ orbit to the naive filling occupancy
$n(\nu g_{9/2})_{0}$, shown by the broken line in Fig. \ref{fig6a}
(a), is considered as a measure of excitation from the $pf_{5/2}$
shell to the $g_{9/2}$ orbit. The magnitude of the additional
occupancy $\Delta n(\nu g_{9/2}) = |n(\nu g_{9/2}) - n(\nu
g_{9/2})_{0}|$ can be seen more clearly in Fig. \ref{fig6a} (b). The
enhanced excitations of the yrast states increase the collectivity
in the isotopes of $N=40$ and 42. In contrast, such excitations are
largely suppressed for the $0_{2}^{+}$ state in these two isotopes.
Thus the magicity of the $N=40$ leads to a lowering of
the $0_{2}^{+}$ state toward $N=40$. Similar conclusions were given
by Ref. \cite{Honma09}.

On the other hand, the proton occupation numbers $n(\pi p_{3/2})$ in
the $p_{3/2}$ orbit are shown in Fig. \ref{fig6b} (a). One sees
again the distinct behavior of the $0_{2}^{+}$ state from the
$0_{1}^{+}$, $2_{1}^{+}$, $4_{1}^{+}$ yrast states. For the yrast
states, $n(\pi p_{3/2})$ are around two for the isotopes below
$N=40$, but start to decrease with increasing neutron number after
$N=40$. As the occupation number for the naive filling configuration
is four, two protons are excited from the $p_{3/2}$ orbit to the
upper orbits below $N=40$, and the proton excitation is gradually
enhanced starting from $N=42$. For the heavier isotopes, additional
protons are excited to the $f_{5/2}$ orbit, generally because of the
deformation and the pairing effect. For the excited $0_{2}^{+}$
state, however, the proton occupancy indicates a sudden jump at
$N=40$ before starting to decrease after that neutron number. Figure
\ref{fig6b} (b) reinforces the discussion in a form of $\Delta n(\pi
p_{3/2}) = |n(\pi p_{3/2}) - n(\pi p_{3/2})_{0}|$.

Thus we can explain the lowering of the excited $0_{2}^{+}$ energy
at $N=40$ (see Fig. \ref{fig5}) as a result of cooperative effects
of magicity at the closed-shell configuration for proton and
neutron. The magicity of the $N=40$ subshell gives rise to the
lowering of the excited $0_{2}^{+}$ state of the Ge isotopes at this
neutron number. When protons occupy the $p_{3/2}$ orbit at $N=40$,
the neutron $p_{1/2}$ orbit is lowered due to the $T=0$ attractive
monopole force $V_{m}^{\rm PMMU}(p_{3/2},p_{1/2},T=0)$ between the
proton $p_{3/2}$ and the neutron $p_{1/2}$ orbits. This results in
both increases of the neutron shell gap between the $p_{1/2}$ and
$g_{9/2}$ orbits and the proton shell gap between the $f_{5/2}$ and
$p_{3/2}$ orbits, which lowers the excitation energy of the first
excited $0_{2}^{+}$ state. 
For the Se isotopes, the calculated $0_{2}^{+}$ energy increases at
$N=40$, in contrast to the experimental data. In this isotopic
chain, the monopole interaction between the proton $f_{5/2}$ orbit
and neutron $g_{9/2}$ orbit results in a decrease of the proton
shell-gap at $N=40$ because the proton $f_{5/2}$ and neutron
$g_{9/2}$ orbits are lowered due to this monopole force. In fact,
for $^{74}$Se the proton $p_{3/2}$ occupation numbers are 3.4 and
2.8 for the $0_{2}^{+}$ and other states, respectively. The
difference is not large and is about half of that for $^{72}$Ge (see
Fig. \ref{fig6b} (a)). This monopole component may be too strong for
the lowering of the $0_{2}^{+}$ state at $N=40$ in the Se isotopes.
For the ground-state band, the neutron occupation number $n(\nu
g_{9/2})$ increases smoothly in Fig. \ref{fig6a} (a) due to the
drastic increase of $\Delta n(\nu g_{9/2})$. This means that the
neutron shell-gap at $N=40$ is washed out due to the strong pairing
and quadrupole-quadrupole force. The ground-state band thus varies
smoothly without showing any shell changes.

Finally, we discuss $E2$ transition probabilities within the
low-lying yrast states. Figure \ref{fig6} compares the calculated
$B(E2,2_{1}^{+}\rightarrow 0_{1}^{+})$ and
$B(E2,4_{1}^{+}\rightarrow 2_{1}^{+})$ with available experimental
data for Ni, Zn, Ge, and Se isotopes. The effective charges in the
$E2$ calculation are taken as $e_{p}=1.5$ and $e_{n}=1.1$ for
protons and neutrons, respectively. The choice of a larger neutron
effective charge than the standard one ($e_{n}=0.5$) gives a better
agreement for most of the data points. There are however
discrepancies. For the Ni isotopes, the calculated $B(E2)$ values
are smaller than the experimental ones except for $N=40$. This could
be due to the polarization effect from the $^{56}$Ni core. The
contributions from the $f_{7/2}$ shell may be important for
enhancement of the $B(E2)$ values. For the Zn isotopes, most
experimental data are reproduced except for those near $N=40$, where
the calculated $B(E2)$ values are too small. As seen from the ESPE
in Fig. \ref{fig3}, the proton shell gap between the $p_{3/2}$ and
the $f_{5/2}$ orbits increases particularly at $N=40$. The
calculated small $B(E2)$ results reflect this increase of the shell
gap. The results for the Ge isotopes are overall in good agreement
with the experimental data except for $N=34$ and $N=36$, where the
calculated $B(E2)$ values are larger than the experiment. This may
indicate that the shell gap between the $p_{3/2}$ and the $f_{5/2}$
orbits is too small in the calculation. For the Se isotopes, the
calculation reproduces the experimental $B(E2)$ values pretty well.

\subsection{Other low-lying states of even-even nuclei}

\begin{figure*}[t]
 \begin{center}
\includegraphics[scale=0.35]{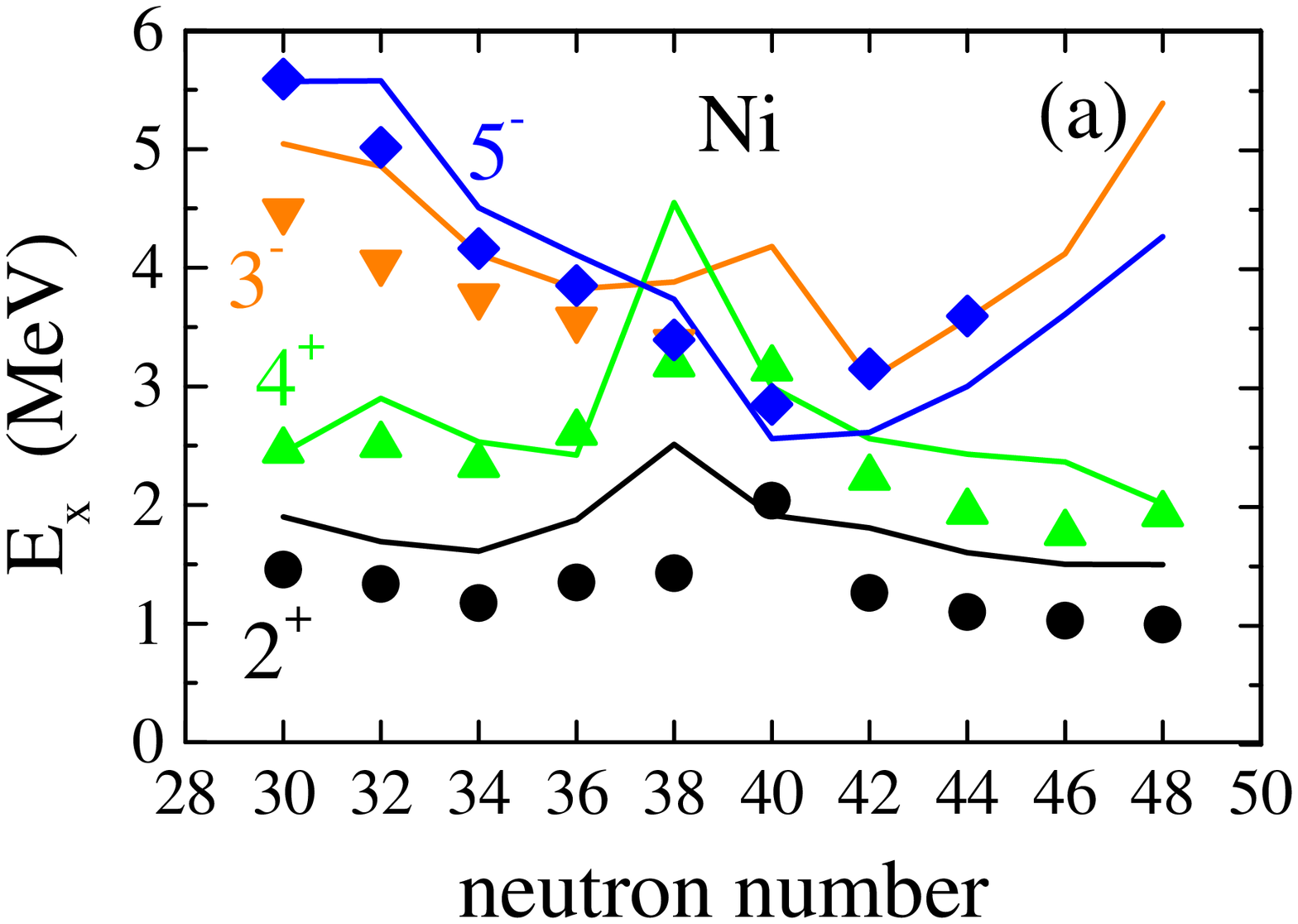}
\includegraphics[scale=0.35]{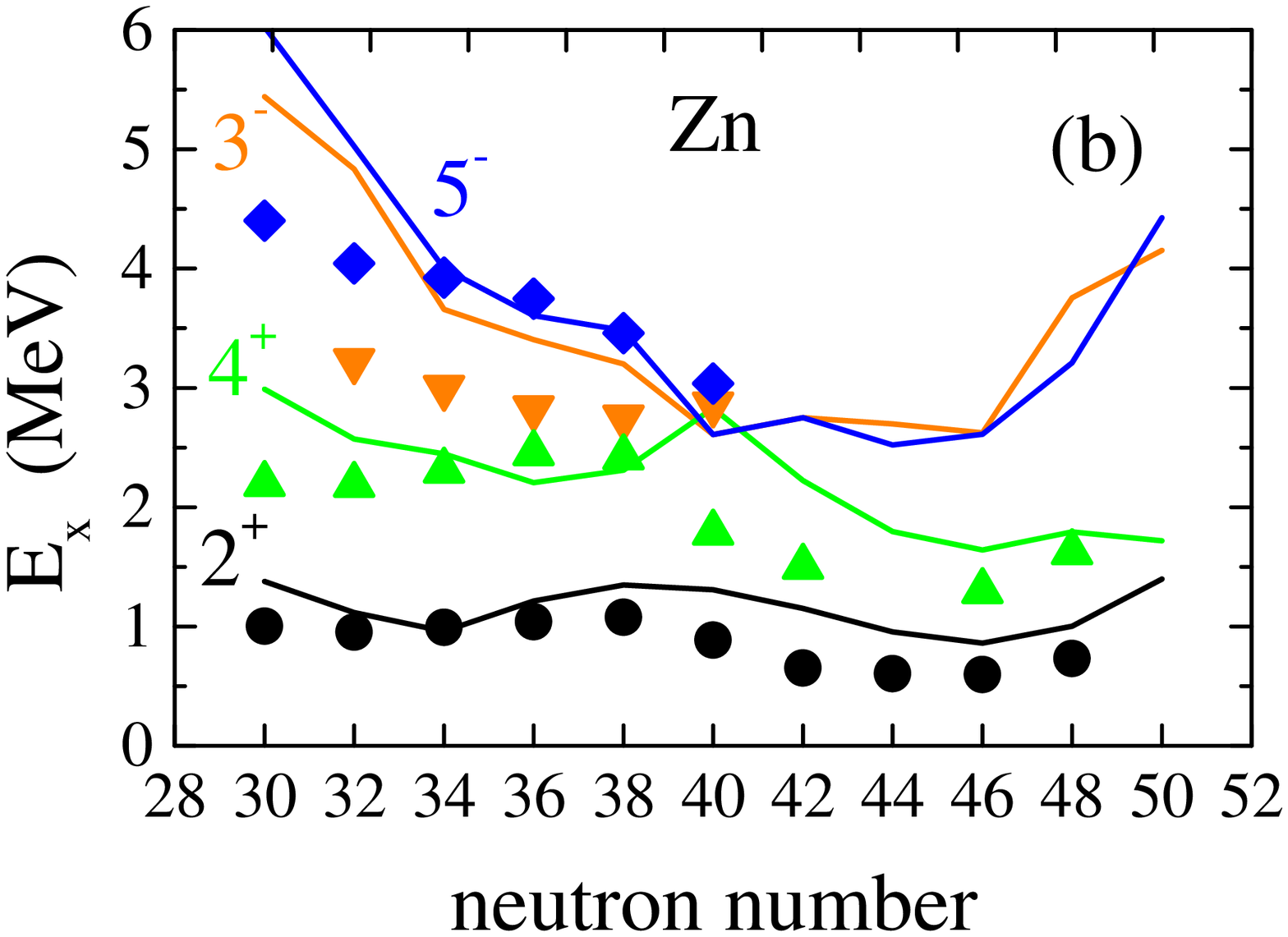}
\includegraphics[scale=0.35]{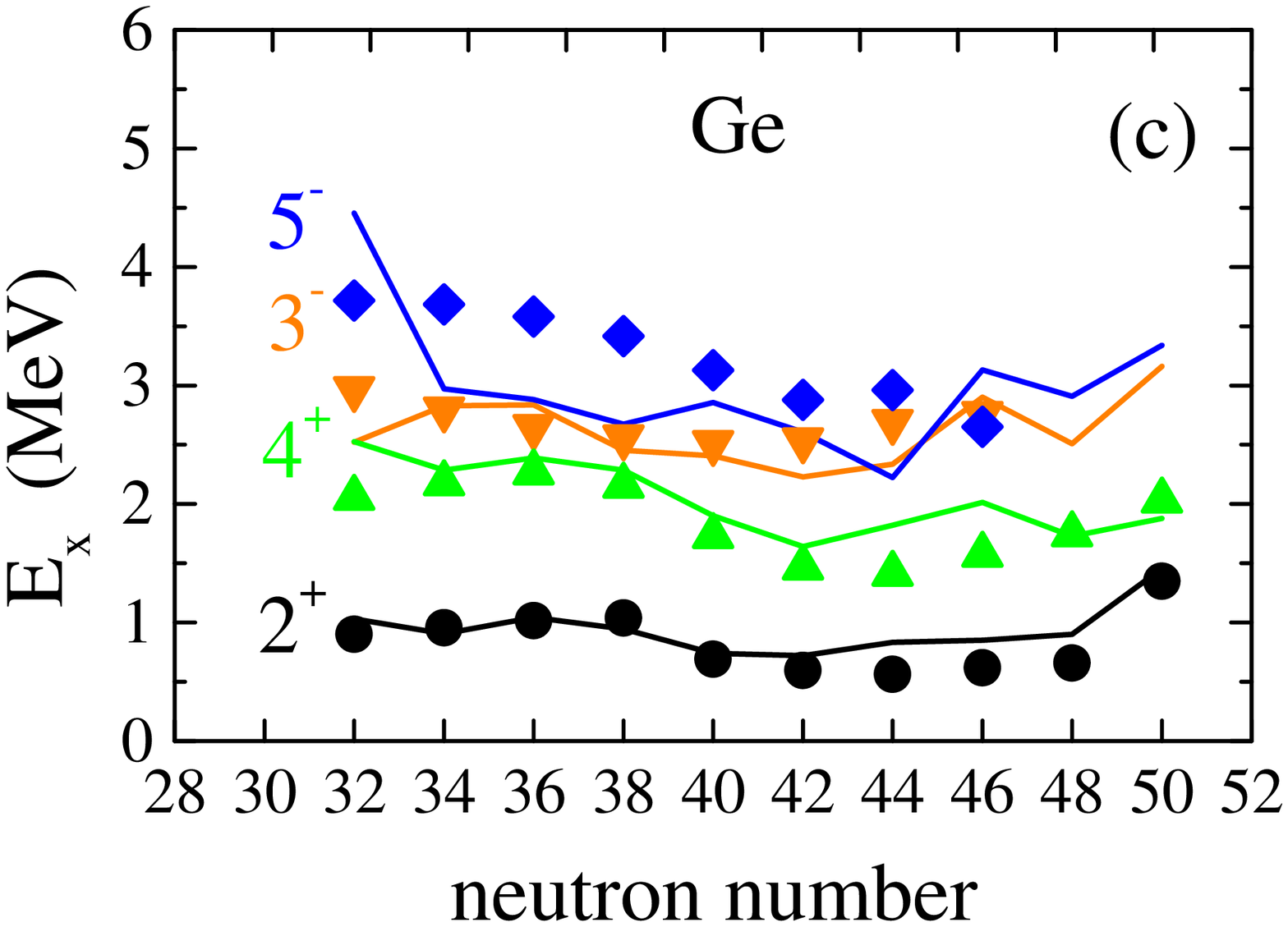}
\includegraphics[scale=0.35]{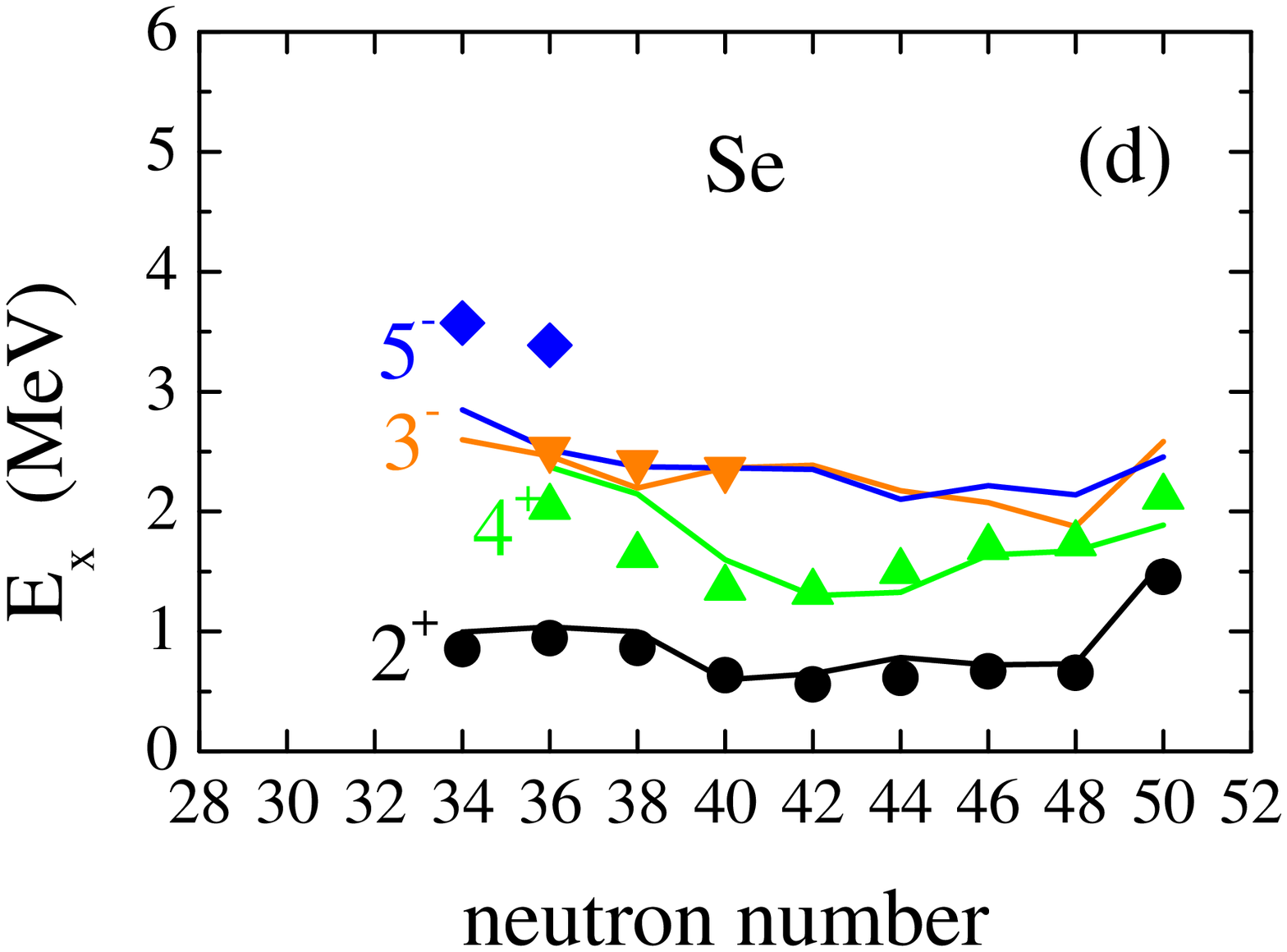}
\caption{(Color online) Energy levels of low-lying states for (a)
Ni, (b) Zn, (c) Ge, and (d) Se isotopes with even number of
neutrons. Calculated energy levels of the first $2^{+}$, $4^{+}$,
$3^{-}$, and $5^{-}$ states are shown with solid, dotted, and
dot-dashed, respectively. They are compared with the experimental
data denoted by circles, squares, and down-triangles. Experimental
data are taken from Ref. \cite{ENDSF}.}
  \label{fig7}
 \end{center}
\end{figure*}

In addition to the systematic behavior of the $2_{1}^{+}$,
$4_{1}^{+}$, and $0_{2}^{+}$ states, other energy levels of the
low-lying states in even-even nuclei are also calculated. In Fig.
\ref{fig7}, the obtained energy levels of the low-lying
negative-parity states $3_{1}^{-}$ and $5_{1}^{-}$ are compared with
the experimental data for Ni, Zn, Ge, and Se isotopes. Although they
have already been discussed in the previous subsection, the states
with $2_{1}^{+}$ and $4_{1}^{+}$ are also given as references to
show the relative energies between states with positive and negative
parities. For the Ni isotopes, the calculated energy levels of the
$3_{1}^{-}$ and $5_{1}^{-}$ states agree reasonably well with the
experimental ones. Only for the light isotopes with $N=30$ and 32,
the calculated $3_{1}^{-}$ states are slightly higher. For the Zn
isotopes, the calculation reproduces the overall trend of the
$3_{1}^{-}$ and $5_{1}^{-}$ states along the isotopic chain.
Nevertheless, the calculated $3_{1}^{-}$ levels are high for the
lighter isotopes as compared to the data. Again for $N=30$ and 32
there are clear deviations of the theoretical results of $5_{1}^{-}$
from the corresponding data. For the Ge isotopes, as already
discussed, the calculation of the $2_{1}^{+}$ and $4_{1}^{+}$ states
reproduces the known data quite well. The calculated $3_{1}^{-}$
energies are found in good agreement with the experimental data,
while the calculated $5_{1}^{-}$ states are generally lower than the
data (except for $N=32$). In Fig. 9, the calculated 5$^{-}$ energies
generally agree with the experimental data for the Ni and Zn
isotopes. An underestimation in the calculation is found only in the
Ge isotopes with $N=34-38$ and the Se ones with $N=34,36$. For the
Se isotopes, the systematics for the $2_{1}^{+}$, $4_{1}^{+}$, and
$3_{1}^{-}$ states are well described by the calculation. However,
the calculation seems to depart from the two data points ($N=34$ and
36) for the $5_{1}^{-}$ states.

\begin{figure*}[t]
 \begin{center}
\includegraphics[scale=0.35]{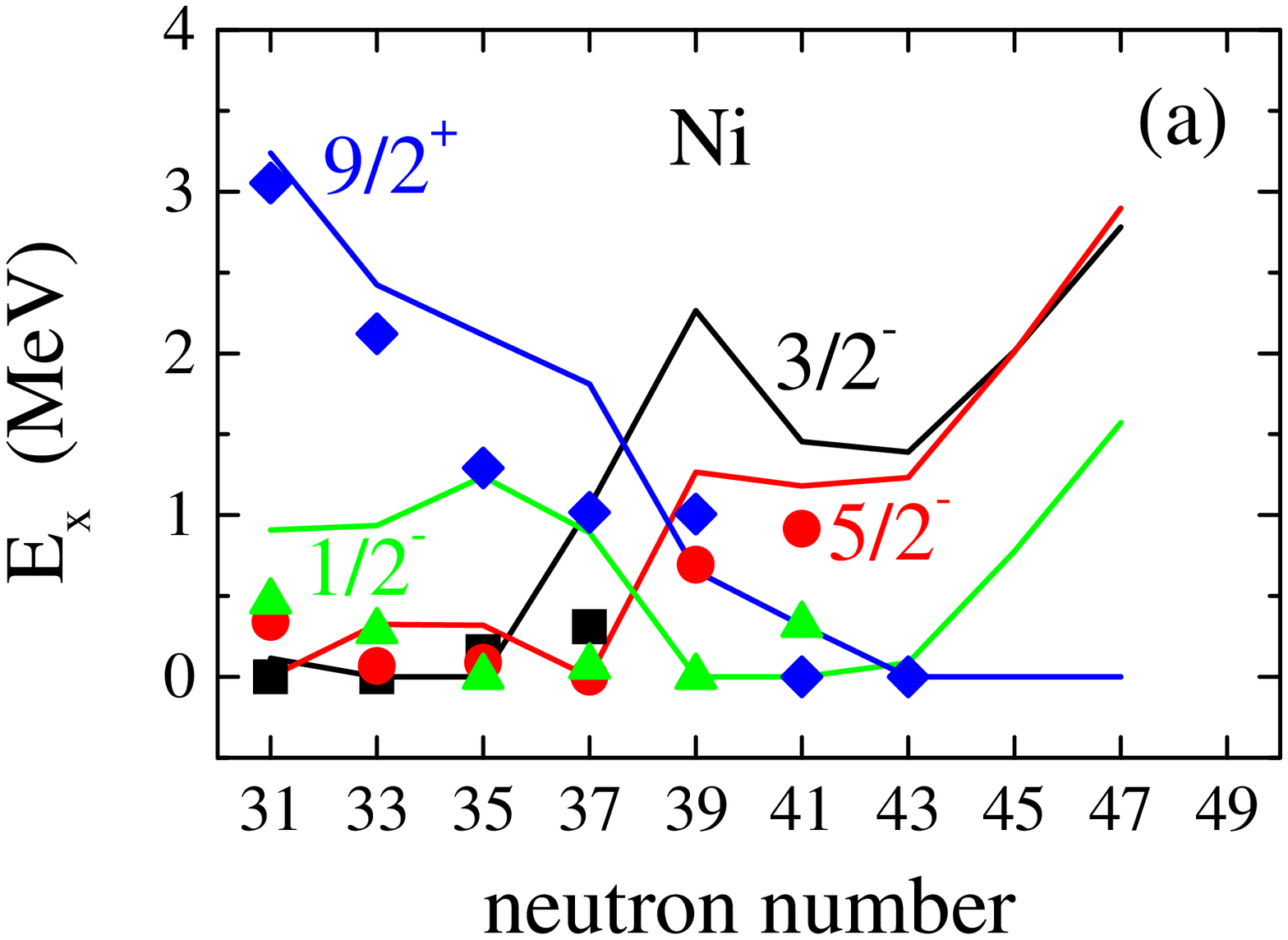}
\includegraphics[scale=0.35]{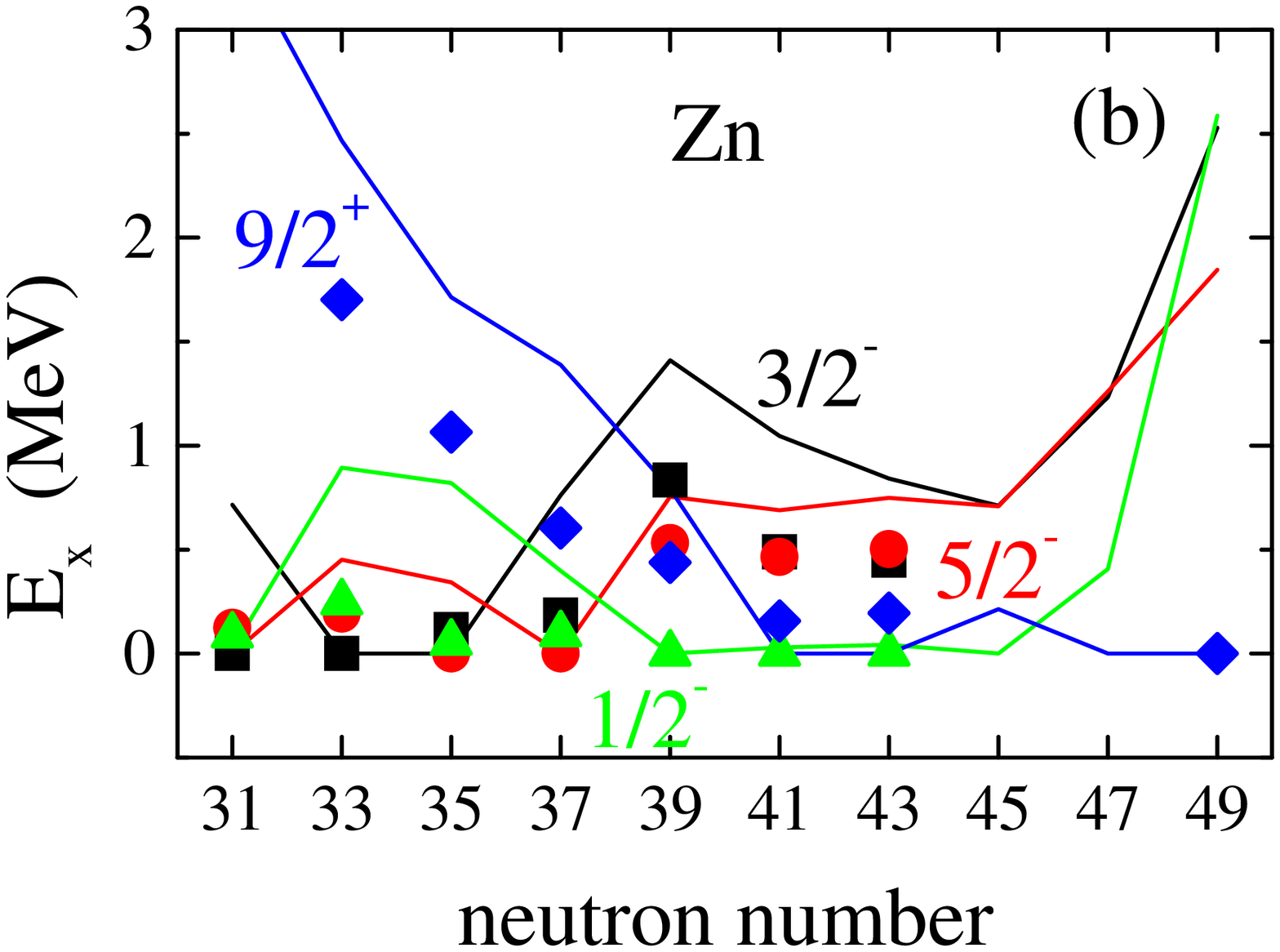}
\includegraphics[scale=0.35]{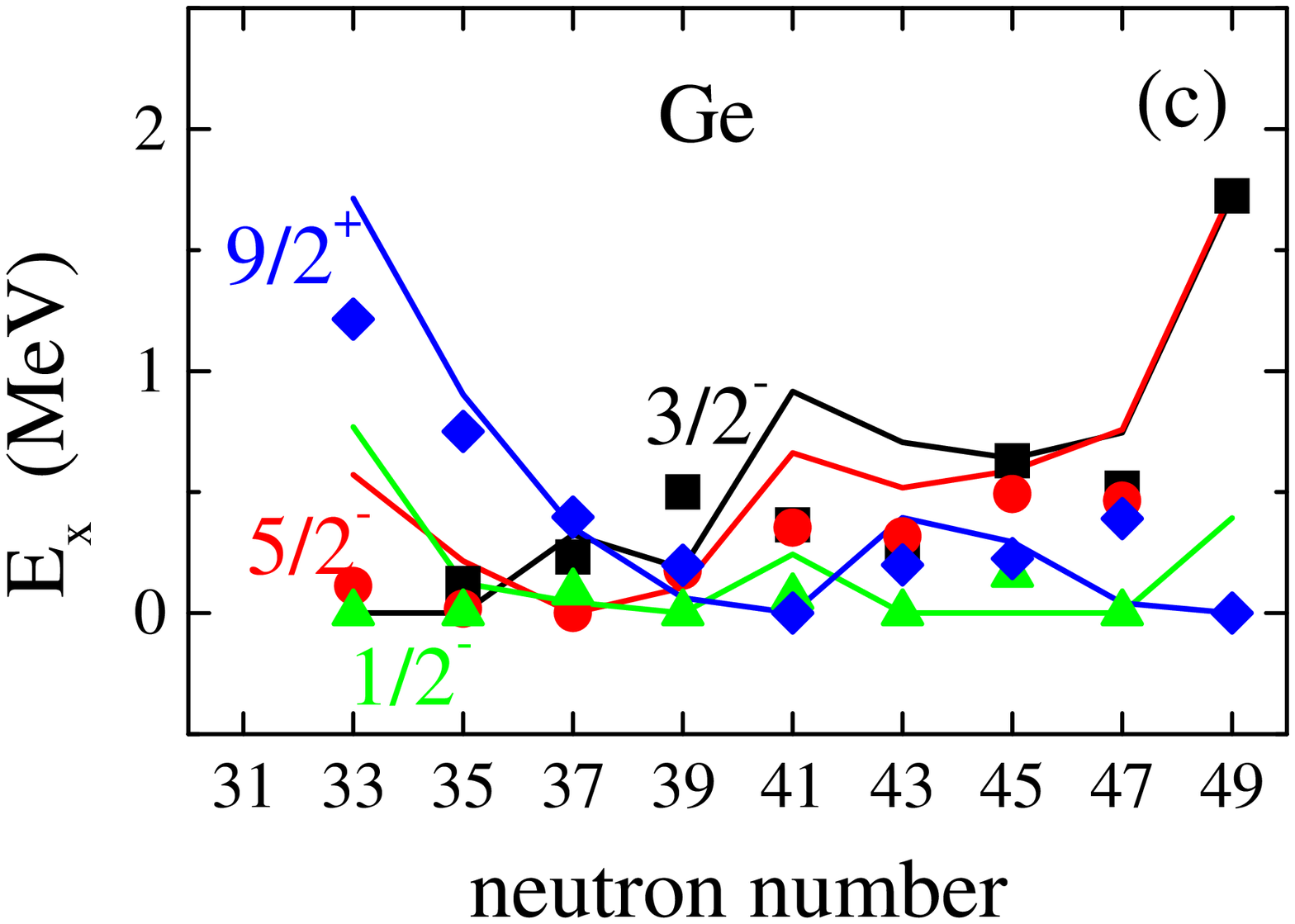}
\includegraphics[scale=0.35]{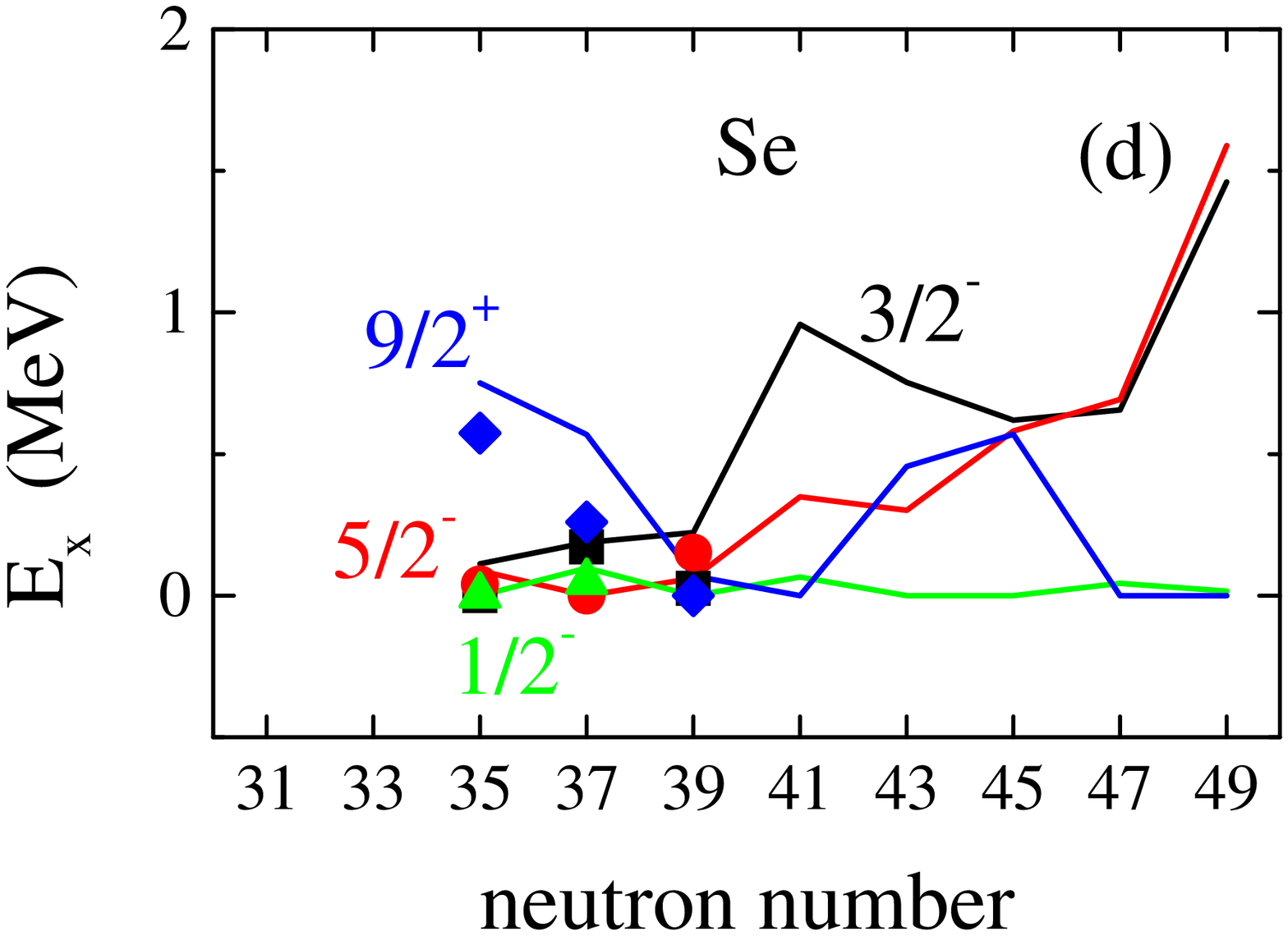}
\caption{(Color online) Energy levels of low-lying states for (a)
Ni, (b) Zn, (c) Ge, and (d) Se isotopes with odd number of neutrons.
Calculated energy levels of the first $3/2^{-}$, $5/2^{-}$,
$1/2^{-}$, and $9/2^{+}$, states are shown with solid, dotted, and
dot-dashed, respectively. They are compared with the experimental
data denoted by circles, squares, and down-triangles. Experimental
data are taken from Ref. \cite{ENDSF}.}
  \label{fig8}
 \end{center}
\end{figure*}

\begin{figure*}
 \begin{center}
\includegraphics[scale=0.33]{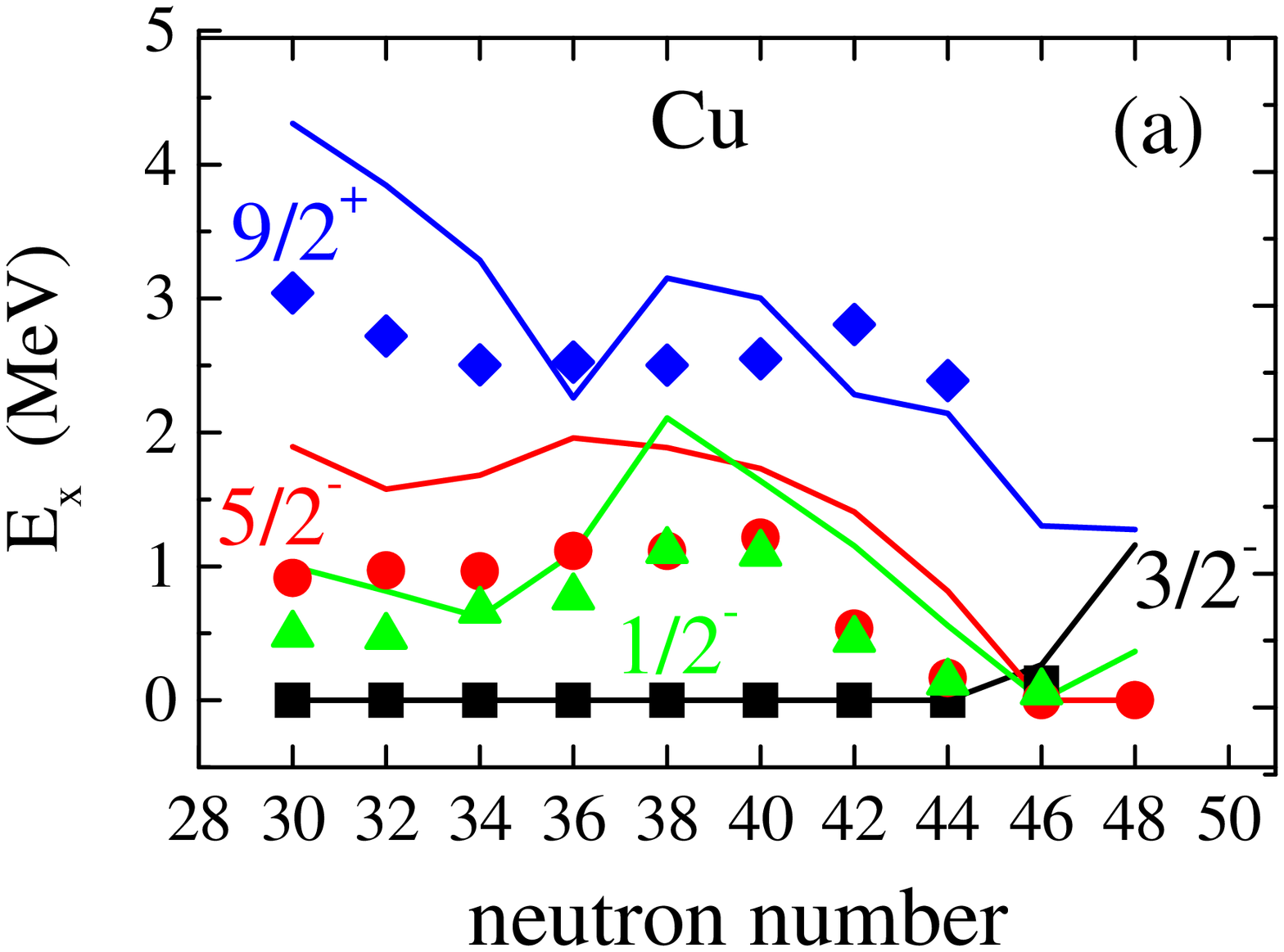}
\includegraphics[scale=0.33]{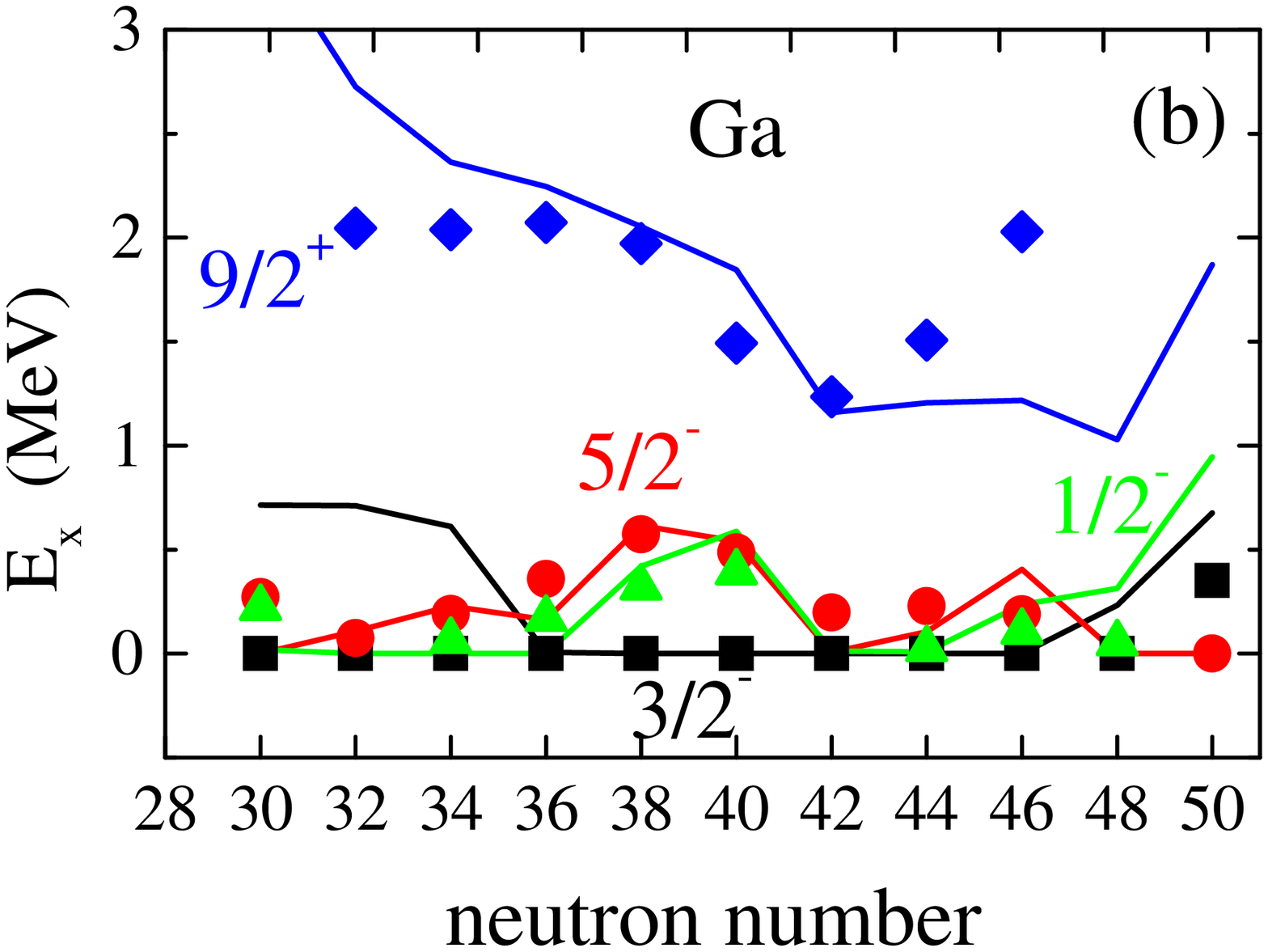}
\includegraphics[scale=0.33]{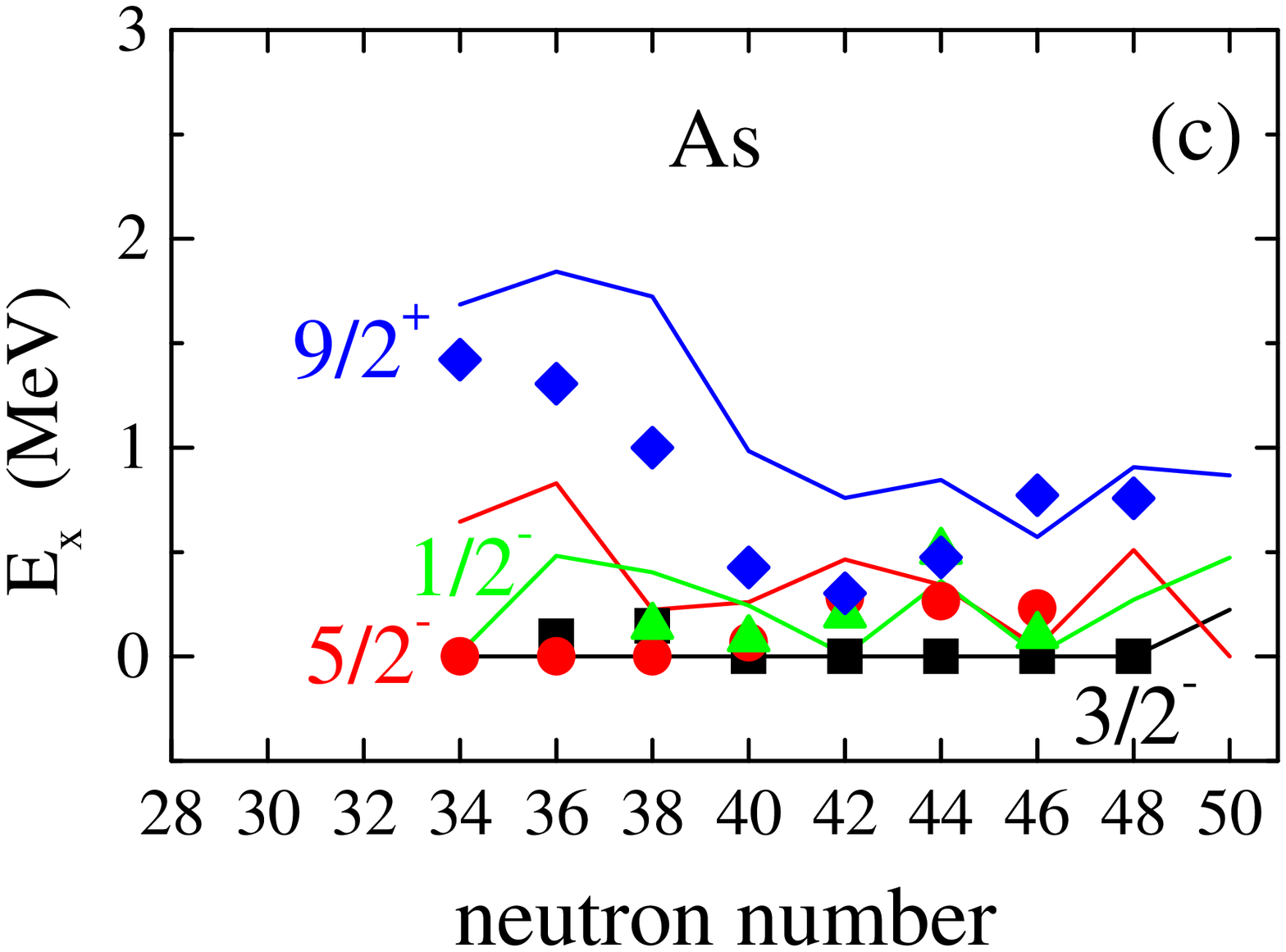}
\caption{(Color online) Energy levels of low-lying states for (a)
Cu, (b) Ga, and (c) As isotopes with even number of neutrons.
Calculated energy levels of the first $3/2^{-}$, $5/2^{-}$,
$1/2^{-}$, and $9/2^{+}$, states are shown with solid, dotted, and
dot-dashed, respectively. They are compared with the experimental
data denoted by circles, squares, and down-triangles. Experimental
data are taken from Ref. \cite{ENDSF}.}
  \label{fig9}
 \end{center}
\end{figure*}

\begin{figure*}
 \begin{center}
\includegraphics[scale=0.30]{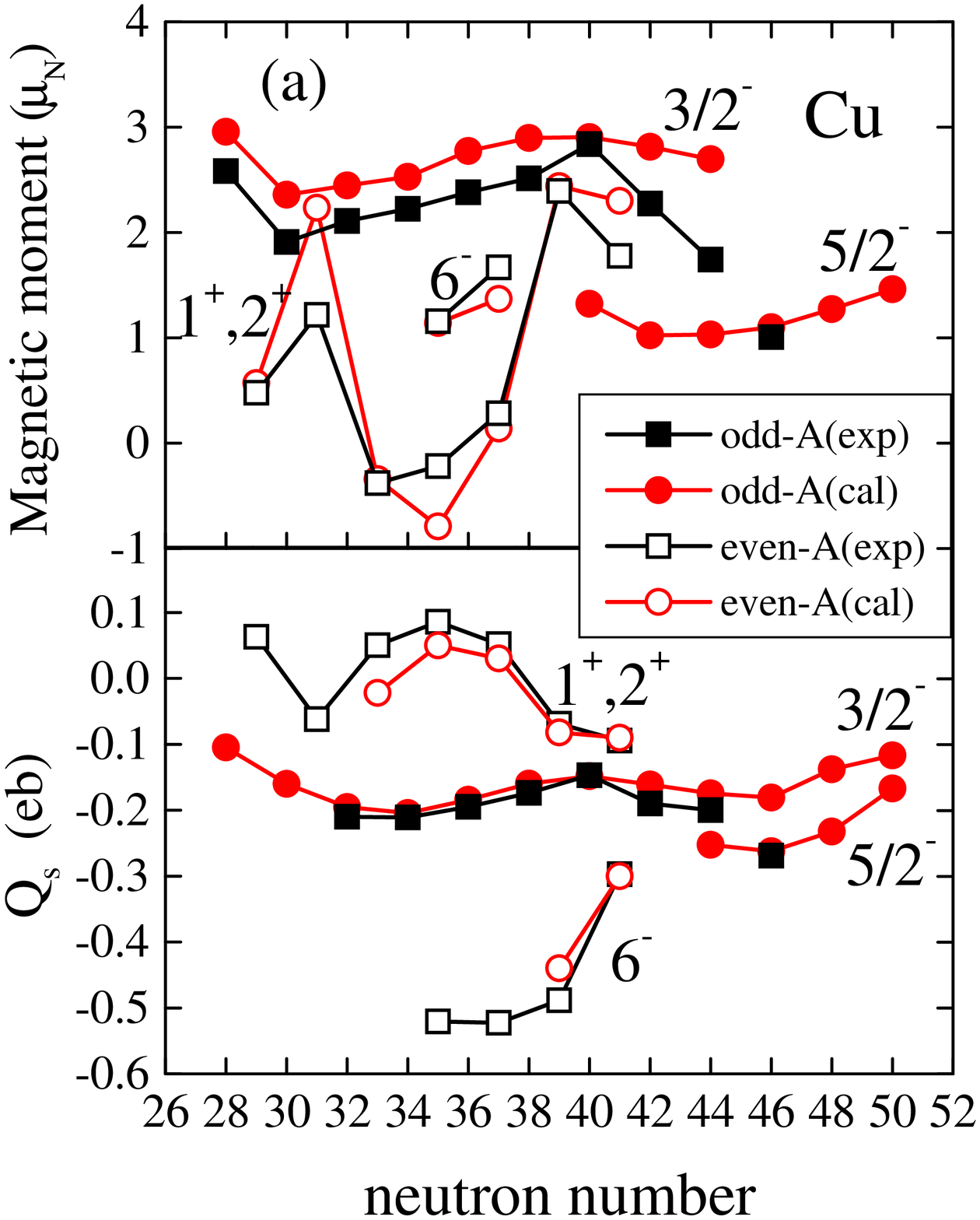}
\includegraphics[scale=0.30]{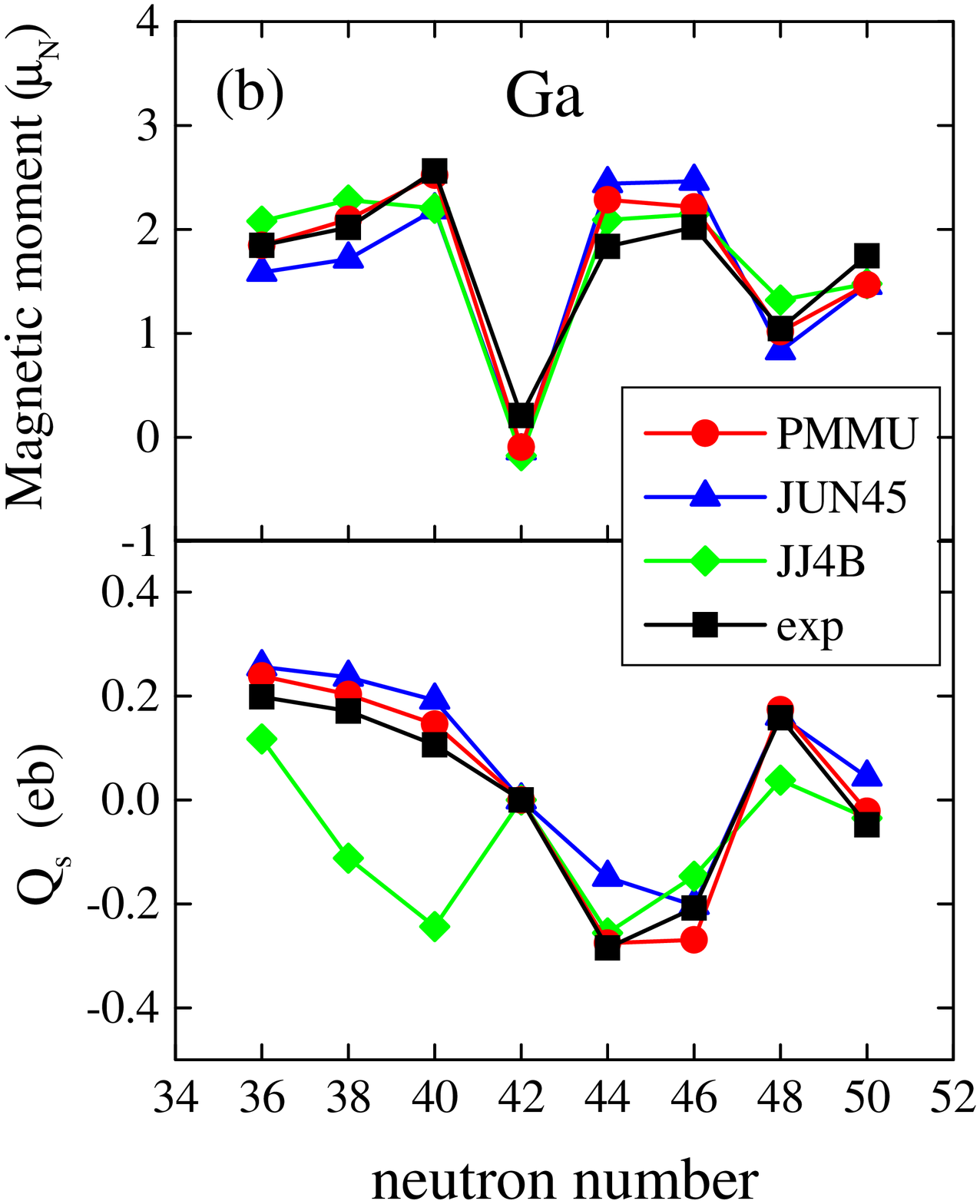}
\includegraphics[scale=0.30]{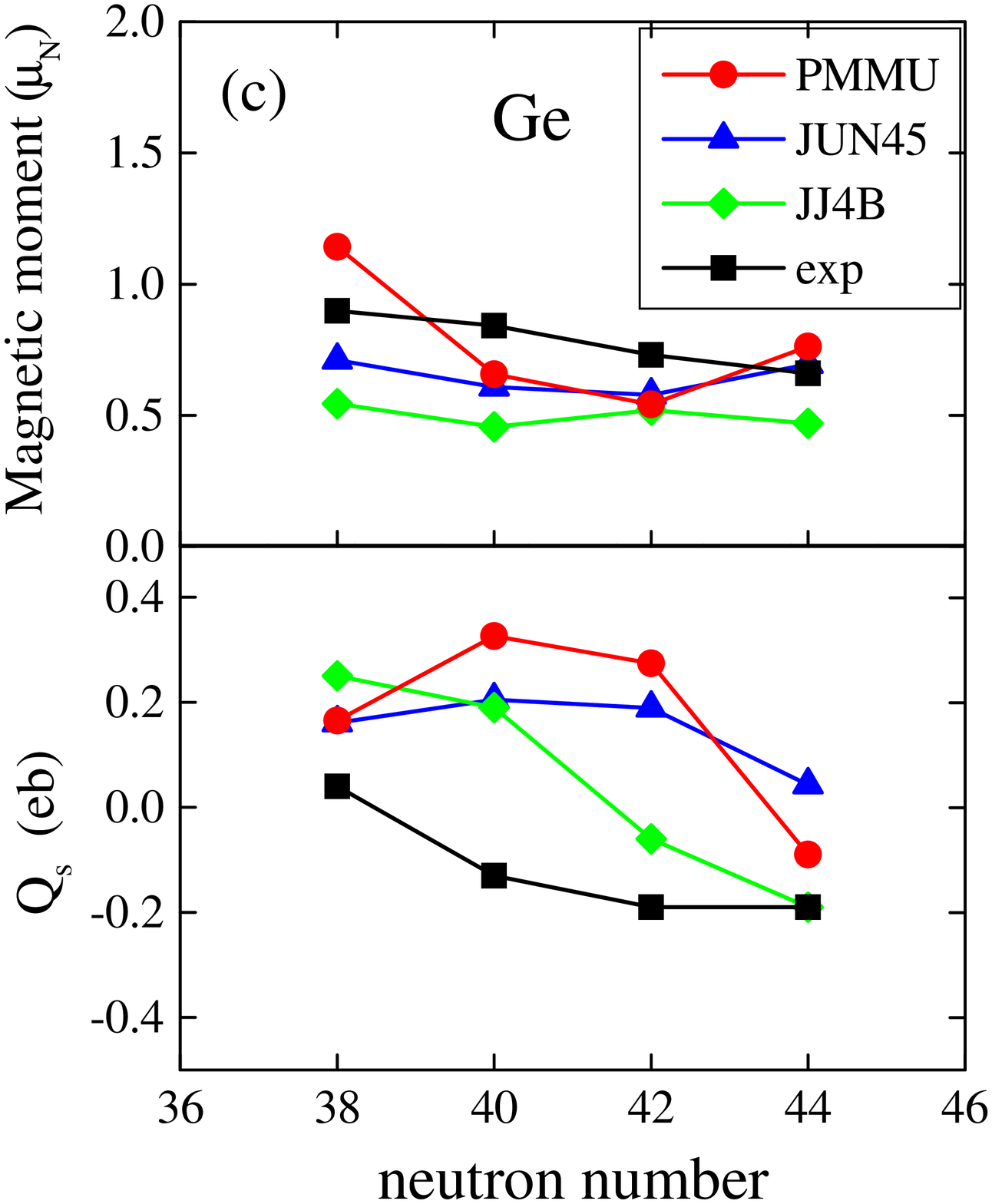}
\caption{(Color online) Magnetic moments and electric quadrupole
moments for (a) Cu, (b) Ga, and (c) Ge isotopes. Magnetic moments
with $g^{eff}_{s}$, $g^{eff}_{s}=0.7g_{s}^{(free)}$, and electric
qudrupole moments with the effective charges $e_{p}=1.5e$ for proton
and $e_{n}=1.1e$ for neutron. Experimental data are taken from
\cite{ENDSF,Gurdal13,Robinson11,Vingerhoets10,Cheal10}.}
  \label{fig10}
 \end{center}
\end{figure*}

\subsection{Low-lying states of odd-mass nuclei}

Low-lying states in odd-mass nuclei contain valuable information on
the interplay between single-particle and collective motion. We now
discuss energy levels of the low-lying states in odd-mass nuclei of
this mass region, and try to extract information on the shell
evolution along the isotopic chains. Figure \ref{fig8} shows the
comparison between the shell-model results and experimental data for
the Ni, Zn, Ge, and Se isotopes with odd number of neutrons. The
common feature for these isotopes is that the three negative-parity
states, $1/2^{-}$, $3/2^{-}$, and $5/2^{-}$, all appear near the
ground state. In particular for the Ni and Zn isotopes, the
experimental data indicate that these three energy levels are nearly
degenerate for $N=31-37$. The degeneracy is lifted for heavier
isotopes with $N\ge 39$. On the other hand, the positive-parity
$9/2^{+}$ state lies high in energy in the lightest isotope with
$N=31$, and dives down quickly as a function of neutron number. It
enters into the ground-state region at $N=41$ (Ni and Zn) or 39 (Ge
and Se).

As one can see in Fig. \ref{fig8}, our calculation reproduces the
global feature correctly although for some states the agreement is
only at a qualitative level. The largest deviation from the experimental
data is seen for the $1/2^{-}$ state in the lightest isotopes, where the
calculated levels are higher than the data. The observed excitation
energy for the $5/2^{-}$ state increases at the neutron number
$N=39$ and $N=41$, which is well described by the calculation. The
calculated $3/2^{-}$ levels rise up drastically from $N=37$, and lie
close to the $5/2^{-}$ levels for $N\ge 41$. For the Ge and Se
isotopes, the overall variations of the low-lying states are not
very drastic when neutron number increases. In all these isotopes,
the calculations predict a near-degeneracy of the $3/2^{-}$ and
$5/2^{-}$ states beyond $N=45$. This degeneracy can be understood by
the influence of the small difference of $T=1$ monopole values
between $V_{m}^{\rm PMMU}(f_{5/2},g_{9/2},T=1)$ and
$V_{m}^{\rm PMMU}(p_{3/2},g_{9/2},T=1)$.

Another notable feature is that, as the neutron $f_{5/2}$ orbit is
occupied, the energy of the positive-parity $9/2^{+}$ state goes
down rapidly. This trend is correctly described by the calculation,
and in particular, the $9/2^{+}$ energy levels for Ge and Se
isotopes are reproduced very well, as seen in Figs. \ref{fig8} (c)
and \ref{fig8} (d). Honma {\it et al.} \cite{Honma09} has explained
this trend as well from their calculations. It can be interpreted as
a result of the monopole effect due to differences between
$V_{m}^{\rm PMMU}(f_{5/2},f_{5/2},T)$ and $V_{m}^{\rm
PMMU}(f_{5/2},g_{9/2},T)$. For the $9/2^{+}$ state, however, the
agreement between our shell-model results and data becomes worse for
nuclei near the closed shell. In the Ni and Zn isotopes, the
calculated $9/2^{+}$ states are considerably higher than the
experimental ones for nuclei with $N=35$ and $N=37$. This may be due
to the large difference between $V_{m}^{\rm
PMMU}(f_{5/2},f_{5/2},T=0)$ and $V_{m}^{\rm
PMMU}(f_{5/2},g_{9/2},T=0)$. In the Ni and Zn nuclei with $N=33-37$,
the monopole contribution disappears and the $g_{9/2}$ ESPE goes up.
This could be the reason why the calculated $9/2^{+}$ states
reproduce the data well for the Ge and Se nuclei, while they fail
for the Ni and Zn nuclei with $N=35$ and $N=37$.

Next, we present the results for the low-lying states of odd-mass
nuclei with odd-proton and even-neutron numbers. Figure \ref{fig9}
shows the energy levels of odd-$Z$ Cu, Ga, and As isotopes as
functions of neutron number. As already discussed in Fig.
\ref{fig3}, the $5/2^{-}$ level comes down rapidly beyond the
neutron number $N=40$ corresponding to the ESPE of the $f_{5/2}$
orbit in Fig. \ref{fig3} (a). The lowering of the $5/2^{-}$ level
can be understood as a result of the large attractive $T=0$ monopole
term $V_{m}^{\rm PMMU}(f_{5/2},g_{9/2},T=0)$ (see Fig. \ref{fig2})
due to the tensor force discussed in Eq. (\ref{eq:3}). In all these
isotopes, the $5/2^{-}$ energy levels are close to the $1/2^{-}$
energy levels. This degeneracy can also be understood by the effects
of the small difference of $T=0$ monopole values between $V_{m}^{\rm
PMMU}(f_{5/2},g_{9/2},T=0)$ and $V_{m}^{\rm
PMMU}(p_{1/2},g_{9/2},T=0)$. The shell-model description is not
successful for the $5/2^{-}$ states of Cu isotopes with $N=30-40$.
This could be due to the $Z=$28 core excitations that are not
included in the present calculation. We can see that the calculated
results for the $9/2^{+}$ states reproduce the experimental data for
$N > 40$ (with an exception for $N=44,46$ Ga isotopes), while those
are worse for $N < 40$. This may be originated from the missing
effects of the $T=0$ monopole term between the $f_{7/2}$ and
$g_{9/2}$ orbits. For the Ga isotopes, the calculated results for
the $1/2^{-}$, $3/2^{-}$, and $5/2^{-}$ states are in good agreement
with the experimental data, while for $N=30-34$ the calculated
$3/2^{-}$ state is higher than the experiment. The calculations for
the $9/2^{+}$ states reproduce well the trend of the experimental
data. For the As isotopes, the energy of the $9/2^{+}$ state comes
down drastically and shows a minimum at $N=42$. The minimum at
$N=42$ here can be associated with the development of deformation
and correlation with the $2_{1}^{+}$ excitation energy of the
doubly-even Ge and Se isotopes, as discussed in Fig. \ref{fig5}. The
present calculations for the odd-proton As isotopes describe well
such a feature.

\subsection{Magnetic dipole moments and electric quadrupole moments
in Cu, Ga, and Ge isotopes}

In addition to energy levels, it is important to test the PMMU model
further with $E2$ transition calculations. The magnetic
moment operator is given as
\begin{eqnarray}
 & {} &  {\bf \mu} = g_{s}{\bf s} + g_{l}{\bf l},  \nonumber \\
\label{eq:11}
\end{eqnarray}
where $g_{s}$ and $g_{l}$ are the spin and the orbital $g$-factors,
respectively. In the calculation, we employ a quenching factor 0.7
($g^{eff}_{s}=0.7g_{s}^{free}$) for the free-nucleon spin $g$-factors,
$g_{s}^{free}=5.586$ for protons and $g_{s}^{free}=-3.826$
for neutrons. The effective charges in the electric quadrupole
moment calculation are taken as $e_{p}=1.5e$ for protons and
$e_{n}=1.1e$ for neutrons. Figure \ref{fig10} shows the calculated
results of magnetic moments and electric quadrupole moments for the
Cu, Ga, and Ge isotopes. For comparison, the theoretical results for
the lowest states obtained from other effective interactions, the
JUN45 \cite{Honma09} and JJ4B interactions \cite{Honma09,jj4b}, are
also plotted. The calculated isotopes range from the magic number
$N=28$ on the neutron-deficient side to $N=50$ on the neutron-rich
side, including the semi-magic subshell closure at $N=40$, where the
parity change between the $pf$ shell and the $g_{9/2}$ orbit occurs.

In the upper graph of Fig. \ref{fig10} (a), the calculated magnetic
moments are shown for the Cu isotopes \cite{Vingerhoets10}. One can
see that overall, the calculation reproduces well the experimental
data. For the odd-$A$ Cu isotopes, the experimental magnetic moments
of $^{57}$Cu and $^{69}$Cu are close to the effective
single-particle value of the $\pi p_{3/2}$ configuration, indicating
clearly the magicity at $N=28$ and $N=40$, respectively. The magic
behavior at $N=40$ originates from the parity exchange between the
$\nu p_{1/2}$ and $\nu g_{9/2}$ single-particle levels, which does
not allow $M1$ excitations from the negative-parity $pf$ shell to
the positive-parity $\nu g_{9/2}$ orbit. Some differences in
magnetic moment between theory and experiment are found below
$N=40$, which may indicate that the $N=28$ shell gap is not very
large, and proton excitations from the $\pi f_{7/2}$ shell become
important. Indeed, the GXPF1A calculation reproduced well the
experimental magnetic moments \cite{Honma04}. The calculated results
begin to differ largely from the experimental data for the
neutron-rich side with $N>40$. This could be due to the influence of
the missing $\pi f_{7/2}$ orbit. When neutrons begin to occupy the
$\nu g_{9/2}$ orbit, proton excitations from the $\pi f_{7/2}$ orbit
for $N>40$ may increase because the gap between the $\pi f_{5/2}$
and $\pi f_{7/2}$ orbits decreases due to the attractive $T=0$
monopole interaction between the $\pi f_{7/2}$ and $\nu g_{9/2}$
orbits. Then $g$ factors would be better reproduced for $N>40$. The
calculated magnetic moments for the even-$A$ isotopes are shown in
the same graph, which are found in good agreement with the data. In
the lower graph of Fig. \ref{fig10} (a), the calculated
spectroscopic quadrupole moments are compared to the known
experimental values. As one can see, a good description of the data
has been achieved by the calculation for all the isotopes.

In the upper graph of Fig. \ref{fig10} (b), the calculated magnetic
moments are shown for the Ga isotopes \cite{Cheal10}. The results
for $^{67,69}$Ga and $^{75,77}$Ga are smaller and larger than the
effective single-particle moments $g^{eff}(\pi p_{3/2})=2.96$ of the
$\pi p_{3/2}$ orbit and $g^{eff}(\pi f_{5/2})=1.46$ of the $\pi
f_{5/2}$ orbit, respectively, which suggests that the ground states
of these isotopes have mixed configurations. For $^{71}$Ga ($N=40$),
the observed magnetic moment is close to the effective moment
$g^{eff}(\pi p_{3/2})=2.96$ of the $\pi p_{3/2}$ orbit, thus the
$\pi p_{3/2}$ configuration with the single-proton is the leading
one in the ground state. For $^{79}$Ga, the calculated magnetic
moment of the lowest $I=3/2^{-}$ state is significantly larger than
the experimental data, and the calculated quadrupole moment has
opposite sign to that of the experiment. This problem can be solved
if we take the calculated magnetic and quadrupole moments for the
first excited $I=3/2^{-}$ state instead of the lowest one to compare
with the experimental data. Then a good agreement with data for the
$I=3/2^{-}$ state of $^{79}$Ga can be achieved, as can be seen in
Fig. 12 (b). Also in the JUN45 and JJ4B calculations, the magnetic
moment of the first excited state was taken to compare with
experimental data. The electric quadrupole moments of the lowest
excited $3/2^{-}$ states are plotted in the lower graph of Fig.
\ref{fig10} (b). One can see that the sign inversion occurs around
$^{73}$Ga with $N=42$. The opposite sign in their quadrupole moments
suggests clearly a drastically different structure. This can be
understood by non-occupation of the $\pi p_{3/2}$ orbit, because the
proton $\pi p_{3/2}$ configuration has the negative quadrupole
moment. The configurations of the ground state in $^{75,77}$Ga have
a dominant $\pi (p_{3/2}f^{2}_{5/2})$ configuration. As one can see,
the JJ4B calculations for quadrupole moment fail to reproduce
experimental data for $^{69,71}$Ga.

The calculated magnetic moments and spectroscopic quadrupole moments
for the first excited $2^{+}$ states of the Ge isotopes are shown in
Fig. \ref{fig10} (c), and compared with experimental data
\cite{Robinson11,Gurdal13}. In the upper graph, the calculated
results underestimate the measured magnetic moments, except for
$^{70}$Ge ($N=38$). It is interesting to notice that our calculation
suggests similar results as compared to the JUN45 calculations. As
one shall see later, the calculated excitation spectrum is in a
reasonable agreement with the experimental data. However, the
quadrupole moments of the $2_{1}^{+}$ states are not at all
accounted for by all the current shell-model calculations, as seen
in the lower graph of Fig. \ref{fig10} (c). This remains to be a
puzzle for future investigations.

\section{Structure of excited states in Ge and Se isotopes}\label{sec5}

In the previous section, we presented systematical results for
low-lying states from the shell-model calculation. In Figs.
\ref{fig8} and \ref{fig9}, we have seen rapid changes in energy
levels occurring around $N=40$. In Fig. \ref{fig6}, the $B(E2)$
values increase rapidly around $N=42$ for Ge and Se isotopes. All
these can be clear signals for transitions from a near-spherical to
a prolate shape. In the transitional region, a shape coexistence
phenomenon would show up. The first excited $0^{+}$ state,
$0_{2}^{+}$ in doubly-even nuclei, is a key measure to indicate
shape coexistence, because the $0_{2}^{+}$ state is usually the
bandhead of the emerging side band. As we have already seen in Fig.
\ref{fig5}, the energy of the $0_{2}^{+}$ state decreases rapidly
and becomes the lowest at $N=40$ among all isotopes. In $^{72}$Ge,
the $0_{2}^{+}$ state becomes notably the first excited state, lower
than $2_{1}^{+}$ . This state has been suggested to have a spherical
shape from the experimental $B(E2)$ data. As already mentioned,
there have been many theoretical approaches based on the mean-field
model to understand this problem \cite{Bender06}. The shell-model
approach described well the $0_{2}^{+}$ state \cite{Honma09}. Thus
we can expect that there are significant structure changes between
the lighter ($N < 40$) and heavier ($N > 40$) isotopes. It is thus
interesting to see whether our PMMU model is applicable to the
description of full spectra including high excitations. In this
section, we will discuss the energy levels and the $E2$
transitions for the Ge and Se isotopes. As already shown in Figs.
\ref{fig7} and \ref{fig8}, our shell-model calculation describes
fairly well the experimental energy levels for both even-even and
odd-mass isotopes. In particular, the anomalous behavior of the
$0_{2}^{+}$ states around $N=40$ is reproduced correctly. Of course,
there are too many excited states in the calculation. Hereafter, we
show only the excited energy levels and bands that may have at least
some experimental indications.

\begin{figure}[b]
\includegraphics[totalheight=7.5cm]{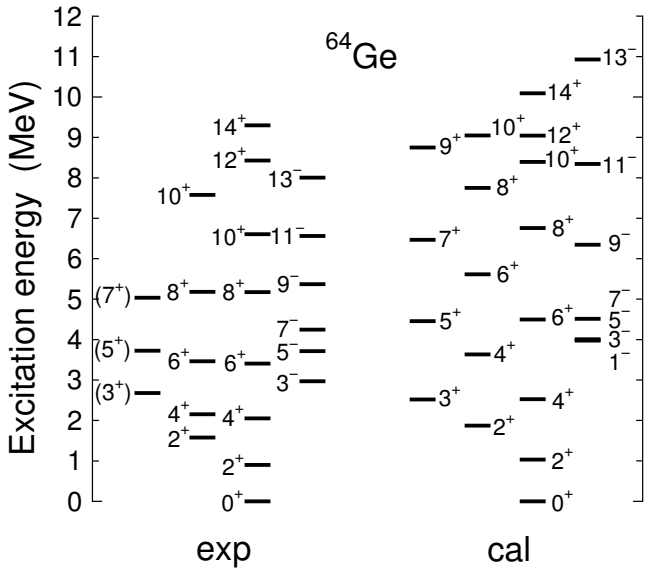}
\caption{Comparison of energy levels between the shell-model results
and the experimental data for $^{64}$Ge.}
  \label{fig28}
\end{figure}
\begin{figure}[t]
\includegraphics[totalheight=7.0cm]{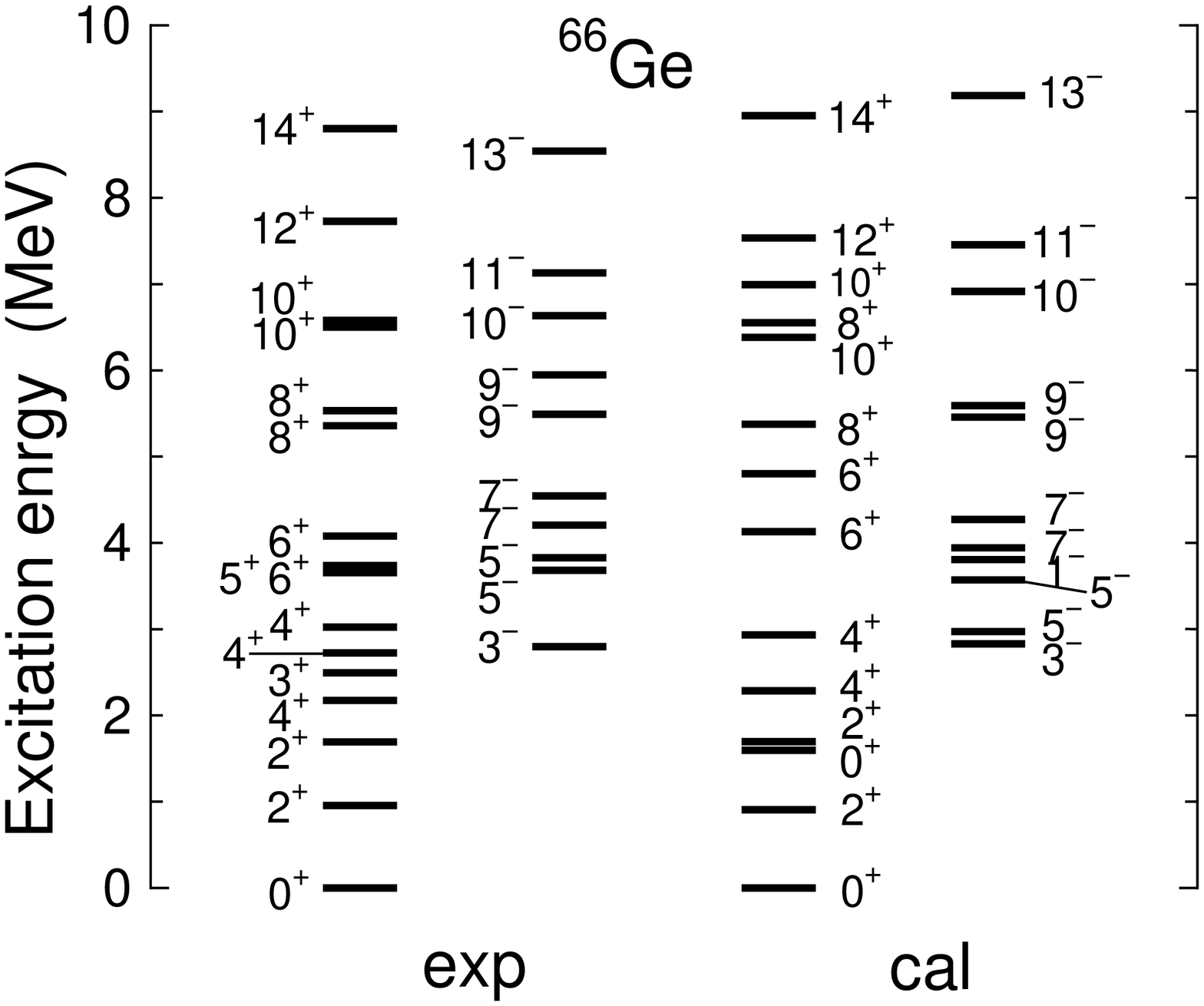}
\caption{Comparison of energy levels between the shell-model results
and the experimental data for $^{66}$Ge. The experimental
(calculated) energy levels are shown in the left (right) side. In
each side, the first (second) line indicates the positive-
(negative-) parity states. Experimental data are taken from
\cite{ENDSF}.}
  \label{fig11}
\end{figure}

\begin{table}[t]
\caption{$B(E2)$ values for the positive-parity yrast states and
         some collective states of $^{66}$Ge and $^{68}$Ge.
         Experimental data are taken from Ref. \cite{ENDSF,Luttke12}.}
\begin{tabular}{lccccc}   \hline \hline
        & \multicolumn{2}{c}{\hspace{1.0cm}$^{66}$Ge  [W.u.]}\hspace{1.0cm}
        & \multicolumn{3}{c}{\hspace{1.0cm}$^{68}$Ge  [W.u.]\hspace{1.0cm}}   \\ \hline
$J_i^\pi \rightarrow J_f^\pi$\hspace{1.0cm} &
\hspace{0.5cm}exp\hspace{0.5cm} & \hspace{0.5cm}cal\hspace{0.5cm}
 & \hspace{0.5cm}exp\hspace{0.5cm} & \hspace{0.5cm}cal\hspace{0.5cm}   \\ \hline
$2_1^+ \rightarrow 0_1^+$  &  12.0(23)  &    29.0       &    15.3(8)   &  25.4      \\
$4_1^+ \rightarrow 2_1^+$  &  >9.6      &    33.2       &    12.8(15)  &  22.4      \\
$6_1^+ \rightarrow 4_1^+$  &  >1.2      &    36.1       &    12(4)     &   1.8      \\
$8_1^+ \rightarrow 6_1^+$  &            &     0.0       &    14(3)     &   0.1      \\
$2_2^+ \rightarrow 2_1^+$  &  16(7)     &    14.5       &    1.0(5)    &  29.5      \\
$2_2^+ \rightarrow 0_1^+$  &  0.13(5)   &      0.5      &    0.40(5)   &   0.1      \\ \hline \hline
\end{tabular}
\label{table1}
\end{table}

\subsection{Even-even Ge isotopes}

We start with the $N=Z=32$ nucleus $_{32}^{64}$Ge. This nucleus has
been known as a typical example showing the $\gamma$-soft structure
among the $N=Z$ nuclei \cite{Ennis91,Hase07}, which was supported by
shell-model calculations with mean-field approximation
\cite{Kaneko04}. The calculation \cite{Kaneko04} predicted $\gamma$
instability in the ground state and triaxial deformation in the
excited states. Comparing our current shell-model calculations for
$^{64}$Ge with the experimental data in Fig. \ref{fig28}, one can
see that the low-lying states below 3 MeV are reasonably reproduced.
However, the side bands for positive- and negative-parity states
cannot be well described. The calculated energy levels are too high
as compared to the experimental ones. This could be due to the
missing $f_{7/2}$ orbit, which is not included in the present model
space.

Figure \ref{fig11} shows the energy levels of the next isotope
$^{66}$Ge. Overall, the calculation describes well the level
distribution for both positive- and negative-parity states. For the
positive-parity states, one finds a one-to-one correspondence
between the theoretical levels and experimental ones. The
$3_{1}^{-}$ excitation energy obtained in the calculation is in good
agreement with data. However, the calculated $5_{1}^{-}$ state is
predicted to be too low by about 0.8 MeV, and is almost degenerate
with the $3_{1}^{-}$ state. The two closely-lying $7^{-}$ and
two $9^{-}$ states seen in the experiment are reasonably well
reproduced. As shown in Table I, the $E2$ transition probabilities
cannot be reproduced by the calculation. The calculated $B(E2,
2_{1}^{+} \rightarrow 0_{1}^{+})$ value is more than twice the
observed one. Nevertheless, the $B(E2, 2_{2}^{+} \rightarrow
2_{1}^{+})$ value is in good agreement with the experimental data.
As seen early in Fig. \ref{fig6}, the theoretical $B(E2, 2_{1}^{+}
\rightarrow 0_{1}^{+})$ for the Ge isotopes reproduce quite well the
experimental data, but those for the $^{66}$Ge and $^{68}$Ge are
larger than the data.

\begin{figure}[t]
\includegraphics[totalheight=7.0cm]{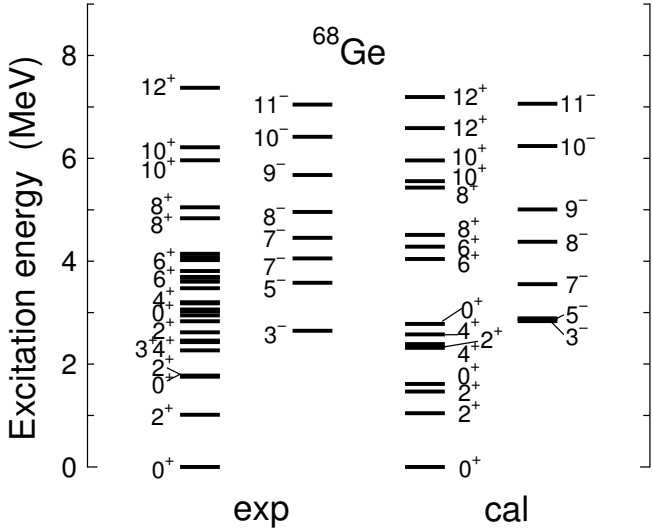}
\caption{Comparison of energy levels between the shell-model results
and the experimental data for $^{68}$Ge. Experimental data are taken
from \cite{ENDSF}. Conventions are the same as those
 in Fig. \ref{fig11}.}
  \label{fig12}
\end{figure}

Figure \ref{fig12} shows the comparison of the $^{68}$Ge energy
levels between our calculation and the experiment. The calculated
results are overall satisfactory. We note that many
experimentally-observed states for the energy range of $3- 4$ MeV
have no spin assignment. The calculation predicts the second excited
$2_{2}^{+}$ state above the known $2_{1}^{+}$ state and the second
excited $6_{2}^{+}$ state above the known $6_{1}^{+}$ state. On the
other hand, the calculated negative-parity energy levels correspond
better to the experimental ones. For electric quadrupole transitions
in the positive-parity states, one sees from Table I that similar to
$^{66}$Ge, the calculated $B(E2, 2_{1}^{+} \rightarrow 0_{1}^{+})$
value is quite large as compared to the corresponding data. The
calculated $B(E2, 2_{2}^{+} \rightarrow 2_{1}^{+})$ is also much
larger than the data. However, the calculated $B(E2, 6_{1}^{+}
\rightarrow 4_{1}^{+})$ and $B(E2, 8_{1}^{+} \rightarrow 6_{1}^{+})$
are too small as compared to the measurement.

\begin{figure}[t]
\includegraphics[totalheight=7.0cm]{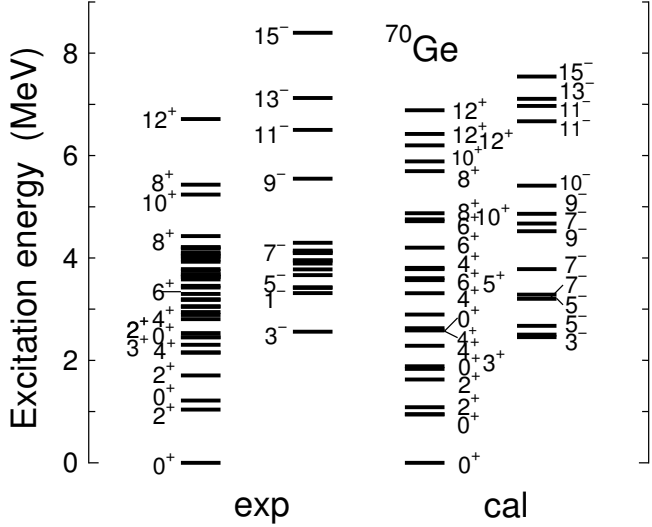}
\caption{Comparison of energy levels between the shell-model results
and the experimental data for $^{70}$Ge. Experimental data are taken
from \cite{ENDSF}. Conventions are the same as those in Fig.
\ref{fig11}.}
  \label{fig13}
\end{figure}
\begin{figure}[b]
\includegraphics[totalheight=7.0cm]{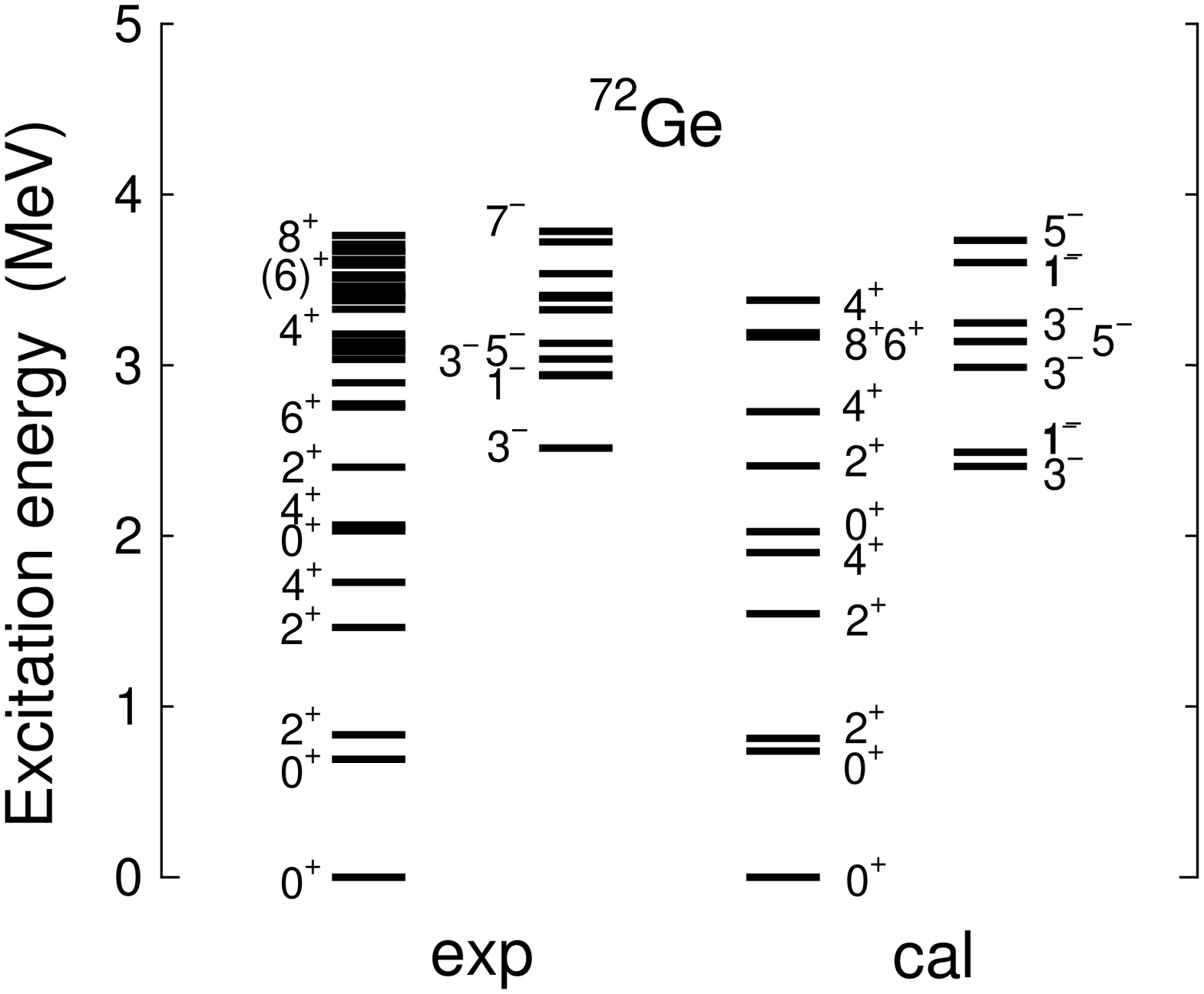}
\caption{Comparison of energy levels between the shell-model results
and the experimental data for $^{72}$Ge. Experimental data are taken
from \cite{ENDSF}. Conventions are the same as those in Fig.
\ref{fig11}.}
  \label{fig14}
\end{figure}
\begin{figure}[b]
\includegraphics[totalheight=7.0cm]{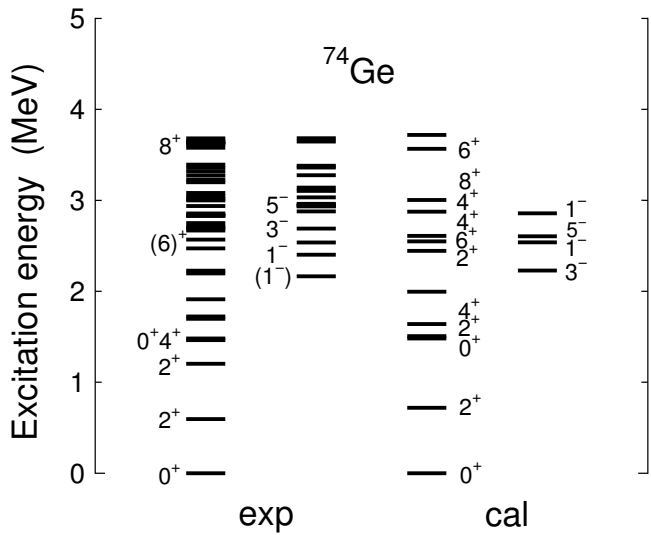}
\caption{Comparison of energy levels between the shell-model results
and the experimental data for $^{74}$Ge. Experimental data are taken
from \cite{ENDSF}. Conventions are the same as those in Fig.
\ref{fig11}.}
  \label{fig15}
\end{figure}

Calculated energy levels of $^{70}$Ge are presented in Fig.
\ref{fig13} and compared with experimental data. The known
experimental levels for both positive- and negative-parities below 7
MeV are taken for comparison. The calculation reproduces correctly
the first excited $0_{2}^{+}$ state as well as the $2^{+}$ and
$4^{+}$ states. Nevertheless, different from the experiment, the
calculated $0_{2}^{+}$ state is lower than the $2_{1}^{+}$ state.
Many experimental levels in the energy range of 2 to 4 MeV have no
spin assignment. The calculated higher spin states above 4 MeV with
$8^{+}$, $10^{+}$, and $12^{+}$ correspond well to the observed
ones. For the negative-parity states, the first excited $3_{1}^{-}$
state is in very good agreement with the experimental data at $\sim$
2.5 MeV, while the higher spin states $5_{1}^{-}$, $7_{1}^{-}$, and
$10_{1}^{-}$ can find correspondence with the observed levels. The
calculated $B(E2)$ values for the positive-parity states are
compared with experimental data in Table II. The agreement with data
is satisfactory, however, with one exception of $B(E2, 2_{2}^{+}
\rightarrow 2_{1}^{+})$, for which the measurement has large
uncertainties.

Figure \ref{fig14} shows energy levels of $^{72}$Ge. For this
isotope, it is expected that the levels reflect the $N=40$ subshell
structure. However, both the observed $2_{1}^{+}$ and $0_{2}^{+}$
states are lower than corresponding ones in $^{70}$Ge. As
already seen in the systematics of Fig. \ref{fig2} and discussed
previously, the $0_{2}^{+}$ state becomes lowest at $N=40$. This
feature is correctly described by the calculation. The calculated
second excited $0_{3}^{+}$ state at about 2 MeV is also in good
agreement with the experimental data. For the positive-parity states,
there is basically a one-to-one correspondence between theory and
experiment up to 4 MeV, although the observed level ordering is not
always reproduced. The predicted $6^{+}$ and $8^{+}$ states are
lower than the experimental data. For the negative-parity states,
the $3_{1}^{-}$ state from the calculation is slightly lower than
the data, and the $1_{1}^{-}$ state is quite low as compared to the
data. The $E2$ transitions of the positive-parity states are shown
in Table II. It is found that the calculated $B(E2)$ values
reproduce the experimental data for $B(E2, 2_{1}^{+} \rightarrow
0_{1}^{+})$, $B(E2, 4_{1}^{+} \rightarrow 2_{1}^{+})$, and $B(E2,
6_{1}^{+} \rightarrow 4_{1}^{+})$, but not for $B(E2, 8_{1}^{+}
\rightarrow 6_{1}^{+})$ and $B(E2, 2_{2}^{+} \rightarrow
2_{1}^{+})$.

\begin{table}[t]
\caption{$B(E2)$ values for the positive-parity yrast states and
         some collective states of $^{70}$Ge and $^{72}$Ge.
         The experimental data are taken from \cite{Robinson11}.}
\begin{tabular}{lccccc}   \hline \hline
        & \multicolumn{2}{c}{\hspace{1.0cm}$^{70}$Ge  [W.u.]}\hspace{1.0cm}
        & \multicolumn{3}{c}{\hspace{1.0cm}$^{72}$Ge  [W.u.]\hspace{1.0cm}}   \\ \hline
$J_i^\pi \rightarrow J_f^\pi$\hspace{1.0cm} &
\hspace{0.5cm}exp\hspace{0.5cm} & \hspace{0.5cm}cal\hspace{0.5cm}
 & \hspace{0.5cm}exp\hspace{0.5cm} & \hspace{0.5cm}cal\hspace{0.5cm}   \\ \hline
$2_1^+ \rightarrow 0_1^+$  &   21.0(4)  &   22.9       &   23.5(4)    &           26.0    \\
$4_1^+ \rightarrow 2_1^+$  &   24(6)    &   29.4       &   37(5)      &           40.0    \\
$6_1^+ \rightarrow 4_1^+$  &   34(7)    &   31.2       &   36$^{+49}_{-31}$ &     41.0    \\
$8_1^+ \rightarrow 6_1^+$  &   6.5(17)  &    7.8       &   4(3)$\times 10^{1}$ &   1.7    \\
$2_2^+ \rightarrow 2_1^+$  &   1.11(60)$\times 10^{2}$ &   8.0        & 62(11) &  19.9    \\
$2_2^+ \rightarrow 0_1^+$  &   25(14)   &    0.2       &   0.130(24)  &            1.3    \\
\hline \hline
\end{tabular}
\label{table2}
\end{table}

\begin{table}[t]
\caption{$B(E2)$ values for the positive-parity yrast states and
         some collective states of $^{74}$Ge and $^{76}$Ge.
         The experimental data are taken from \cite{Robinson11}.}
\begin{tabular}{lccccc}   \hline \hline
        & \multicolumn{2}{c}{\hspace{1.0cm}$^{74}$Ge  [W.u.]}\hspace{1.0cm}
        & \multicolumn{3}{c}{\hspace{1.0cm}$^{76}$Ge  [W.u.]\hspace{1.0cm}}  \\ \hline
$J_i^\pi \rightarrow J_f^\pi$\hspace{1.0cm} &
\hspace{0.5cm}exp\hspace{0.5cm} & \hspace{0.5cm}cal\hspace{0.5cm}
 & \hspace{0.5cm}exp\hspace{0.5cm} & \hspace{0.5cm}cal\hspace{0.5cm}   \\ \hline
$2_1^+ \rightarrow 0_1^+$   &  33.0(4)  &   36.7   &  29(1)     &  31.7   \\
$4_1^+ \rightarrow 2_1^+$   &  41(3)    &   51.3   &  38(9)     &  42.4   \\
$6_1^+ \rightarrow 4_1^+$   &           &   50.7   &            &   0.7   \\
$8_1^+ \rightarrow 6_1^+$   &           &   44.3   &            &  27.6   \\
$2_2^+ \rightarrow 2_1^+$   &  43(6)    &   39.6   &  42(9)     &  33.7   \\
$2_2^+ \rightarrow 0_1^+$   &  0.71(11) &    1.7   &  0.90(22)  &   0.0   \\ \hline \hline
\end{tabular}
\label{table3}
\end{table}

\begin{figure}[t]
\includegraphics[totalheight=7.0cm]{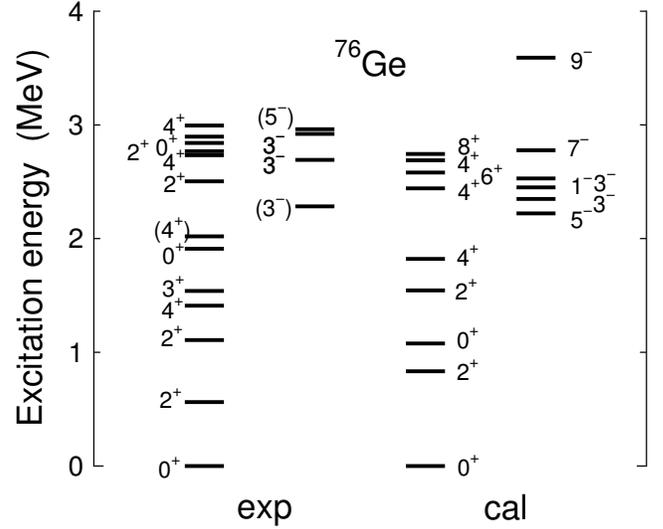}
\caption{Comparison of energy levels between the shell-model results
and the experimental data for $^{76}$Ge. Experimental data are taken
from \cite{ENDSF}. Conventions are the same as those in Fig.
\ref{fig11}.}
  \label{fig16}
\end{figure}

Energy levels of $^{74}$Ge are shown in Fig. \ref{fig15}. The
shell-model description for this isotope is found to be reasonable.
Comparing the energy levels between $^{74}$Ge and $^{72}$Ge, the
$0_{2}^{+}$ state in $^{74}$Ge is considerably higher than that in
$^{72}$Ge. This feature is successfully reproduced by the
calculation. For the negative-parity levels, the $1_{1}^{-}$ state
is the lowest in experiment while the theory gives $3_{1}^{-}$.
Comparing the $E2$ transitions between the calculation and
experiment in Table III, one can see that a reasonable agreement is
obtained, although there is again a problem for $B(E2, 2_{2}^{+}
\rightarrow 2_{1}^{+})$. The large $B(E2)$ values in this isotope
suggest an increasing collectivity beyond $N=40$.

Energy levels of $^{76}$Ge are shown in Fig. \ref{fig16}. The
observed $0_{2}^{+}$ state lies at about 1.9 MeV, but the calculated
one is found much lower than that. The first two $2^{+}$ states are
calculated too high as compared to the data. Nevertheless, the
$B(E2)$ values for this isotope are successfully reproduced by the
calculation, as can be seen in Table III.

\begin{figure}[t]
\includegraphics[totalheight=9.0cm]{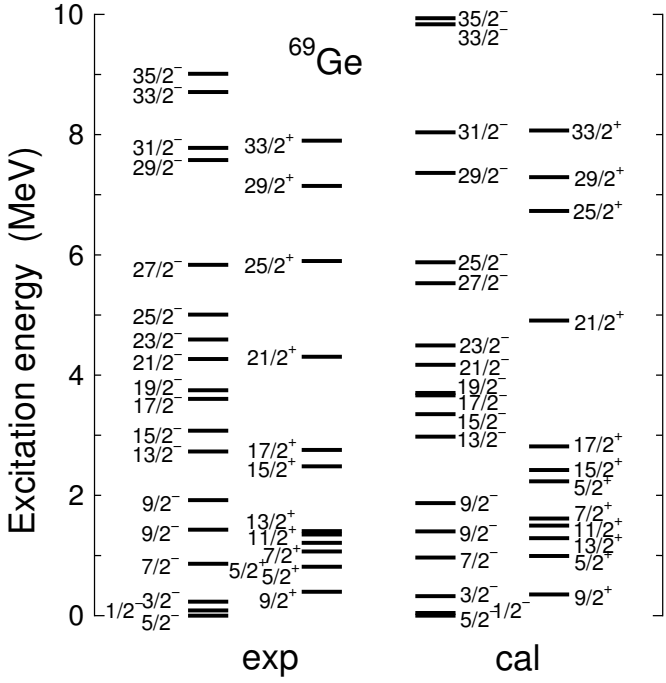}
\caption{Comparison of energy levels between the shell-model results
and the experimental data for $^{69}$Ge. The experimental (
calculated) energy levels are shown in the left (right) side. In
each side, the first (second) line indicates the negative-
(positive-) parity states. Experimental data are taken from
\cite{ENDSF}.}
  \label{fig17}
\end{figure}
\begin{figure}
\includegraphics[totalheight=8.0cm]{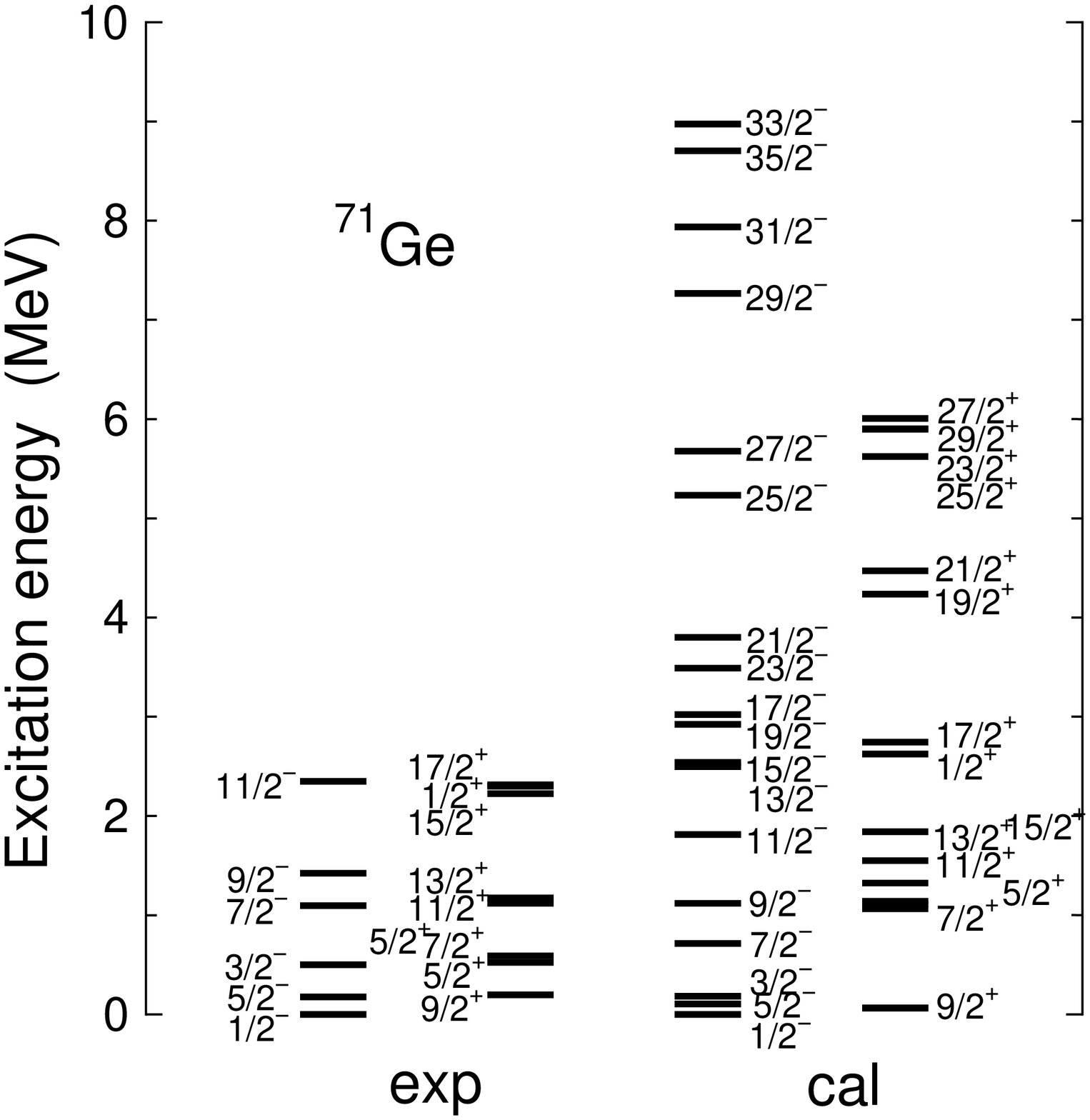}
\caption{Comparison of energy levels between the shell-model results
and the experimental data for $^{71}$Ge. Experimental data are taken
from \cite{ENDSF}. Conventions are the same as those in Fig.
\ref{fig17}.}
  \label{fig18}
\end{figure}

\begin{table}[t]
\caption{$B(E2)$ values for the positive-parity yrast states and
         some collective states of $^{69}$Ge and $^{71}$Ge.
         The experimental data are taken from \cite{ENDSF}.}
\begin{tabular}{lccccc}   \hline \hline
        & \multicolumn{2}{c}{\hspace{1.0cm}$^{69}$Ge  [W.u.]}\hspace{1.0cm}
        & \multicolumn{3}{c}{\hspace{1.0cm}$^{71}$Ge  [W.u.]\hspace{1.0cm}}  \\ \hline
$J_i^\pi \rightarrow J_f^\pi$\hspace{1.0cm} &
\hspace{0.5cm}exp\hspace{0.5cm} & \hspace{0.5cm}cal\hspace{0.5cm}
 & \hspace{0.5cm}exp\hspace{0.5cm} & \hspace{0.5cm}cal\hspace{0.5cm}   \\ \hline
$1/2_1^- \rightarrow 5/2_1^-$     &   0.583(25)   &   5.7   &  6.87(18) &   2.1    \\
$3/2_1^- \rightarrow 1/2_1^-$     &  32(23)       &   7.4   &           &   4.3    \\
$3/2_1^- \rightarrow 5/2_1^-$     &  65(19)       &  21.4   &           &  15.5    \\
$7/2_1^- \rightarrow 3/2_1^-$     &  12(5)        &   6.8   &           &  21.2    \\
$11/2_1^+ \rightarrow 9/2_1^+$    &  27(4)        &  30.2   &  23(6)    &  29.4    \\
$13/2_1^+ \rightarrow 9/2_1^+$    &  23(4)        &  19.1   &  42(9)    &  10.4    \\
$9/2_1^- \rightarrow 5/2_1^-$     &   9.0(17)     &   8.0   &  8.7(18)  &  21.8    \\
$3/2_1^+ \rightarrow 5/2_1^+$     &   5$^{+7}_{-5}\times 10^{1}$  &  1.3  &    &   0.1   \\
$15/2_1^+ \rightarrow 13/2_1^+$   &   2.4(11)     &  18.0   &           &    3.7   \\
$17/2_1^+ \rightarrow 13/2_1^+$   &  12.4(21)     &  18.3   &           &    1.8   \\
$15/2_1^- \rightarrow 13/2_1^-$   &   3$^{+4}_{-3}\times 10^{1}$ & 0.0  &      &   0.2   \\
$17/2_1^+ \rightarrow 15/2_1^+$   &  10(6)        &  0.2    &           &   12.1   \\
$19/2_1^- \rightarrow 15/2_1^-$   &   8.5(10)     &  5.2    &           &    8.2   \\
$21/2_1^- \rightarrow 19/2_1^-$   &   1.2(4)      &  1.1    &           &   20.4   \\
$23/2_1^- \rightarrow 19/2_1^-$   &   1.86(15)    & 11.7    &           &    0.9   \\
 \hline \hline
\end{tabular}
\label{table4}
\end{table}

\subsection{Odd-mass Ge isotopes}

We now turn our discussion on the odd-mass Ge isotopes. Figure
\ref{fig17} shows the energy levels of $^{69}$Ge. A reasonable
correspondence between theory and experiment can be seen for both
the negative- and positive-parity states, and up to about 8 MeV. The
calculation reproduces correctly the ground state $5/2^{-}$ and
other nearby states. The positive-parity states are also in good
agreement with data, while the second excited $5/2_{2}^{+}$ state is
calculated higher than the observed one. This may suggest that this
state has contributions from the $d_{5/2}$ orbit, which is not
included in the present model space. The calculated highest spin
states in Fig. \ref{fig17}, the states with $25/2^{-}$, $33/2^{-}$,
and $35/2^{-}$, are by about 1 MeV higher than the experimental
data. Moreover, we find in Table IV that the most known $E2$
transitions can be reasonably described by the current theoretical
results.

For $^{71}$Ge, energy levels between theory and experiment are
compared in Fig. \ref{fig18}. The calculation describes correctly
the change of the ground state from the $5/2^{-}$ state in $^{69}$Ge
to the $1/2^{-}$ state in $^{71}$Ge. The calculated level sequences
are essentially similar to those of $^{69}$Ge, although there is no
spin-parity assigned levels above 3 MeV in the experimental data.
While the yrast states with negative-parity are predicted at
reasonable excitation energies, the positive-parity states are not
reproduced well. For example, the excited states above the lowest
$9/2^{+}$ state are predicted too high by the calculation. However,
the behavior of the $1/2^{+}$ and $17/2^{+}$ doublet states are
nicely described. In Table IV, the calculated $B(E2)$ values are
compared with the known experimental data.

\begin{figure}[t]
\includegraphics[totalheight=8.5cm]{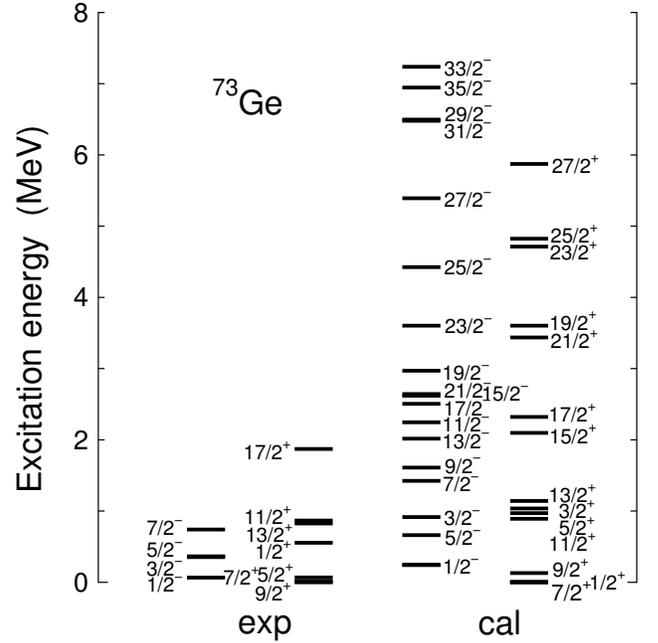}
\caption{Comparison of energy levels between the shell-model results
and the experimental data for $^{73}$Ge. Experimental data are taken
from \cite{ENDSF}. Conventions are the same as those in Fig.
\ref{fig17}.}
  \label{fig19}
\end{figure}
\begin{figure}[t]
\includegraphics[totalheight=8.5cm]{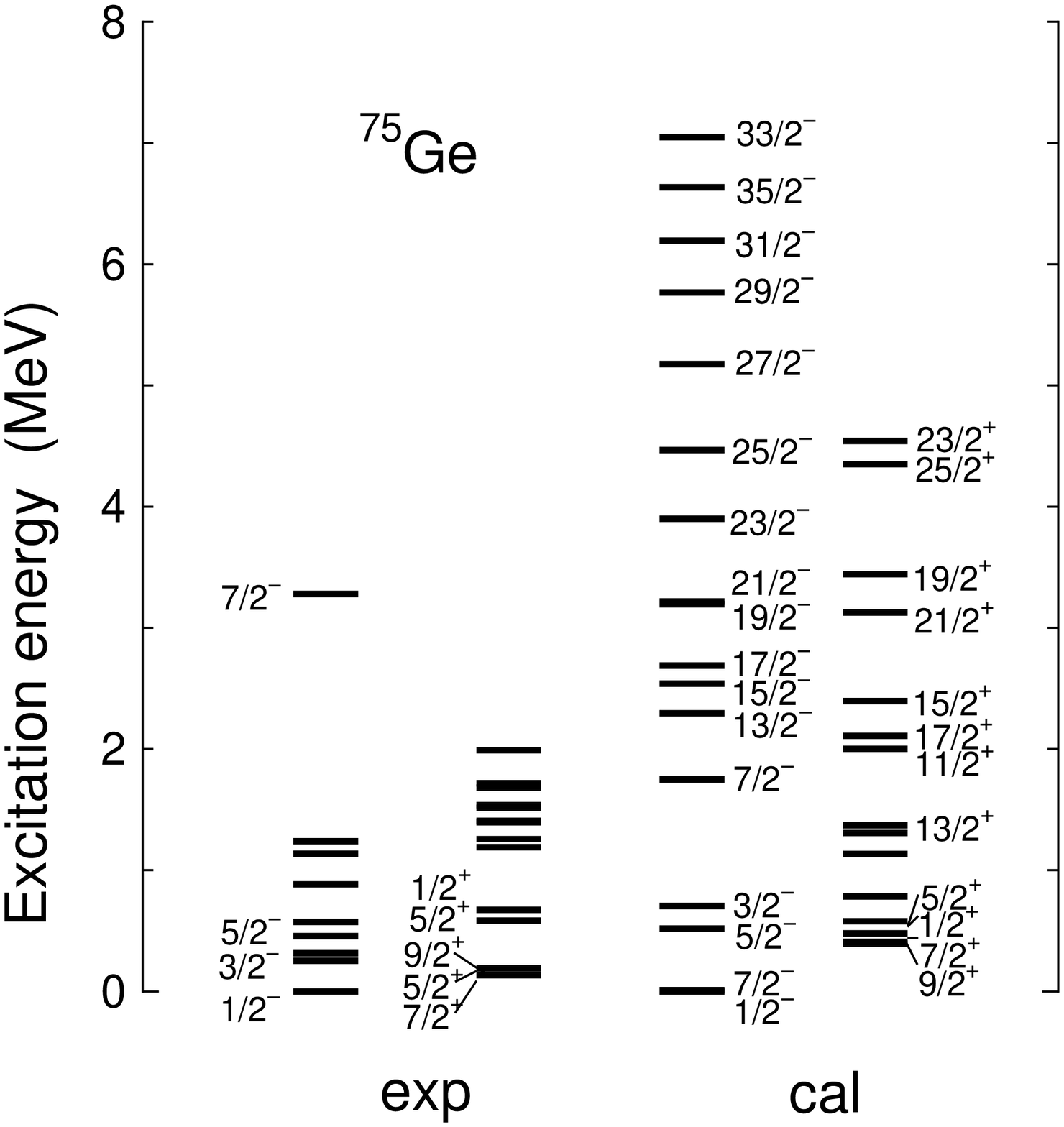}
\caption{Comparison of energy levels between the shell-model results
and the experimental data for $^{75}$Ge. Experimental data are taken
from \cite{ENDSF}. Conventions are the same as those in Fig.
\ref{fig17}.}
  \label{fig20}
\end{figure}

\begin{table}[t]
\caption{$B(E2)$ values for the positive-parity yrast states and
         some collective states of $^{73}$Ge and $^{75}$Ge.
         Non-yrast states are distinguished with their subscripts
         from the yrast states with no subscript.
         The subscript denotes a serial number for each spin $J$.}
\begin{tabular}{lccccc}   \hline \hline
        & \multicolumn{2}{c}{\hspace{1.0cm}$^{73}$Ge  [W.u.]}\hspace{1.0cm}
        & \multicolumn{3}{c}{\hspace{1.0cm}$^{75}$Ge  [W.u.]\hspace{1.0cm}}  \\ \hline
$J_i^\pi \rightarrow J_f^\pi$\hspace{1.0cm} &
\hspace{0.5cm}exp\hspace{0.5cm} & \hspace{0.5cm}cal\hspace{0.5cm}
 & \hspace{0.5cm}exp\hspace{0.5cm} & \hspace{0.5cm}cal\hspace{0.5cm}   \\ \hline
$5/2_1^+ \rightarrow 9/2_1^+$     &   23.1(8)   &  20.5    &           &   28.4     \\
$7/2_1^+ \rightarrow 9/2_1^+$     &   41(8)     &  48.4    &           &   44.0     \\
$13/2_1^+ \rightarrow 9/2_1^+$    &   30(2)     &  29.5    &           &   37.8     \\
$5/2_1^+ \rightarrow 7/2_1^+$     &             &   4.8    & 30(24)    &  18.4  \\
 \hline \hline
\end{tabular}
\label{table5}
\end{table}

As seen in Fig. \ref{fig19}, the shell-model results for $^{73}$Ge
suggest the lowest-lying $7/2^{+}$, $1/2^{+}$ and $9/2^{+}$ states,
however, fail to reproduce the experimental ground-state spin
$9/2^{+}$. The calculated $5/2^{+}$ ($1/2^{+}$) state is higher
(lower) than the experimental one. The failure to describe the
$5/2^{+}$ state can be systematically seen in $^{69}$Ge and
$^{71}$Ge, which could be due to the missing $d_{5/2}$ orbit in the
present model space. The higher-lying yrast states $11/2^{+}$,
$13/2^{+}$, and $17/2^{+}$ are reasonably reproduced, while there
is no observed spectrum above 2 MeV for this isotope. For the
negative-parity states, the calculated $7/2^{-}$ is higher than the
experimental one. The calculated $B(E2)$ values are shown in Table
V.

In Fig. \ref{fig20}, energy levels of the heaviest odd-mass isotope
studied in this paper, $^{75}$Ge, are shown. Energy levels for this
isotope are experimentally observed below 2 MeV. The calculation
reproduces correctly the spin-parity for the ground state $1/2^{-}$.
The low-lying negative-parity states $3/2^{-}$ and $5/2^{-}$ are
reproduced reasonably. For the positive-parity states, the triplet
$9/2^{+}$, $7/2^{+}$ and $5/2^{+}$ states are also in good agreement
with the experimental data. Since a precise level ordering for
states with high density can be sensitive to the interaction matrix
elements, the reproduction of the data may be accidental. As for
electric quadrupole transitions, only the $B(E2)$ value between
$7/2^{+}$ and $5/2^{+}$ has been observed, and is in good agreement
with our calculation, as seen in Table V. The other $E2$ transition
probabilities are predicted.

\begin{figure}[t]
\includegraphics[totalheight=9.0cm]{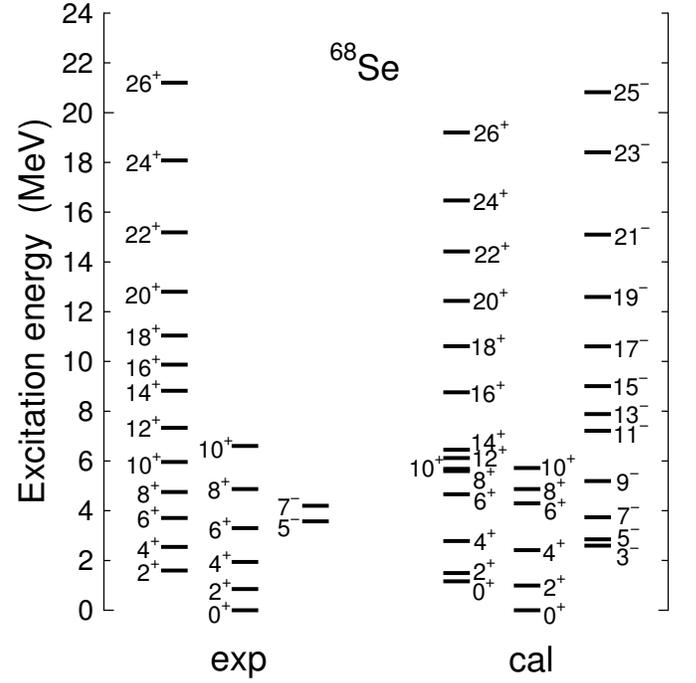}
\caption{Comparison of energy levels between the shell-model results
and the experimental data for $^{68}$Se. Experimental data are taken
from \cite{ENDSF}. Conventions are the same as those in Fig.
\ref{fig11}.}
  \label{fig30}
\end{figure}

\begin{figure}[t]
 \begin{center}
\includegraphics[totalheight=6.0cm]{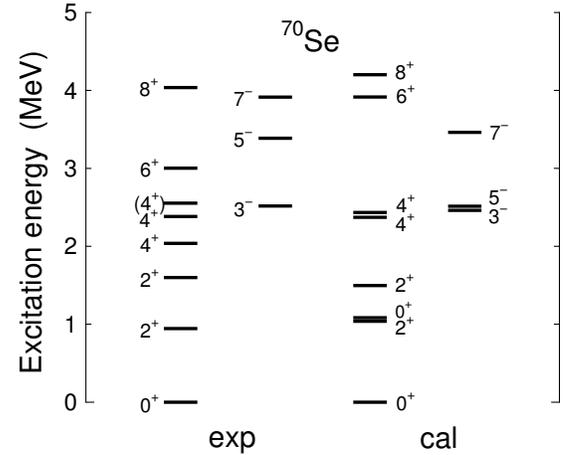}
\caption{Comparison of energy levels between the shell-model results
and the experimental data for $^{70}$Se. Experimental data are taken
from \cite{ENDSF}. Conventions are the same as those in Fig.
\ref{fig11}.}
  \label{fig21}
 \end{center}
\end{figure}

\subsection{Even-even Se isotopes}

The $A\sim 70$ nuclei with $N\sim Z$ are known to exhibit a variety
of nuclear shapes. Experimental data have provided clear evidence
for shape coexistence in this mass region. The experimental
signature of shape coexistence is the presence of a low-lying
$0_{2}^{+}$ state. Determination of shapes was inferred indirectly
from the study of rotational bands, while direct quadrupole
measurements are difficult for short-lived states. The expectation
of oblate deformation in this mass region was confirmed by the
observation of an oblately-deformed ground-state band in $^{68}$Se
\cite{Fischer00}. Figure \ref{fig30} shows the calculated energy
levels up to high-spin states, which are compared with the known
experimental data. The yrast states are fairly well described up to
spin 10. The calculation predicts the first excited $0_{2}^{+}$
state just above the second excited $2^{+}$ state, while the
$0_{2}^{+}$ state is not yet observed in experiment. Our calculation
supports the interpretation of an oblately-deformed rotational
ground-state band in $^{68}$Se. The calculated $E2$ transition
strength $B(E2;2_{1}^{+}\rightarrow 0_{1}^{+})$ = 27.5 W.u. is in
good agreement with the experimental value 27(4) W.u.. The
negative-parity $5^{-}$ and $7^{-}$ levels are well described.

There have been experiments that propose an oblate shape for the
ground state in $^{70}$Se \cite{Ljungvall08}, while the others have
reported a prolate shape \cite{Hurst07}. There have been a number of
microscopic calculations \cite{Ljungvall08,Petrovici02,Hinohara09},
among which the Hartree-Fock-Bogoliubov (HFB) calculation
\cite{Ljungvall08} predicted an oblate shape for the $2^{+}$ and
$4^{+}$ states in the yrast band that changes to a prolate shape for
the $6^{+}$ state. The self-consistent collective coordinate method
\cite{Hinohara09} provided a similar picture. These calculations
show that the ground $0_{1}^{+}$ state has oblate shape, the first
excited $2_{1}^{+}$ and $4_{1}^{+}$ states have mixed the
oblate-prolate configurations, and the $6_{1}^{+}$ state has a
prolate shape. Indeed, the spectroscopic quadrupole moments in the
HFB calculations suggest positive values for the $2_{1}^{+}$ and
$4_{1}^{+}$ states but a negative value for $6_{1}^{+}$ in $^{70}$Se
\cite{Ljungvall08}. Figure \ref{fig21} compares the energy levels
between experiment and calculation. The calculated yrast states are
in good agreement with the experimental ones, except for the $6^{+}$
and $5^{-}$ states. The calculation predicts the low-lying
$0_{2}^{+}$ state above the first excited $2_{1}^{+}$ state, while
the experimental one is not yet observed. In Table VI, the $E2$
transition strengths for the $2_{1}^{+}$, $4_{1}^{+}$, and
$6_{1}^{+}$ states are listed. The calculated $B(E2)$ values explain
fairly well the trend of variation with increasing spin $J$. The
calculated spectroscopic quadrupole moments are 0.17, 0.49, and 0.70
eb, respectively for the $J^{\pi}=2_{1}^{+}, 4_{1}^{+},$ and
$6_{1}^{+}$ states. Using these values and assuming an axial
deformation, we can estimate the quadrupole deformation $\beta_{2}$
as $-0.10$, $-0.22$, and $-0.28$ for the $J^{\pi}=2_{1}^{+},
4_{1}^{+},$ and $6_{1}^{+}$ states, respectively, suggesting that
the yrast states in $^{70}$Se are oblately deformed. This is
consistent with the calculations in Ref. \cite{Ljungvall08}, except
for $J^{\pi}=6_{1}^{+}$.

\begin{figure}[t]
 \begin{center}
\includegraphics[totalheight=6.0cm]{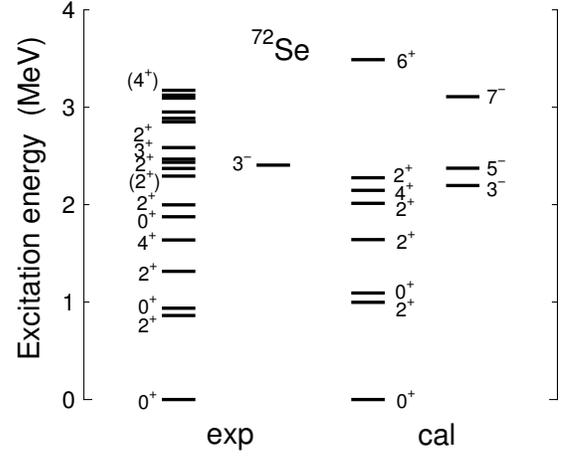}
\caption{Comparison of energy levels between the shell-model results
and the experimental data for $^{72}$Se. Experimental data are taken
from \cite{ENDSF}. Conventions are the same as those in Fig.
\ref{fig11}.}
  \label{fig22}
 \end{center}
\end{figure}

\begin{table}[t]
\caption{$B(E2)$ values for the positive-parity yrast states and
         some collective states of $^{70}$Se and $^{72}$Se.
        The experimental data are taken from \cite{Ljungvall08,ENDSF}.}
\begin{tabular}{lccccc}   \hline \hline
        & \multicolumn{2}{c}{\hspace{1.0cm}$^{70}$Se  [W.u.]}\hspace{1.0cm}
         & \multicolumn{3}{c}{\hspace{1.0cm}$^{72}$Se  [W.u.]\hspace{1.0cm}}  \\ \hline
$J_i^\pi \rightarrow J_f^\pi$\hspace{1.0cm} &
\hspace{0.5cm}exp\hspace{0.5cm} & \hspace{0.5cm}cal\hspace{0.5cm}
 & \hspace{0.5cm}exp\hspace{0.5cm} & \hspace{0.5cm}cal\hspace{0.5cm}   \\ \hline
$2_1^+ \rightarrow 0_1^+$   &  44(9)  & 27.1  &  23.7(17)  &  21.9  \\
$4_1^+ \rightarrow 2_1^+$   &  21(5)  & 34.7  &  55(5)     &  36.5  \\
$6_1^+ \rightarrow 4_1^+$   &  15(4)  & 42.9  &  65(5)     &  48.5
\\  \hline \hline
\end{tabular}
\label{table6}
\end{table}

\begin{figure}[t]
 \begin{center}
\includegraphics[totalheight=6.0cm]{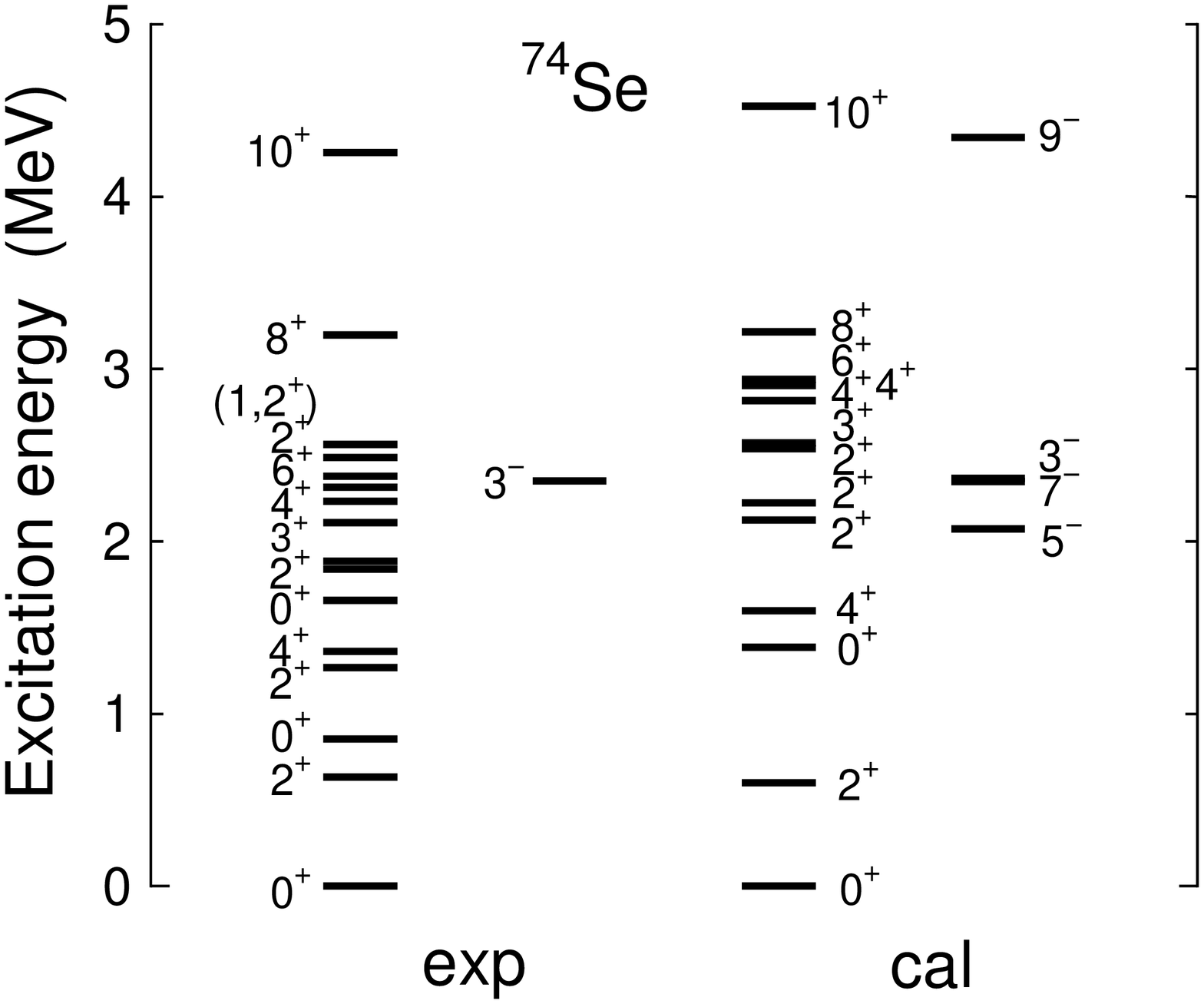}
\caption{Comparison of energy levels between the shell-model results
and the experimental data for $^{74}$Se. Experimental data are taken
from \cite{ENDSF}. Conventions are the same as those in Fig.
\ref{fig11}.}
  \label{fig23}
 \end{center}
\end{figure}

\begin{figure}[t]
 \begin{center}
\includegraphics[totalheight=7.0cm]{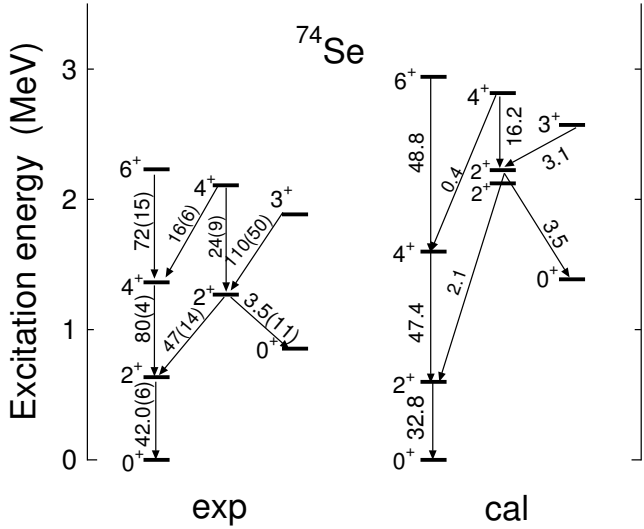}
\caption{Comparison of level scheme between the shell-model results
and the experimental data for $^{74}$Se. Experimental data are taken
from \cite{ENDSF}. The numbers with the arrows represent the
absolute $B(E2)$ transition strengths (in W.u.).}
  \label{fig24}
 \end{center}
\end{figure}

$^{72}$Se is another well-known example that exhibits an
oblate-prolate shape coexistence. The ground-state band in $^{72}$Se
in the HFB-based configuration mixing method shows a positive value
of spectroscopic quadrupole moment for the $2_{1}^{+}$ state that
turns into negative for states above $4_{1}^{+}$
\cite{Ljungvall08,McCutchan11}. The present calculation suggests
positive values for $J^{\pi}=2^{+}-6^{+}$ but a negative value for
$J^{\pi}=8^{+}$. The moment of inertia in the ground-state band
exhibits anomalous behavior at the low-spin states of $^{70,72}$Se,
while the high-spin states with $J>8$ follow a smoothly varying
moment of inertia interpreted as collective rotation. Thus as
$^{70,72}$Se rotate, their ground-state band evolves quickly into a
prolate collective-rotation, while the $2_{1}^{+}$ state can be
associated with an oblate shape. The signature of shape coexistence
is the presence of a low-lying first excited $0_{2}^{+}$ state. In
$^{72}$Se, the observed $0_{2}^{+}$ state lies only 75 keV above the
$2_{1}^{+}$ state. As seen in Fig. \ref{fig22}, the calculation
produces exactly the $0_{2}^{+}$ state just above the $2_{1}^{+}$
state. The calculated $B(E2;2_{1}^{+}\rightarrow 0_{1}^{+})$ is also
in good agreement with the experimental data. Thus the present
calculations describe well the shape coexistence in $^{72}$Se.

There have been many discussions on the structure of $^{74}$Se.
Quite recently, it has been shown that the low-lying states of
$^{74}$Se can be described as a coexistence of near-spherical and
prolate shapes \cite{McCutchan13}. As seen in Figs. \ref{fig23} and
\ref{fig24}, the experimental data indicate that the $4_{1}^{+}$,
$2_{2}^{+}$, and $0_{2}^{+}$ levels may be regarded as candidates of
two-phonon states. Indeed, the $B(E2)$ transition strengths agree
well with the predictions for a spherical structure, as can be seen
in Fig. \ref{fig24}. The small $B(E2,2_{2}^{+}\rightarrow
0_{2}^{+})=3.5$ W.u. is consistent with the interpretation that both
states are members of the two-phonon states. At high spins, the
yrast states become highly deformed with a prolate shape, as the
moment of inertia becomes larger. Our calculations are compared with
the experimental data in Fig. \ref{fig23}. The results reproduce
qualitatively the experimental data. Large deviations are seen for
the calculated first excited $0_{2}^{+}$ state and the first excited
$6^{+}$ state, which are higher in energy than the data. The
$B(E2,J\rightarrow J-2)$ values for the $E2$ transitions de-exciting
the $J^{\pi}=2_{1}^{+}, 4_{1}^{+},$ and $6_{1}^{+}$ states are shown
in Fig. \ref{fig24}. These $E2$ transitions indicate a trend of
increasing magnitude with increasing spin, while the calculated
magnitudes are smaller than the experimental data reported in Ref.
\cite{Ljungvall08,McCutchan13}.

\begin{figure}[t]
\includegraphics[totalheight=10cm]{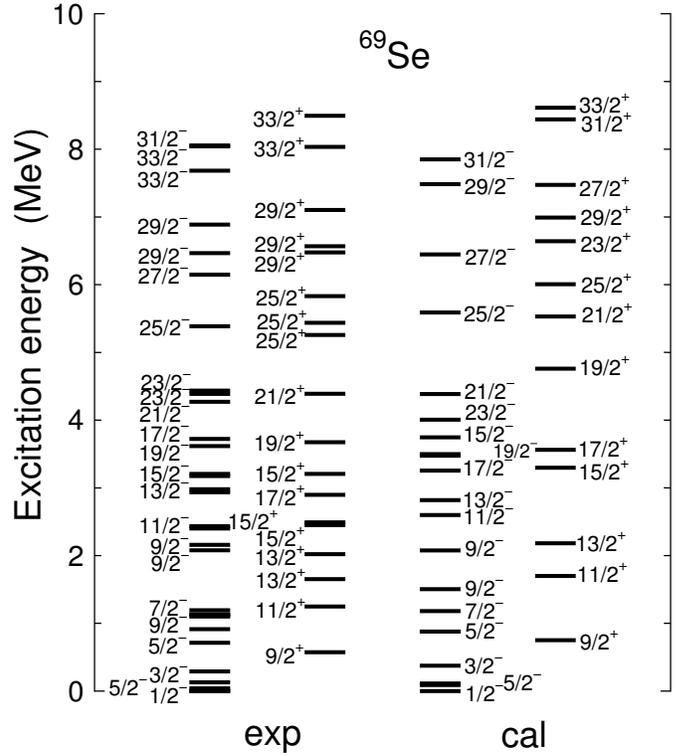}
\caption{Comparison of energy levels between the shell-model results
and the experimental data for $^{69}$Se. Experimental data are taken
from \cite{ENDSF}. Conventions are the same as those in Fig.
\ref{fig11}.}
  \label{fig25}
\end{figure}

\begin{figure}[t]
\includegraphics[totalheight=10cm]{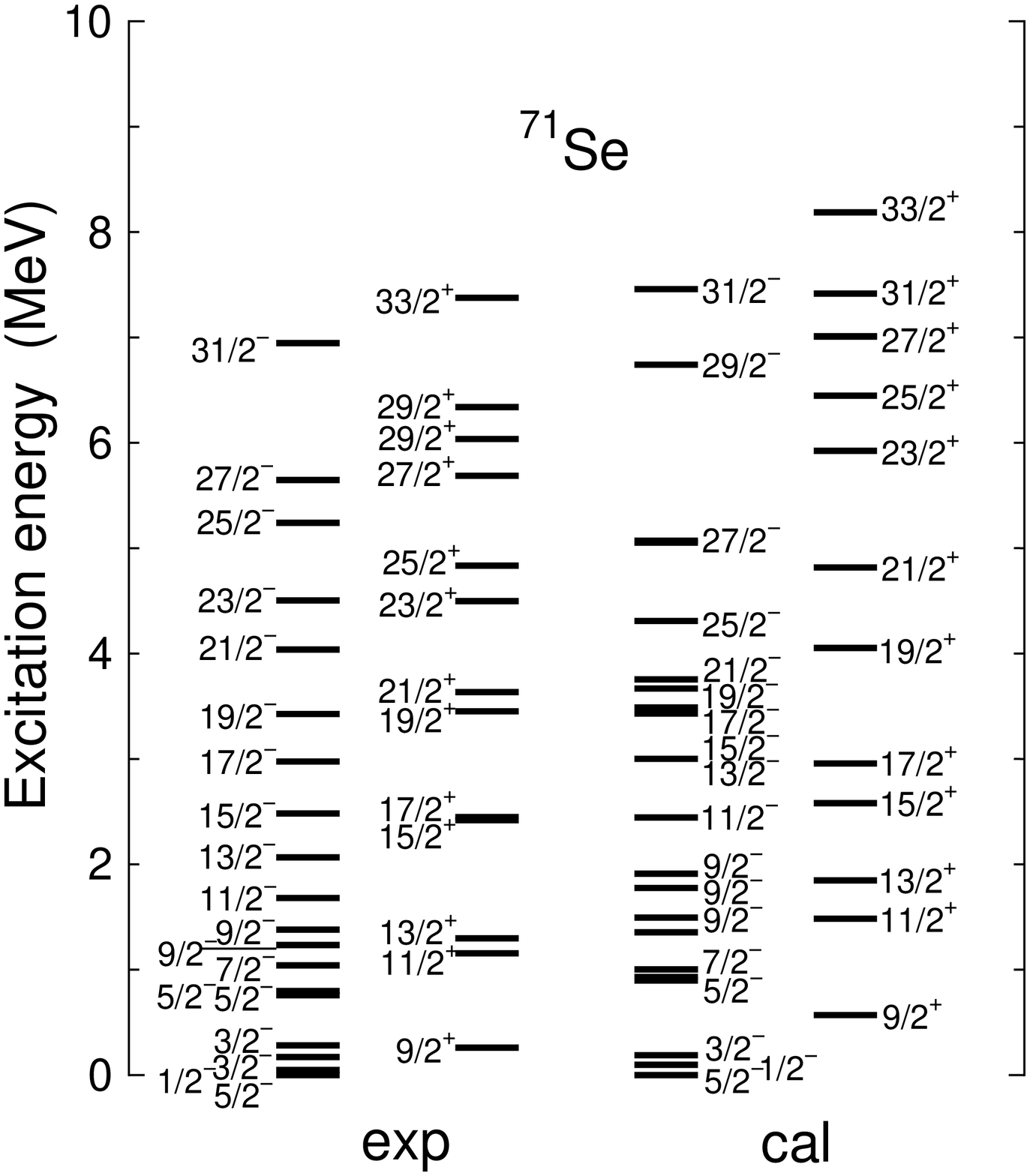}
\caption{Comparison of energy levels between the shell-model results
and the experimental data for $^{71}$Se. Experimental data are taken
from \cite{ENDSF}. Conventions are the same as those in Fig.
\ref{fig11}.}
  \label{fig26}
\end{figure}

\begin{figure}[t]
\includegraphics[totalheight=10.5cm]{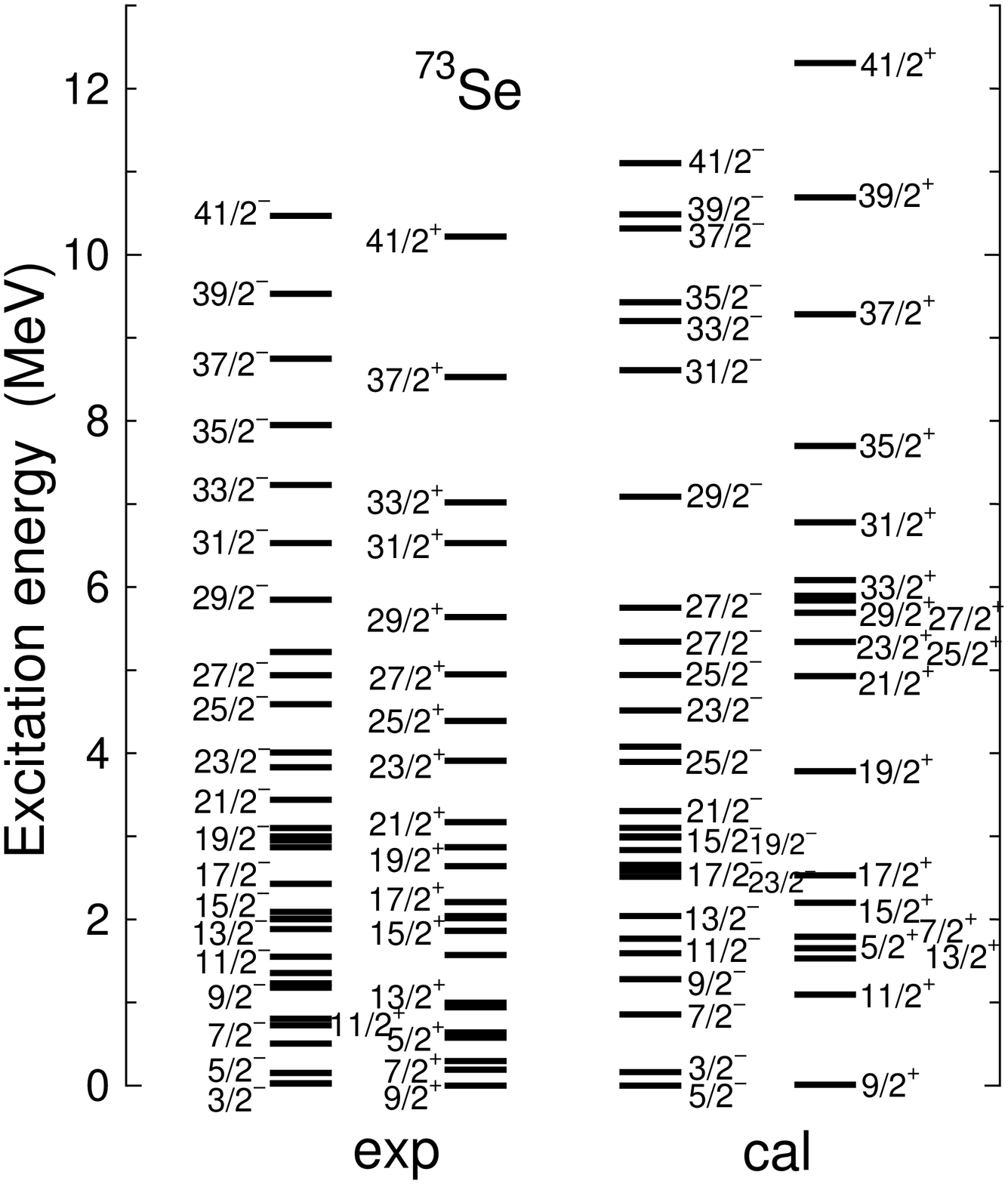}
\caption{Comparison of energy levels between the shell-model results
and the experimental data for $^{73}$Se. Experimental data are taken
from \cite{ENDSF}. Conventions are the same as those in Fig.
\ref{fig11}.}
  \label{fig27}
\end{figure}

\begin{table}[t]
\caption{$B(E2)$ values for the positive-parity yrast states and
         some collective states of $^{71}$Se and $^{73}$Se.
      The experimental data are taken from \cite{ENDSF,Howe12,Loritz99}.}
\begin{tabular}{lccccc}   \hline \hline
        & \multicolumn{2}{c}{\hspace{1.0cm}$^{71}$Se  [W.u.]}\hspace{1.0cm}
        & \multicolumn{3}{c}{\hspace{1.0cm}$^{73}$Se  [W.u.]\hspace{1.0cm}}  \\ \hline
$J_i^\pi \rightarrow J_f^\pi$\hspace{1.0cm} & \hspace{0.5cm}exp\hspace{0.5cm} & \hspace{0.5cm}cal\hspace{0.5cm}
 & \hspace{0.5cm}exp\hspace{0.5cm} & \hspace{0.5cm}cal\hspace{0.5cm}   \\ \hline
$1/2_1^- \rightarrow 5/2_1^-$      &    1.63(21)   &  1.6    &           &  6.0   \\
$13/2_1^+ \rightarrow 9/2_1^+$     &   30(10)      &  8.7    &   41(4)   &  9.1   \\
$17/2_1^+ \rightarrow 13/2_1^+$    &               &  8.5    &  140(40)  &  3.8   \\
$21/2_1^+ \rightarrow 17/2_1^+$    &               &  9.4    &  110(30)  &  1.3   \\
$5/2_1^+ \rightarrow 9/2_1^+$      &               & 10.6    &   22(6)   &  6.7   \\
 \hline \hline
\end{tabular}
\label{table7}
\end{table}

\subsection{Odd-mass Se isotopes}

Odd-mass Se nuclei in this mass region exhibit also rapidly-changing
structures \cite{Stefanescu04,Howe12,Loritz99}. The study of
odd-mass nuclei completes the information on the structural
properties of shape-coexisting nuclei, and thus can be helpful in
understanding the effects produced by coupling of the last particle
with an even-even core, which gives more insight into the interplay
between oblate and prolate shapes. As one can see from the levels in
Fig. \ref{fig25}, for $^{69}$Se the low-lying negative-parity states
suggest a small deformation and appear to be dominated by
single-particle excitations, where the $1/2^{-}$ state is the ground
state \cite{Stefanescu04}. These states have irregular
level-spacings that do not follow a rotational pattern. Our
calculation reproduces nicely the ground state with correct spin and
parity $1/2^{-}$ and the low-lying states with $5/2^{-}$ and
$3/2^{-}$, as well as $9/2^{+}$. The reported experimental results
\cite{Stefanescu04} have confirmed that the oblate $g_{9/2}$ band in
$^{69}$Se is generated by coupling of the odd neutron to $^{68}$Se,
which also has an oblate shape in the ground state. The $9/2^{+}$
level is an isomeric state and the bandhead of the positive-parity
band. The one-quasiparticle states of the yrast and near yrast
positive-parity bands have been discussed in terms of a particle
coupled to a rigid triaxial rotor model \cite{Stefanescu04}. Above
spin $15/2^-$ around 3 MeV, collective excitations predominate, and
can be regarded as the $g_{9/2}$ neutron coupled to the $3^{-}$,
$5^{-}$, and $7^{-}$ states of the neighboring even-even nucleus
$^{68}$Se. The high-lying negative-parity states $25/2^{-}$,
$27/2^{-}$, $29/2^{-}$, and $31/2^{-}$ above 5 MeV can be
interpreted as the $g_{9/2}$ neutron coupled to $8^{-}-11^{-}$
states in $^{68}$Se.

The $^{71}$Se nucleus, lying between $^{69}$Se and the collective
$^{73}$Se, represents a transitional feature in the level scheme
\cite{Howe12}. The spin-parity assignment of $5/2^{-}$ was
established for the ground state, and the first excited state was
observed as $1/2^{-}$. The calculations reproduce well the low-lying
energy scheme. As seen in Fig. \ref{fig26}, the $5/2^{-}$ ground
state found in experiment can be reproduced by the calculation, and
the $1/2^{-}$ and $3/2^{-}$ states are also well described. The
lowest positive-parity $9/2^{+}$ state is an isomer with a half-life
($T_{1/2}=19\mu s$). The high-lying states above 2 MeV are
interpreted as the coupled states of a $g_{9/2}$ neutron to the
lowest $3^{-}$, $5^{-}$, and $7^{-}$ states of the $^{70}$Se core.
Thus the $^{71}$Se nucleus shows transitional characters between
neighboring nuclei that indicate single-particle excitations and
collectivity. The $E2$ transition probabilities between the
low-lying positive-parity states are shown in Table VII. The
shell-model calculation predicts smaller $B(E2)$ values than the
experimental data.

In contrast to $^{71}$Se, $^{73}$Se has a different picture with
highly collective behavior \cite{Loritz99}. The calculation predicts
well the ground state $9/2^{+}$. The yrast positive-parity states
with high spins show a collective character with large $B(E2)$
values, as seen in Table VII. However, the calculation cannot
describe these features. This may be due to the missing $d_{5/2}$
orbit in the present model space, because the attractive $T=0$
monopole interaction between the $\pi f_{5/2}$ and $\nu d_{5/2}$
orbits would push down the $\nu d_{5/2}$ orbit when protons occupy
the $\pi f_{5/2}$ orbit and neutrons occupy the $\pi f_{5/2}$ orbit
at high spins. The total Routhian surface (TRS) calculation for this
band \cite{Kaplan91} indicated the $\gamma$-softness and some
potential minima. Thus the shape coexistence picture found in the
neighboring even-even isotopes $^{70,72,74}$Se persists in
$^{73}$Se.

\section{Summary and conclusions}
\label{sec6}

We have recently proposed a new effective interaction PMMU
\cite{Kaneko14} for shell-model calculations within the
$pf_{5/2}g_{9/2}$ model space. We employ the pairing-plus-multipole
Hamiltonian and adopt the monopole interaction obtained by empirical
fits starting from the monopole-based universal force. For this
shell model space, 14 monopole terms of $V_{m}^{MU}$ are modified so
as to fit energy levels, $E2$ transitions, and binding energies for
91 nuclei in the $pf_{5/2}g_{9/2}$ space. In the present paper, this
new PMMU interaction has been examined systematically with a wide
range of nuclei of the $pf_{5/2}g_{9/2}$ mass region, from the
neutron-deficient $N=Z$ nuclei to neutron-rich ones. The calculation
has been performed not only for the low-lying states, but also for
high excitations for which collective rotations are dominant.

It has been demonstrated that the proposed shell model works well
for most nuclei considered in the calculation, except for the Ni
isotopes. Large discrepancies with the experimental data are found
for the Ni isotopes. This would be due to the effects of the
$f_{7/2}$ orbit, which is not included in the present model space.
The effects of the missing $d_{5/2}$ orbit have also been discussed
with the shell-model results for nuclei with neutron numbers beyond
$N=36$. The $d_{5/2}$ orbit could be necessary for generating a
strong collectivity in the heavier $N=Z$ nuclei beyond the mass
$A=72$.

In the present paper, we have particularly studied the $0_{2}^{+}$
state, which is the key measure of shape-coexistence. It has been
found that $0_{2}^{+}$ has an irregular behavior around the subshell
at $N=40$. Our shell-model calculation using the PMMU interaction
has well explained the behavior of this subshell closure. We have
also discussed the excited spectra of the Ge and Se isotopes in
detail, and concluded that the PMMU interaction also describe the
high-spin states.

As mentioned in our previous paper \cite{Kaneko14}, the PMMU model
could be a feasible method to unify different shell models and could
extend shell-model calculations to heavier systems. In this way, one
may begin to talk about universality for shell models. With the
inclusion of the $d_{5/2}$ orbit, shell-model calculations in the
$pf_{5/2}g_{9/2}d_{5/2}$ model space are in progress. We expect that
the PMMU model could be applicable for the neutron-rich nuclei in
the $fpg$-shell region \cite{Kaneko08}, the $sd$-$pf$ shell region
\cite{Kaneko11}, and even the heavy neutron-rich nuclei around
$^{132}$Sn \cite{Jin11,Wang13,Wang14}, where the
pairing-plus-multipole Hamiltonian have been applied and monopole
corrections have been found to be important.

\begin{acknowledgments}
Research at SJTU was supported by the National Natural Science
Foundation of China (Nos. 11135005, 11575112), by the 973 Program of
China (No. 2013CB834401), and by the Open Project Program of State
Key Laboratory of Theoretical Physics, Institute of Theoretical
Physics, Chinese Academy of Sciences, China (No. Y5KF141CJ1).
\end{acknowledgments}




\begin{thebibliography}{00}

\bibitem{Bernas82} M. Bernas, Ph. Dessagne, M. Langevin, J. Payet, F. Pougheon,
and P. Roussel, Phys. Lett. {\bf 113B}, 279 (1982).

\bibitem{Broda95} R. Broda, B. Fornal, W. Kr\'olas, T. Pawlat, D. Bazzacco,
S. Lunardi, C. Rossi-Alvarez, R. Menegazzo, G. de Angelis, P.
Bednarczyk, J. Rico, D. De Acu\~na, P. J. Daly, R. H. Mayer, M.
Sferrazza, H. Grawe, K. H. Maier, and R. Schubart , Phys. Rev. Lett.
{\bf 74}, 868 (1995).

\bibitem{Guenaut07} C. Gu\'enaut, G. Audi, D. Beck, K. Blaum, G. Bollen, P. Delahaye,
F. Herfurth, A. Kellerbauer, H.-J. Kluge, J. Libert, D. Lunney, S.
Schwarz, L. Schweikhard, and C. Yazidjian, Phys. Rev. C {\bf 75},
044303 (2007).

\bibitem{Mueller99} W. F. Mueller, B. Bruyneel, S. Franchoo, H. Grawe, M. Huyse,
U. K\"oster, K.-L. Kratz, K. Kruglov, Y. Kudryavtsev, B. Pfeiffer,
R. Raabe, I. Reusen, P. Thirolf, P. Van Duppen, J. Van Roosbroeck,
L. Vermeeren, W. B. Walters, and L. Weissman, Phys. Rev. Lett. {\bf
83}, 3613 (1999).

\bibitem{Sorlin02} O. Sorlin, S. Leenhardt, C. Donzaud, J. Duprat, F. Azaiez,
F. Nowacki, H. Grawe, Zs. Dombr\'adi, F. Amorini, A. Astier, D.
Baiborodin, M. Belleguic, C. Borcea, C. Bourgeois, D. M. Cullen, Z.
Dlouhy, E. Dragulescu, M. G\'orska, S. Gr\'evy, D.
Guillemaud-Mueller, G. Hagemann, B. Herskind, J. Kiener, R. Lemmon,
M. Lewitowicz, S. M. Lukyanov, P. Mayet, F. de Oliveira Santos, D.
Pantalica, Yu.-E. Penionzhkevich, F. Pougheon, A. Poves, N. Redon,
M. G. Saint-Laurent, J. A. Scarpaci, G. Sletten, M. Stanoiu, O.
Tarasov, and Ch. Theisen, Phys. Rev. Lett. {\bf 88}, 092501 (2002).

\bibitem{Langanke03} K. Langanke, J. Terasaki, F. Nowaki, D. J. Dean,
and W. Nazarewicz, Phys. Rev. C {\bf 67}, 044314 (2003).

\bibitem{Luttke12} R. L\"uttke, E. A. McCutchan, V. Werner, K. Aleksandrova,
S. Atwater, H. Ai, R. J. Casperson, R. F. Casten, A. Heinz, A. F.
Mertz, J. Qian, B. Shoraka, J. R. Terry, E. Williams, and R.
Winkler, Phys. Rev. C {\bf 85}, 017301 (2012).

\bibitem{Gurdal13} G. G\"urdal, E. A. Stefanova, P. Boutachkov, D. A. Torres,
G. J. Kumbartzki, N. Benczer-Koller, Y. Y. Sharon, L. Zamick, S. J.
Q. Robinson, T. Ahn, V. Anagnostatou, C. Bernards, M. Elvers, A.
Heinz, G. Ilie, D. Radeck, D. Savran, V. Werner, and E. Williams,
Phys. Rev. C {\bf 88}, 014301 (2013).

\bibitem{Hurst07} A. M. Hurst, P. A. Butler, D. G. Jenkins, P. Delahaye,
F.Wenander, F. Ames, C. J. Barton, T. Behrens, A. B\"urger, J.
Cederk\"all, E. Cl\'ement, T. Czosnyka, T. Davinson, G. de Angelis,
J. Eberth, A. Ekstr\"om, S. Franchoo, G. Georgiev, A. G\"orgen,
R.-D. Herzberg, M. Huyse, O. Ivanov, J. Iwanicki, G. D. Jones, P.
Kent, U. K\"oster, T. Kr\"oll, R. Kr\"ucken, A. C. Larsen, M.
Nespolo, M. Pantea, E. S. Paul, M. Petri, H. Scheit, T. Sieber, S.
Siem, J. F. Smith, A. Steer, I. Stefanescu, N. U. H. Syed, J. Van de
Walle, P. Van Duppen, R. Wadsworth, N. Warr, D. Weisshaar, and M.
Zieli\'nska, Phys. Rev. Lett. {\bf 98}, 072501 (2007).

\bibitem{Ljungvall08} J. Ljungvall, A. G\"orgen, M. Girod, J.-P. Delaroche,
A. Dewald, C. Dossat, E. Farnea, W. Korten, B. Melon, R. Menegazzo,
A. Obertelli, R. Orlandi, P. Petkov, T. Pissulla, S. Siem, R. P.
Singh, J. Srebrny, Ch. Theisen, C. A. Ur, J. J. Valiente-Dob\'on, K.
O. Zell, and M. Zieli\'nska , Phys. Rev. Lett. {\bf 100}, 102502
(2008).

\bibitem{McCutchan11} E. A. McCutchan, C. J. Lister, T. Ahn, R. J. Casperson,
A. Heinz, G. Ilie, J. Qian, E. Williams, R. Winkler, and V. Werner,
Phys. Rev. C {\bf 83}, 024310 (2011).

\bibitem{McCutchan13} E. A. McCutchan, C. J. Lister, T. Ahn, V. Anagnostatou,
N. Cooper, M. Elvers, P. Goddard, A. Heinz, G. Ilie, D. Radeck, D.
Savran, and V. Werner, Phys. Rev. C {\bf 87}, 014307 (2013).

\bibitem{Lister87} C. J. Lister, M. Campbell, A. A. Chishti, W. Gelletly,
L. Goettig, R. Moscrop, B. J. Varley, A. N. James, T. Morrison, H.
G. Price, J. Simpson, K. Connel, and O. Skeppstedt, Phys. Rev. Lett.
{\bf 59}, 1270 (1987).

\bibitem{Doring98} J. Doring, G. D. Johns, M. A. Riley, S. L. Tabor, Y. Sun, and
J. A. Sheikh, Phys. Rev. C {\bf 57}, 2912 (1998).

\bibitem{Ennis91} P. J. Ennis, C. J. Lister, W. Gelletly, H. G. Price, B. J. Varley,
P. A. Butler, T. Hoare, S. Cwoik, and W. Nazarewicz, Nucl. Phys. {\bf A535}, 392 (1991).

\bibitem{Fischer00} S. M. Fischer, D. P. Balamuth, P. A. Hausladen, C. J. Lister,
M. P. Carpenter, D. Seweryniak, and J. Schwartz, Phys. Rev. Lett. {\bf 84}, 4064 (2000).

\bibitem{Kaneko04} K. Kaneko, M. Hasegawa, and T. Mizusaki,
Phys. Rev. C {\bf 70}, 051301 (2004).

\bibitem{Obertelli09} A. Obertelli, T. Baugher, D. Bazin, J. -P. Delaroche,
F. Flavigny, A. Gade, M. Girod, T. Glasmacher, A. Goergen, G. F.
Grinyer, W. Korten, J. Ljungvall, S. McDaniel, A. Ratkiewicz, B.
Sulignano, and D. Weisshaar, Phys. Rev. C {\bf 80}, 031304(R)
(2009).

\bibitem{Kaneko02} K. Kaneko, M. Hasegawa, and T. Mizusaki,
Phys. Rev. C {\bf 66}, 051306(R) (2002).

\bibitem{Hasegawa07} M. Hasegawa, T. Mizusaki, K. Kaneko, and Y. Sun, Nucl. Phys.
{\bf A 789}, 46 (2007).

\bibitem{Honma09} M. Honma, T. Otsuka, T. Mizusaki, and M. Hjorth-Jensen,
Phys. Rev. C {\bf 80}, 064323 (2009).

\bibitem{Robinson11} S. J. Q. Robinson, L. Zamick, and Y. Y. Sharon, Phys. Rev. C
{\bf 83}, 027302 (2011).

\bibitem{Petrovici02} A. Petrovici, K. W. Schmid, and A. Faessler, Nucl. Phys. {\bf A710}
246 (2002); {\bf 728} 396 (2003).

\bibitem{Sun04} Y. Sun, Eur. Phys. J. A {\bf 20}, 133 (2004).

\bibitem{Sun05} Y. Sun, M. Wiescher, A. Aprahamian, and J. Fisker,
Nucl. Phys. {\bf A 758}, 765 (2005).

\bibitem{Hinohara09} N. Hinohara, T. Natatsukasa, M. Matsuo, and K. Matsuyanagi,
Phys. Rev. C {\bf 80}, 041305 (2009).

\bibitem{Yang10} Y.-C. Yang, Y. Sun, K. Kaneko, and M. Hasegawa, Phys. Rev. C
{\bf 82}, 031304(R) (2010).

\bibitem{Padilla05} E. Padilla-Rodal, A. Galindo-Uribarri, C. Baktash, J. C.
Batchelder, J. R. Beene, R. Bijker, B. A. Brown, O. Casta\~nos, B.
Fuentes, J. Gomez del Campo, P. A. Hausladen, Y. Larochelle, A. F.
Lisetskiy, P. E. Mueller, D. C. Radford, D. W. Stracener, J. P.
Urrego, R. L. Varner, and C.-H. Yu, Phys. Rev. Lett. {\bf 94},
122501 (2005).

\bibitem{Lisetskiy04} A. F. Lisetskiy, B. A. Brown, M. Horoi, and H. Grawe,
Phys. Rev. C {\bf 70}, 044314 (2004).

\bibitem{jj4b} B. A. Brown and A. F. Lisetskiy (unpublished).

\bibitem{Kuo68} T. T. S. Kuo and G. E. Brown, Nucl. Phys. A {\bf 114}, 241 (1968).

\bibitem{Jensen95} M. Hjorth-Jensen, T. T. S. Kuo, and E. Osnes,
 Phys. Rep. {\bf 261}, 125 (1995).

\bibitem{Epelbaum09} E. Epelbaum, H.-W. Hammer and U.-G. Meisner,
Rev. Mod. Phys. {\bf 81}, 1773 (2009).

\bibitem{Honma04} M. Honma, T. Otsuka, B. A. Brown, and T. Mizusaki,
Phys. Rev. C {\bf 69}, 034335 (2004).

\bibitem{Brown88} B. A. Brown and B. H. Wildenthal, Ann. Rev. Nucl. Part. Sci.
{\bf 38}, 29 (1988).

\bibitem{Brown06} B. A. Brown and W. A. Richter, Phys. Rev. C {\bf 74}, 034315 (2006).

\bibitem{Poves01} A. Poves, J. Sanchez-Solano, E. Caurier, and F. Nowacki,
 Nucl. Phys. A {\bf 694}, 157 (2001).

\bibitem{LNPS} S. M. Lenzi, F. Nowacki, A. Poves, and K. Sieja, Phys. Rev.
C {\bf 82}, 054301 (2010).

\bibitem{Kahana69} S. Kahana, H. C. Lee, and C. K. Scott, Phys. Rev. {\bf 180}, 956
(1969).

\bibitem{Kisslinger63} L. S. Kisslinger and R. A. Sorensen, Rev. Mod. Phys.
{\bf 35}, 853 (1963).

\bibitem{Bes69} D. R. Bes and R. A. Sorensen, `Advances in Nuclear Physics'
(Plenum Press) vol. 2, 129 (1969).

\bibitem{Dufour96} M. Dufour and A. P. Zuker, Phys. Rev. C {\bf 54}, 1641 (1996).

\bibitem{Hasegawa01} M. Hasegawa, K. Kaneko, and S. Tazaki, Nucl.
Phys. A {\bf 688}, 765 (2001).

\bibitem{Kaneko08} K. Kaneko, Y. Sun, M. Hasegawa, and T. Mizusaki,
 Phys. Rev. C {\bf 78}, 064312 (2008).

\bibitem{Kaneko11} K. Kaneko, Y. Sun, M. Hasegawa, and T. Mizusaki,
 Phys. Rev. C {\bf 83}, 014320 (2011).

\bibitem{Jin11} H. Jin, M. Hasegawa, S. Tazaki, K. Kaneko, and Y. Sun, Phys. Rev.
C {\bf 84}, 044324 (2011).

\bibitem{Wang13} H.-K. Wang, Y. Sun, H. Jin, K. Kaneko, and S. Tazaki,
Phys. Rev. C {\bf 88}, 054310 (2013).

\bibitem{Wang14} H.-K. Wang, K. Kaneko, and Y. Sun, Phys. Rev. C {\bf 89}, 064311 (2014).

\bibitem{Wang15} H.-K. Wang, K. Kaneko, and Y. Sun, Phys. Rev. C {\bf 91}, 021303(R) (2015).

\bibitem{Otsuka01} T. Otsuka, R. Fujimoto, Y. Utsuno, B. A. Brown, M. Honma, and
T. Mizusaki, Phys. Rev. Lett. {\bf 87}, 082502 (2001).

\bibitem{Otsuka05} T. Otsuka, T. Suzuki, R. Fujimoto, H. Grawe, and Y. Akaishi,
Phys. Rev. Lett. {\bf 95}, 232502 (2005).

\bibitem{Otsuka10b} T. Otsuka, T. Suzuki, M. Honma, Y. Utsuno, N. Tsunoda,
K. Tsukiyama, and M. Hjorth-Jensen, Phys. Rev. Lett. {\bf 104}, 012501 (2010).

\bibitem{Kaneko14} K. Kaneko, T. Mizusaki, Y. Sun, and S. Tazaki,
 Phys. Rev. C {\bf 89}, 011302(R) (2014).

\bibitem{Poves81} A. Poves and A. Zuker, Phys. Rep. {\bf 70}, 235 (1981).

\bibitem{Zuker03} A. P. Zuker, Phys. Rev. Lett. {\bf 90}, 042502 (2003).

\bibitem{Sieja12} K. Sieja and F. Nowaki, Phys. Rev. C {\bf 85}, 051301(R) (2012).

\bibitem{Otsuka10a} T. Otsuka, T. Suzuki, J. D. Holt, A. Schwenk, and Y. Akaishi,
Phys. Rev. Lett. {\bf 105}, 032501 (2010).

\bibitem{Yuan12} C. Yuan, T. Suzuki, T. Otsuka, F. Xu, and N. Tsunoda, Phys.
 Rev. C {\bf 85}, 064324 (2012).

\bibitem{Utsuno12} Y. Utsuno, T. Otsuka, B. A. Brown, M. Honma, T. Mizusaki,
 and N. Shimizu, Phys. Rev. C {\bf 86}, 051301 (2012).

\bibitem{Togashi15} T. Togashi, N. Shimizu, Y. Utsuno, T. Otsuka, and M. Honma,
Phys. Rev. C {\bf 891}, 024320 (2015).

\bibitem{Bansal64} R. K. Bansal and J. B. French, Phys. Lett. {\bf 11}, 145 (1964).

\bibitem{Mizusaki} T. Mizusaki, N. Shimizu, Y. Utsuno, and M. Honma,
code MSHELL64 (unpublished).

\bibitem{ENDSF} Data extracted using the NNDC World Wide Web site from
the ENDSF data base.

\bibitem{Audi03} G. Audi, A. H. Wapstra, and C. Thibault, Nucl. Phys. A
{\bf 729}, 337 (2003).

\bibitem{Cole99} B. J. Cole, Phys. Rev. C {\bf 59}, 726 (1999).

\bibitem{Vingerhoets10} P. Vingerhoets, K. T. Flanagan, M. Avgoulea, J. Billowes,
M. L. Bissell, K. Blaum, B. A. Brown, B. Cheal, M. De Rydt, D. H.
Forest, Ch. Geppert, M. Honma, M. Kowalska, J. Kr\"amer, A. Krieger,
E. Man\'e, R. Neugart, G. Neyens, W. N\"ortersh\"auser, T. Otsuka,
M. Schug, H. H. Stroke, G. Tungate, and D. T. Yordanov, Phys. Rev. C
{\bf 82}, 064311 (2010).

\bibitem{Cheal10} B. Cheal, E. Man\'e, J. Billowes, M. L. Bissell, K. Blaum,
B. A. Brown, F. C. Charlwood, K. T. Flanagan, D. H. Forest, C.
Geppert, M. Honma, A. Jokinen, M. Kowalska, A. Krieger, J. Kr\"amer,
I. D. Moore, R. Neugart, G. Neyens, W. N\"ortersh\"auser, M. Schug,
H. H. Stroke, P. Vingerhoets, D. T. Yordanov, and M. Z\'akov\'a,
Phys. Rev. Lett. {\bf 104}, 252502 (2010).

\bibitem{Bender06} M. Bender, P. Bonche, and P.-H. Heenen,
Phys. Rev. C {\bf 74}, 024312 (2006).

\bibitem{Hase07} M. Hasegawa, K. Kaneko, T. Mizusaki, and Y. Sun,
Phys. Lett. {\bf B 656}, 51 (2007).

\bibitem{Stefanescu04} I. Stefanescu, J. Eberth, G. de Angelis, N. Warr,
G. Gersch, T. Steinhardt, O. Thelen, D. Weisshaar, T. Martinez, A.
Jungclaus, R. Schwengner, K. P. Lieb, E. A. Stefanova, D. Curien,
and A. Gelberg, Phys. Rev. C {\bf 69}, 034333 (2004).

\bibitem{Howe12} A. R. Howe, R. A. Haring-Kaye, J. D\"oring, N. R. Baker,
S. J. Kuhn, S. L. Tabor, S. R. Arora, J. K. Bruckman, and C. R.
Hoffman, Phys. Rev. C {\bf 86}, 014328 (2012).

\bibitem{Loritz99} R. Loritz, O. Iordanov, E. Galindo, A. Jungclaus,
D. Kast, K. P. Lieb, C. Teich, F. Cristancho, Ch. Ender, T.
H\"artlein, F. K\"ock, D. Schwalm, Eur. Phys. J. A. {\bf 6}, 257
(1999).

\bibitem{Kaplan91} M. S. Kaplan, J. X. Saladin, D. F. Winchell, H. Takai,
and J. Dudek, Phys. Rev. C {\bf 44}, 668 (1991).

\end{thebibliography}
\end{document}